\documentclass[12pt]{article}
\pdfoutput=1

\usepackage{putex}
\usepackage{graphicx}
\usepackage{caption}
\usepackage{amsmath}
\usepackage{amssymb,bm}
\usepackage{array}
\usepackage{multirow}
\usepackage{mathtools}
\usepackage{comment}
\usepackage{subcaption}
\usepackage{epstopdf}
\usepackage{enumerate}
\usepackage{cite}
\usepackage{youngtab}
\usepackage{tensor}
\usepackage{slashed}
\usepackage[aligntableaux=center]{ytableau}
\usepackage{textcomp}
\usepackage[utf8]{inputenc}
\usepackage{rotating}
\usepackage{bigfoot}
\usepackage[
colorlinks=true,
linkcolor=blue,
urlcolor=blue,
filecolor=black,
citecolor=red,
linktocpage=true
]{hyperref}

\usepackage{tikz-cd}
\usepackage{tikz}

\usetikzlibrary{calc}
\usetikzlibrary{snakes}
\usetikzlibrary{arrows}
\usetikzlibrary{positioning}
\usetikzlibrary{decorations.pathmorphing}
\usetikzlibrary{ decorations.markings}
\tikzset{snake it/.style={decorate, decoration={snake, segment length=2mm, amplitude=1mm}}}
\usetikzlibrary{shapes.misc}
\tikzset{cross/.style={cross out, draw=black, minimum size=2*(#1-\pgflinewidth), inner sep=0pt, outer sep=0pt},
	cross/.default={1pt}}
\usetikzlibrary{shapes.geometric}
\definecolor{bleudefrance}{rgb}{0.19, 0.55, 0.91}
\definecolor{candyapplered}{rgb}{1.0, 0.03, 0.0}

\newcommand{\abs}[1]{\left\lvert #1 \right\rvert}

\newcommand {\be} {\begin {equation}}
\newcommand {\ee} {\end {equation}}

\newcommand {\bes} {\begin {equation*}}
\newcommand {\ees} {\end {equation*}}

\newcommand{\es}[2]{%
	\begin{equation}
		\begin{aligned}
			#2
		\end{aligned}
		\phantomsection\label{#1}%
	\end{equation}%
}

\newcommand{\Z}{\mathbb{Z}}

\newcommand{\R}{\mathbb{R}}
\newcommand{\C}{\mathbb{C}}

\newcommand{\cA}{{\mathcal A}}
\newcommand{\cB}{{\mathcal B}}
\newcommand{\cC}{{\mathcal C}}
\newcommand{\cD}{{\mathcal D}}

\newcommand{\cF}{{\mathcal F}}
\newcommand{\cG}{{\mathcal G}}

\newcommand{\cN}{{\mathcal N}}
\newcommand{\cO}{{\mathcal O}}
\newcommand{\cP}{{\mathcal P}}

\newcommand{\cM}{{\mathcal M}}

\newcommand{\cZ}{{\mathcal Z}}

\def\Tr{\mop{Tr}}

\newcommand{\bea}{\begin{equation}\begin{aligned}}
		\newcommand{\eea}[1]{\label{#1}\end{aligned}\end{equation}}

\newcommand{\beq}{\begin{equation}}
	\newcommand{\eeq}{\end{equation}}

\newcommand\ep{\epsilon}

\def\ie{\begin{equation}\begin{aligned}}
		\def\fe{\end{aligned}\end{equation}}

\newcommand{\la}{\langle}
\newcommand{\ra}{\rangle}

\newcommand{\m}{\mu}
\newcommand{\n}{\nu}

\newcommand{\A}{{\alpha}}
\newcommand{\B}{{\beta}}
\newcommand{\D}{{\delta}}

\newcommand{\mC}{{\mathbb C}}

\newcommand{\mZ}{{\mathbb Z}}
\newcommand{\mR}{{\mathbb R}}

\newcommand{\mf}{\mathfrak }

\numberwithin{equation}{section}

\def\<{\langle}
\def\>{\rangle}

\begin{document}
	
	\preprint{PUPT-2651}

	\institution{Imp}{Blackett Laboratory, Imperial College, Prince Consort Road, London, SW7 2AZ, U.K.}
	\institution{PU}{Joseph Henry Laboratories, Princeton University, Princeton, NJ 08544, USA}
	\institution{PCTS}{Princeton Center for Theoretical Science, Princeton University, Princeton, NJ 08544, USA}
	\institution{IAS}{Institute for Advanced Study, Princeton, NJ 08540, USA}
	\institution{NYU}{Center for Cosmology and Particle Physics, New York University, New York, NY 10003, USA}
	\institution{HU}{Jefferson Physical Laboratory, Harvard University, Cambridge, MA 02138, USA}
	
	\title{ Bootstrapping M-theory Orbifolds  }

	\authors{
		Shai M. Chester,\worksat{\Imp}\footnote{e-mail: {\tt iahs81@gmail.com}}
		Silviu S. Pufu,\worksat{\PU,\PCTS,\IAS}\footnote{e-mail: {\tt spufu@princeton.edu}}
		Yifan Wang,\worksat{\NYU}\footnote{e-mail: {\tt yw6417@nyu.edu}}
		and
		Xi Yin\worksat{\HU}\footnote{e-mail: {\tt xiyin@fas.harvard.edu}}}

	\abstract{We analyze correlation functions of $SU(k) \times SU(2)_F$ flavor currents in a family of three-dimensional ${\cal N}=4$ superconformal field theories, combining analytic bootstrap methods with input from supersymmetric localization. Via holographic duality, we extract gluon and graviton scattering amplitudes of M-theory on ${\rm AdS}_4\times S^7/\mathbb{Z}_k$ which contains a $\mathbb{C}^2/\mathbb{Z}_{k}$ orbifold singularity. From these results, we derive aspects of the effective description of M-theory on the orbifold singularity beyond its leading low energy limit. We also determine a threshold correction to the holographic correlator, which vanishes due to equal and opposite contributions from the two-loop gluon and the tree-level bulk graviton exchange.
	}
	\date{}

	\maketitle
	
	\tableofcontents
	
	\section{Introduction}
	\label{intro}
	
	M-theory is believed to be an ultraviolet (UV)-complete theory of quantum gravity that reduces to eleven-dimensional supergravity in its low-energy limit \cite{Witten:1995ex}.  Starting with the work of Ho\v{r}ava and Witten \cite{Horava:1995qa,Horava:1996ma}, it was recognized that M-theory also admits spacetimes with certain mild singularities, where, typically, the singular loci carry additional light degrees of freedom.   A major, still unanswered, question is to classify possible singularities admitted by M-theory, as well as to understand the dynamics of their excitations.  This question goes beyond the low-energy supergravity limit, and it is also not accessible in string perturbation theory. Presently, the only potentially systematic approach to it is through holographic dualities.  
	
	In this paper, we will focus on codimension-four orbifold singularities, particularly M-theory in the spacetime $\mathbb{C}^2/\mathbb{Z}_k \times \R^{1,6}$, which is known to support $SU(k)$ non-Abelian gauge fields at the seven-dimensional orbifold singularity locus \cite{Acharya:1998pm,Acharya:2004qe, Anderson:2006pb}.  We will develop a framework for extracting the dynamics of these excitations through a family of three-dimensional ${\cal N}=4$ superconformal gauge theories, whose holographic duals involve M-theory backgrounds that contain such orbifold singularities \cite{Ferrara:1998vf,Gomis:1998xj,Entin:1998ub,Pelc:1999ms,Cherkis:2002ir}.  
	
	Our starting point will be the holographic duality \cite{Maldacena:1997re,Gubser:1998bc,Witten:1998qj} between M-theory in the spacetime ${\rm AdS}_4 \times S^7 / \Z_k$, where the $\Z_k$ action fixes an $S^3 \subset S^7$\@,\footnote{This is different from the fixed-point-free $\Z_k$ quotient of $S^7$ that appears in the holographic dual of Aharony-Bergman-Jafferis-Maldacena (ABJM) theory \cite{Aharony:2008ug}. See \cite{Pelc:1999ms,Cherkis:2002ir} for the precise supergravity background in M-theory and also its reduction in type IIA string theory (where the 11d $\mZ_k$ orbifold singularity reduces to a stack of $k$ D6-branes).} and the three-dimensional ${\cal N} = 4$ superconformal field theory (SCFT) that is reached in the infrared (IR) limit of a $U(N)$ gauge theory coupled to an adjoint hypermultiplet and $k$ fundamental hypermultiplets \cite{Benini:2009qs,Bashkirov:2010kz}. As is usually the case in the anti-de Sitter / Conformal Field Theory (AdS/CFT) correspondence, in the large $N$ limit, single-trace operators in the field theory are dual to fluctuation modes of the M-theory background.  Of particular interest are the $SU(k)$ flavor symmetry currents $j_\mu^a$, with $a=1, \ldots, k^2-1$, which act nontrivially on the fundamental hypermultiplets and are dual to massless ``gluons'' in AdS localized at the 7-dimensional singular locus (namely the ${\rm AdS}_4 \times S^3$ that is fixed by the $\Z_k$ action), and the $SU(2)_F$ flavor symmetry currents ${\widetilde j}_\mu^A$, with $A=1,2,3$, which act nontrivially on the adjoint hypermultiplet and are dual to a particular set of graviton and three-form modes in the bulk ${\rm AdS}_4 \times S^7 / \Z_k$ spacetime (see Table~\ref{tab:fieldsops} for details).
	
	The M-theory orbifold spacetime $\C^2 / \Z_k \times \R^{1,6}$ can be recovered from the large radius limit of the ${\rm AdS}_4 \times S^7 / \Z_k$ background. In particular, certain scattering amplitudes in $\C^2 / \Z_k \times \R^{1,6}$, involving modes supported both at the orbifold singularity as well as in the bulk, can be obtained from the large radius limit of the Mellin amplitudes in the dual CFT, which are related to correlation functions via the Mellin transform \cite{Mack:2009mi,Mack:2009gy}, through the prescription of \cite{Penedones:2010ue}.
	For example, from the four-point function of the $SU(k)$ flavor currents $j^a_\mu$, one can extract the four-point scattering amplitude of the gluons $g^a$ living on the $\R^{1,6}$ located at the orbifold singularity locus:
	\es{JJJJ}{
		\langle j_\mu^a(\vec{x}_1) j_\nu^b(\vec{x}_2) j_\rho^c(\vec{x}_3) j_\sigma^d(\vec{x}_4)  \rangle 
		\xrightarrow[\text{limit}]{\text{flat space}} {\cal A}(g^ag^bg^cg^d) \,,
	}
	where the scattering amplitude depends on the momenta and polarizations of the gluons, which originate from the kinematic configurations in the CFT correlator.  
	
	We can also consider the correlation function of a pair of $SU(k)$ flavor currents $j^a_\mu$ with a pair of $SU(2)_F$ flavor currents $\widetilde j_\mu^A$.  From the mixed four-point function of $SU(k)$ and $SU(2)_F$ currents, one can extract, in a way to be made precise in Section~\ref{largeN}, the scattering amplitude of two 7d gluons $g$ and two 11d gravitons $h$:\footnote{The flat space limit for CFT operators with spin has not been worked out explicitly, and therefore we will not use it.  Instead, we will make use of supersymmetry to relate the conserved currents to scalar operators in the same multiplets, for which one can use the flat space limit formula of \cite{Penedones:2010ue}.}${}^{,}$\footnote{Note that in the limiting procedure of \eqref{JJJJ} and \eqref{JJTT}, the 11d momenta of the gluons and gravitons are restricted to lie within the 4d subspace of the ${\rm AdS}_4$ directions before taking the limit.  For the four-point amplitude of four 7d gluons, such a restriction can be made without loss of generality, so one can then fully reconstruct the full amplitude ${\cal A}(g^ag^bg^cg^d)$.  For the amplitude of two gluons and two gravitons, this is indeed a restricted kinematic configuration, so from \eqref{JJTT} one cannot reconstruct the full amplitude ${\cal A}(g^ag^b hh)$.}
	\es{JJTT}{
		\langle j_\mu^a(\vec{x}_1) j_\nu^b(\vec{x}_2) \widetilde j_\rho^A (\vec{x}_3) \widetilde j_\sigma^B (\vec{x}_4)  \rangle 
		\xrightarrow[\text{limit}]{\text{flat space}} {\cal A}(g^ag^b hh) \,.
	}

	The essential step in implementing this program is, of course, computing the SCFT correlation functions.  While the aforementioned SCFT can be realized as the (strongly coupled) infrared limit of a Lagrangian gauge theory, thus far the Lagrangian description has had limited use in the determination of infrared observables, except for a class of supersymmetric observables that can be computed exactly along the Renormalization Group (RG) flow  \cite{Kapustin:2009kz,Jafferis:2010un,Dedushenko:2016jxl,Dedushenko:2017avn,Dedushenko:2018icp}.  In particular, there are two distinct integrated 4-point functions of conserved current operators  $j_\mu^a(\vec{x})$ in a 3d ${\cal N} = 4$ SCFT,\footnote{For 3d ${\cal N} = 4$ SCFTs, the supeconformal algebra has two distinct representations that contain conserved currents (see, for instance, \cite{Cordova:2016emh}).  The statement that follows assumes that all four operators belong to multiplets of the same type, which is indeed the case for the $SU(k)$ and $SU(2)_F$ currents in our theory of interest (also known as the Higgs branch type).  Note, however, that our theory is also invariant under a $U(1)$ topological symmetry that we do not consider here, and the associated $U(1)$ current multiplet is of the opposite type (also known as the Coulomb branch type).}  schematically of the form
	\es{IntCorr}{
		\int \left( \prod_{i=1}^4 d^3 \vec{x}_i\right) \mu^{\mu\nu\rho\sigma}_{abcd} (\vec{x}_1, \vec{x}_2, \vec{x}_3, \vec{x}_4) \, 
		\langle j_\mu^a(\vec{x}_1) j_\nu^b(\vec{x}_2) j_\rho^c (\vec{x}_3) j_\sigma^d (\vec{x}_4)  \rangle 
	} 
	for suitable measure factors $\mu$, that can be determined from supersymmetric localization.  See \cite{Binder:2018yvd,Binder:2019jwn,Chester:2020dja,Binder:2019mpb,Binder:2021cif,Chester:2022sqb} for analogous examples of such integrated correlators in supersymmetric gauge theories in various dimensions.
	
	On the other hand, the SCFT correlators at sufficiently generic kinematic configurations, from which we wish to extract scattering amplitudes in the large radius limit of the holographic dual, cannot be determined from supersymmetric localization alone.   Additional ingredients, such as crossing equations and analytic properties of the corresponding Mellin amplitudes, have recently been combined with an extensive set of supersymmetric localization results to constrain holographic correlators and extract M-theory or string theory observables beyond the supergravity limit \cite{Chester:2018aca,Binder:2019jwn,Binder:2018yvd,Binder:2019mpb,Chester:2020dja,Chester:2019jas,Chester:2020vyz,Binder:2020ckj,Behan:2023fqq,Alday:2022rly,Alday:2021vfb,Alday:2021ymb,Chester:2019pvm}.\footnote{Similar localization constraints \cite{Chester:2020jay,Chester:2021gdw} have also been used to constrain higher derivative corrections to the effective supergravity action \cite{Bobev:2020egg}.} The same strategy will be adopted in this paper to extract the dynamics of gluons at the M-theory orbifold singularity.
	
	One of our main results is that the $\Tr F^4$ and $(\Tr F^2)^2$ couplings, whose precise definitions will be addressed in Section~\ref{sec:amps},\footnote{We will adopt a scheme-independent definition of such effective couplings through particular terms in the momentum expansion of the four-gluon amplitude, which are unambiguously separated from the non-analytic terms in momenta.} which are a priori allowed in a 7d supersymmetric effective gauge theory, are absent for the gluons supported at the $\C^2 / \Z_k \times \R^{1,6}$ orbifold singularity of M-theory. This is reminiscent of the absence of $D^4 R^4$ term in the bulk M-theory (graviton) effective action, recently extracted from correlators of ABJM theory \cite{Binder:2018yvd}. Let us note that, previously, the bulk BPS-protected higher derivative couplings had also been determined through a different chain of logic, namely, by considering toroidal compactification of M-theory, and constraining the moduli dependence thereof via supersymmetry non-renormalization theorems \cite{Green:1998by, Green:2005ba, Wang:2015jna, Wang:2015aua}. A similar argument applies to the Abelian effective gauge theory on the Coulomb branch of the M-theory orbifold, where the moduli dependence of the BPS-protected Abelian $F^4$ effective coupling can be determined \cite
	{Paban:1998ea,Sethi:1999qv,Maxfield:2012aw,Lin:2015ixa}. The relation between this Abelian $F^4$ coupling, which becomes singular at the origin of the Coulomb branch, and the non-Abelian $\Tr F^4$ and $(\Tr F^2)^2$ couplings determined in our holographic approach, will be discussed in Section \ref{sings}.
	
	Going in the other direction, we can also combine what we know about the flat space scattering amplitude with supersymmetric localization data in the holographic gauge theory to learn more about the holographic correlators, or equivalently the Mellin amplitudes, in AdS\@. 
	In particular, we will determine a certain logarithmic threshold contribution to the large $N$ expansion of the correlator \eqref{JJJJ} coming from two-loop gluon exchange and tree-level bulk graviton exchange combined. We find that the contributions of each of these two terms is equal and opposite, so that the logarithmic threshold in fact vanishes.

	The rest of this paper is organized as follows.  In Section~\ref{sings}, we review the construction of the $\C^2 / \Z_k$ orbifold background of M-theory, the general structure of the scattering amplitudes involving the gluons supported at the orbifold singularity, and supersymmetry constraints for the effective theory on the Coulomb branch where the orbifold singularity is resolved. We also review the AdS counterpart, which we refer to as the AdS orbifold, and its holographic dual 3d ${\cal N}=4$ SCFT\@. In Section~\ref{4point}, we describe the kinematic structure and the superconformal block expansion of the correlators for the flavor current multiplets of interest. The structure of the large radius expansion of the corresponding Mellin amplitudes and its connection to the flat space amplitudes are analyzed in Section~\ref{largeN}. Sections~\ref{loc} and \ref{sec:backtom} contain the key technical results of this paper, namely the supersymmetric localization and the subsequent determination of (the absence of) the four-derivative effective couplings in the M-theory orbifold, as well as the determination of threshold contributions to the Mellin amplitude. We conclude in Section~\ref{conc} with a discussion of future directions. Some details concerning 7d superamplitudes, the superconformal blocks, the algorithm of extracting OPE data from the Mellin amplitudes, and matrix model technique used to extract supersymmetric localization data, are presented in the Appendices.

	\section{The M-theory orbifold}
	\label{sings}

	\begin{figure}[!htb]
		\centering
		\scalebox{.8}{ 	\begin{tikzpicture} 
				\filldraw [color=bleudefrance, thick, fill=bleudefrance!30] (0,0) -- (0,6) -- (2,8) -- (2,2) -- (0,0); 
				\fill[
				top color=red!50,
				bottom color=red!10,
				shading=axis,
				opacity=0.25
				] 
				(5,4) circle (.4 and 1);
				\fill[
				left color=pink!10!red,
				right color=pink!50!red,,
				middle color=pink!50,
				shading=axis,
				opacity=0.25
				] 
				(5,5) -- (1,4) -- (5,3) arc (-90:90:.4 and 1);
				\node [label={$\mathbb{R}^{1,6}$}]   at (1,5) {};
				\node [label={$\mathbb{C}^2/\mathbb{Z}_k$}]   at (4,3.8) {};
			\end{tikzpicture}
		}
		\caption{A poor man's portrait of the $\mC^2/\mZ_k$ orbifold singularity in M-theory.}
		\label{fig:orbifoldsing}
	\end{figure}
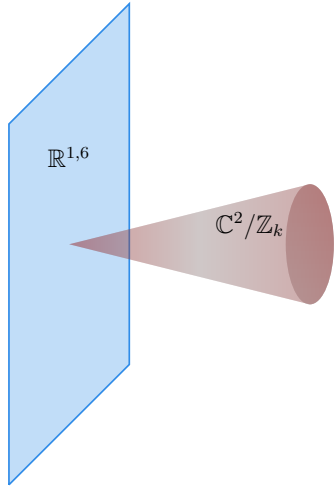

	The $\mathbb{C}^2/\mathbb{Z}_k$ or $A_{k-1}$ orbifold is the quotient space of $\mathbb{C}^2$, parametrized  by complex coordinates $(z_1, z_2)$, with respect to the discrete group action
	\ie
	(z_1, z_2) \mapsto (e^{2\pi i n/k} z_1, e^{-{2\pi i n/k}} z_2)\,, \qquad n=0,1,\ldots, k-1 \,.
	\label{Zkorb}
	\fe
	M-theory in the orbifold spacetime $\mathbb{C}^2/\mathbb{Z}_k \times \mathbb{R}^{1,6} $ (portrayed schematically in Figure \ref{fig:orbifoldsing}) may be approached either as the strong coupling limit of type IIA string theory in the orbifold background, which admits (at least in the weak coupling regime) a well-known worldsheet CFT description \cite{Dixon:1985jw, Dixon:1986jc}, or as a geometrically singular limit of M-theory in the smooth ALE spacetime \cite{Acharya:1998pm,Acharya:2004qe}, namely $\mathbb{R}^{1,6}$ times the four-manifold equipped with the metric
	\ie\label{alemetric}
	& ds_4^2 = U(\vec x) d\vec x^2 + U(\vec x)^{-1} (dy + \vec\omega\cdot d\vec x) \,,
	\\
	& U(\vec x) = \sum_{i=1}^k {1\over |\vec x-\vec x_i|} \,, \qquad  \nabla\times\vec\omega = \nabla U \,,
	\fe
	where $\vec x\in \mathbb{R}^3$, $y\sim y+4\pi$, in the limit where $\vec x_1,\ldots, \vec x_k$ collide. The agreement of these two different descriptions and the existence of the singular orbifold as an M-theory background are highly nontrivial and lie at the foundation of a wealth of constructions of M-theory vacua \cite{Acharya:1998pm,Acharya:2004qe,Anderson:2006pb}.

	\subsection{Effective theory and scattering amplitudes}
	\label{sec:amps}
	
	The massless degrees of freedom of M-theory supported at the orbifold singularity are expected to be those of $SU(k)$ gluons and their superpartners \cite{Acharya:1998pm,Acharya:2004qe}, governed in the low-energy limit by the 7-dimensional ${\cal N}=1$ super-Yang-Mills theory coupled to the bulk 11d supergravity. The bulk orbifold spacetime $M_{11}^{(k)}$ can be represented as a quotient $\widetilde M_{11}/\mathbb{Z}_k$, such that the metric $G_{MN}$ lifts to a smooth $\mathbb{Z}_k$-invariant metric on the covering space $\widetilde M_{11}$. The 11d supergravity action takes the form
	\es{11dR}{
		S_\text{grav}[G_{MN},\ldots] = \int_{M_{11}^{(k)}} d^{11}x\, \sqrt{-G}\left( \frac{R}{2\kappa^2}+ {\rm SUSY\,completion} \right)\,,
	}
	where $\kappa$ is related to the 11d Planck length $\ell_p$ by the convention $2\kappa^2=(2\pi)^8\ell_p^9$. 
	The 7d $SU(k)$ gauge theory action takes the form of ${\cal N}=1$ super-Yang-Mills covariantized with respect to the restriction $g_{\mu\nu}$ of the bulk metric $G_{MN}$ to the 7d orbifold singularity locus $M_7$, 
	\es{7d}{
		S_{\rm SYM}[A_\mu,\ldots; G_{MN},\ldots]=\int_{M_7} d^7 x\,  \sqrt{-g}\left(  -{1\over 2g_{\rm YM}^2} \Tr F_{\m\n}F^{\m\n} + \cdots \right)\,,
	}
	where the Lie algebra trace $\Tr$ is defined by the one in the fundamental representation (i.e. $\Tr\equiv \Tr_{\rm fund}$) throughout the paper and 
	the Yang-Mills coupling $g_{\rm YM}$ is related to the 11d Planck length $\ell_p$ by $g_{\rm YM}^2=2(2\pi)^4 \ell_p^3$.\footnote{This can be determined by the identification of the 2+1 dimensional Yang-Mills instanton-membranes with M2-branes located at the orbifold singularity. By matching the instanton tension with that of the M2-brane, $T_{\rm M2} = T_{\rm inst}$, where $T_{\rm M2}={1\over (2\pi)^2\ell_p^3}$, and $T_{\rm inst}={8\pi^2\over g_{\rm YM}^2}$. 
	}  
	
	It is nontrivial to construct the supersymmetric completion $S_{\rm SYM}$ of the covariantized 7d Yang-Mills action  (see \cite{Anderson:2006pb}). As is familiar in the context of brane-bulk effective theories \cite{Horava:1996ma, Goldberger:2001tn, Michel:2014lva}, a fully consistent treatment of the coupling between the localized and the bulk degrees of freedom is only expected at the quantum level.  However, note that in contrast to typical brane-bulk effective theories, the coupling between the localized degrees of freedom at the orbifold singularity and the bulk ones does not admit non-linearly realized translation symmetries in the transverse directions, and there is no backreaction in the form of brane tension. 
	
	In principle, one may formulate the effective theory of the M-theory orbifold in a Wilsonian framework, by working with a Wilsonian effective action that contains an infinite series of higher derivative corrections to $S_{\rm SYM} + S_{\rm grav}$, along with a UV cutoff scheme. In practice, it can be exceedingly complicated to implement gauge invariance and supersymmetry constraints on the effective Lagrangian beyond the leading order in the derivative expansion. It is often more convenient to instead formulate the effective theory in terms of the small momentum expansion of scattering amplitudes, where gauge redundancies are eliminated, and supersymmetry constraints are realized linearly through Ward identities \cite{Wang:2015jna, Lin:2015ixa, Wang:2015aua}. Loosely speaking, the analytic terms in the momentum expansion of the amplitudes are in correspondence with the higher-derivative couplings in the effective Lagrangian, while the non-analytic terms of the momentum expansion are constrained by unitarity and locality. We will primarily adopt the latter approach in this paper, and we will organize the effective couplings of the M-theory orbifold in terms of the on-shell supervertices of the localized gluons and the bulk gravitons.
	
	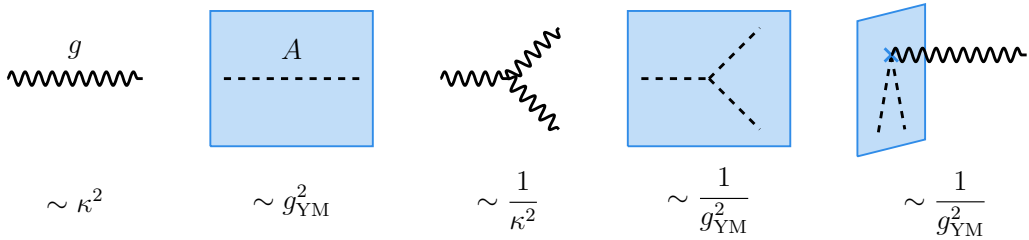
\begin{figure}[!htb]
		\centering 
		\scalebox{.9}{	\begin{tikzpicture}[baseline={(0,0)}]
				\node[above] at (0,0.15) {$\displaystyle g$};
				\draw [very thick, snake it] (-1,0) -- (1,0);
				\node at (0,-1.8) {$\displaystyle \sim \kappa^2$};
			\end{tikzpicture} 
			~~~~~
			\begin{tikzpicture}[baseline={(0,0)}]
				\filldraw [color=bleudefrance,thick, fill=bleudefrance!30] (-1.2,-1) -- (-1.2,1) -- (1.2,1) -- (1.2,-1) -- (-1.2,-1);
				\node[above] at (0,0.15) {$\displaystyle A$};
				\draw [very thick, dashed] (-1,0) -- (1,0);
				\node at (0,-1.8) {$\displaystyle \sim {g_{\rm YM}^2}$};
			\end{tikzpicture} 
			~~~~~
			\begin{tikzpicture}[baseline={(0,0)}]
				\draw [very thick, snake it] (-1,0) -- (0,0);
				\draw [very thick, snake it] (0,0) -- (0.75,0.75);
				\draw [very thick, snake it] (0,0) -- (0.75,-0.75);
				\node at (0,-1.8) {$\displaystyle \sim \frac{1}{\kappa^2}$};
			\end{tikzpicture} 
			~~~~~
			\begin{tikzpicture}[baseline={(0,0)}]
				\filldraw [color=bleudefrance,thick, fill=bleudefrance!30] (-1.2,-1) -- (-1.2,1) -- (1.2,1) -- (1.2,-1) -- (-1.2,-1);
				\draw [very thick, dashed] (-1,0) -- (0,0);
				\draw [very thick,  dashed] (0,0) -- (0.75,0.75);
				\draw [very thick,  dashed] (0,0) -- (0.75,-0.75);
				\node at (0,-1.8) {$\displaystyle \sim \frac{1}{g_{\rm YM}^2}$};
			\end{tikzpicture} 
			~~~~~
			\begin{tikzpicture}[baseline={(0,-.35)}]
				\filldraw [color=bleudefrance,, thick, fill=bleudefrance!30] (-1.5,-1.5) -- (-1.5,0.5) -- (-0.5,0.75) -- (-0.5,-1.25) -- (-1.5,-1.5);
				\draw [very thick, dashed] (-1,0) -- (-1.2,-1.2);
				\draw [very thick, dashed] (-1,0) -- (-0.8,-1.1);
				\node[cross=4pt, color=bleudefrance,, very thick] at (-1,0) {};
				\draw [very thick, snake it] (-1,0) -- (1,0);
				\node at (-.2,-2.2) {$\displaystyle \sim \frac{1}{g_{\rm YM}^2}$};
			\end{tikzpicture} 
		}
		\caption{Propagators and minimal coupling vertices for the gluons localized at the singularity (represented as the blue-shaded subspace) and the bulk graviton. In a Wilsonian effective theory, there are infinitely many more higher derivative vertices, e.g.~those of Figures \ref{fig:higherderAAAA} and \ref{fig:higherderAAgg}.}
		\label{fig:Feynmanrules}
	\end{figure}
	
	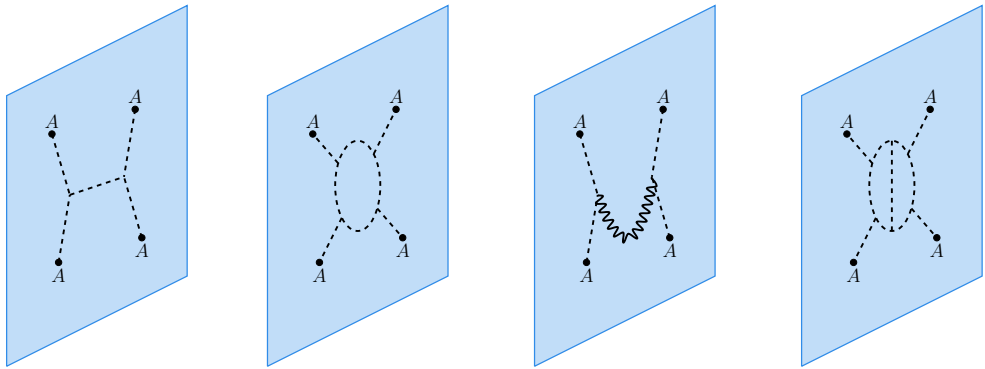
\begin{figure}[!htb]
		\centering
		\scalebox{.6}{	\begin{minipage}[b]{0.2\textwidth}
				\centering
				\begin{tikzpicture} 
					\filldraw [color=bleudefrance, thick, fill=bleudefrance!30] (0,0) -- (0,6) -- (4,8) -- (4,2) -- (0,0); 
					\coordinate[label={$A$}] (A2)  at ($(2,4)+1.2*(45: 1 and 2)$) []{};
					\coordinate[label={$A$}]  (B2) at ($(2,4)+1.15*(150: 1 and 2)$) {};
					\coordinate [label={below:$A$}]  (C2) at ($(2,4)+1.2*(225: 1 and 2)$) {};
					\coordinate [label={below:$A$}]  (D2) at ($(2,4)+1.15*(330: 1 and 2)$) {};
					\coordinate (V1) at ($(2,4)+(0.6,0.2)$) {};
					\coordinate (V2) at ($(2,4)-(0.6,0.2)$) {};
					\draw [very thick,dashed] (A2)--(V1);
					\draw [very thick,dashed] (D2)--(V1);
					\draw [very thick,dashed] (B2)--(V2);
					\draw [very thick,dashed] (C2)--(V2);
					\draw [very thick,dashed] (V2)--(V1);
					\filldraw (A2) circle (2pt);
					\filldraw (B2) circle (2pt);
					\filldraw (C2) circle (2pt);
					\filldraw (D2) circle (2pt);
				\end{tikzpicture}
			\end{minipage}\qquad\qquad\qquad
			\begin{minipage}[b]{0.2\textwidth}
				\centering
				\begin{tikzpicture} 
					\filldraw [color=bleudefrance, thick, fill=bleudefrance!30] (0,0) -- (0,6) -- (4,8) -- (4,2) -- (0,0); 
					\draw [very thick, dashed] (2,4) ellipse (0.5 and 1); 
					\coordinate [] (A1) at ($(2,4)+(45: 0.5 and 1)$) {};
					\coordinate [label={$A$}] (A2)  at ($(2,4)+1.2*(45: 1 and 2)$) []{};
					\coordinate [] (B1) at ($(2,4)+(150: 0.5 and 1)$) {};
					\coordinate [label={$A$}]  (B2) at ($(2,4)+1.15*(150: 1 and 2)$) {};
					\coordinate[] (C1) at ($(2,4)+(225: 0.5 and 1)$) {};
					\coordinate[label={below:$A$}]  (C2) at ($(2,4)+1.2*(225: 1 and 2)$) {};
					\coordinate [] (D1) at ($(2,4)+(330:0.5 and 1)$) {};
					\coordinate[label={below:$A$}]  (D2) at ($(2,4)+1.15*(330: 1 and 2)$) {};
					\draw [very thick,dashed] (A1)--(A2);
					\draw [very thick,dashed] (B1)--(B2);
					\draw [very thick,dashed] (C1)--(C2);
					\draw [very thick,dashed] (D1)--(D2);
					
					\filldraw (A2) circle (2pt);
					\filldraw (B2) circle (2pt);
					\filldraw (C2) circle (2pt);
					\filldraw (D2) circle (2pt);
				\end{tikzpicture}
			\end{minipage}
			\qquad\qquad\qquad
			\begin{minipage}[b]{0.2\textwidth}
				\centering
				\begin{tikzpicture} 
					\filldraw [color=bleudefrance, thick, fill=bleudefrance!30] (0,0) -- (0,6) -- (4,8) -- (4,2) -- (0,0); 
					\coordinate[label={$A$}] (A2)  at ($(2,4)+1.2*(45: 1 and 2)$) []{};
					\coordinate[label={$A$}]  (B2) at ($(2,4)+1.15*(150: 1 and 2)$) {};
					\coordinate [label={below:$A$}]  (C2) at ($(2,4)+1.2*(225: 1 and 2)$) {};
					\coordinate [label={below:$A$}]  (D2) at ($(2,4)+1.15*(330: 1 and 2)$) {};
					\coordinate (V1) at ($(2,4)+(0.6,0.2)$) {};
					\coordinate (V2) at ($(2,4)-(0.6,0.2)$) {};
					\coordinate (V3) at ($(2,4)+(0.0,2)$) {};
					\draw [very thick,dashed] (A2)--(V1);
					\draw [very thick,dashed] (D2)--(V1);
					\draw [very thick,dashed] (B2)--(V2);
					\draw [very thick,dashed] (C2)--(V2);
					\draw [decorate, very thick, decoration={snake, segment length=2mm, amplitude=1mm}] (V2)   parabola[bend pos=0.2, bend={+(0.4, -1)}] (V1);
					\filldraw (A2) circle (2pt);
					\filldraw (B2) circle (2pt);
					\filldraw (C2) circle (2pt);
					\filldraw (D2) circle (2pt);
				\end{tikzpicture}
			\end{minipage}
			\qquad\qquad\qquad
			\begin{minipage}[b]{0.2\textwidth}
				\centering
				\begin{tikzpicture} 
					\filldraw [color=bleudefrance, thick, fill=bleudefrance!30] (0,0) -- (0,6) -- (4,8) -- (4,2) -- (0,0); 
					\draw [very thick, dashed] (2,4) ellipse (0.5 and 1); 
					\coordinate [] (A1) at ($(2,4)+(45: 0.5 and 1)$) {};
					\coordinate [label={$A$}] (A2)  at ($(2,4)+1.2*(45: 1 and 2)$) []{};
					\coordinate [] (B1) at ($(2,4)+(150: 0.5 and 1)$) {};
					\coordinate [label={$A$}]  (B2) at ($(2,4)+1.15*(150: 1 and 2)$) {};
					\coordinate[] (C1) at ($(2,4)+(225: 0.5 and 1)$) {};
					\coordinate[label={below:$A$}]  (C2) at ($(2,4)+1.2*(225: 1 and 2)$) {};
					\coordinate [] (D1) at ($(2,4)+(330:0.5 and 1)$) {};
					\coordinate[label={below:$A$}]  (D2) at ($(2,4)+1.15*(330: 1 and 2)$) {};
					\draw [very thick,dashed]  ($(2,4)+(0,1)$)--($(2,4)-(0,1)$);
					\draw [very thick,dashed] (A1)--(A2);
					\draw [very thick,dashed] (B1)--(B2);
					\draw [very thick,dashed] (C1)--(C2);
					\draw [very thick,dashed] (D1)--(D2);
					
					\filldraw (A2) circle (2pt);
					\filldraw (B2) circle (2pt);
					\filldraw (C2) circle (2pt);
					\filldraw (D2) circle (2pt);
				\end{tikzpicture}
			\end{minipage}
		}
		\caption{Diagrams that exhibit unitarity cuts of the four-gluon amplitude into minimal coupling vertices, of momentum scalings $s^0,s^{3\over 2},s^3,s^3$ respectively. Note that the graviton propagator in the third diagram is not restricted to the orbifold locus, leading to an enhanced momentum scaling.}
		\label{fig:AAAA}
	\end{figure}

	To leading order in the low-energy limit, the scattering amplitudes can be evaluated via tree-level Feynman diagrams (built out of the propagators and vertices of Figure \ref{fig:Feynmanrules}) of the 7d SYM coupled to 11d supergravity. In particular, the leading low energy superamplitude of four gluon multiplets is given, in the super-spinor-helicity convention described in Appendix \ref{sec:superspinorhelicity}, by \cite{Bern:1998ug}
	\ie\label{af2super}
	\mathfrak{A}_{F^2} \left(\{p_i, \lambda_i, \eta_i, a_i\}_{i=1}^4\right) =  2 g_{\rm YM}^2 \delta^7(P) \delta^8(Q) \left[ {{\mathtt B}^{a_1a_2a_3a_4} \over st}+ ({\rm cyclic~permutations~on~}2,3,4) \right].
	\fe
	Here $s,t,u=-s-t$ are 7d Mandelstam variables defined below,
	\ie 
	s=-(p_1+p_2)^2\,,\quad t=-(p_1+p_4)^2\,,\quad u=-(p_1+p_3)^2\,,
	\fe 
	$a_i$ are adjoint gauge indices, $P$ is the total momentum, $Q$ is the total super-momentum, and ${\mathtt B}^{abcd}$ is a color factor defined by
	\ie 
	{\mathtt B}^{abcd}\equiv\Re\Tr(T^a T^bT^cT^d)\,,
	\label{colorB}
	\fe
	where $T^a$ are hermitian $SU(k)$ generators normalized according to $\Tr (T^a T^b)={1\over 2}\D^{ab}$.

\begin{figure}[!htb]
		\centering
		\scalebox{.6}{		
			\begin{minipage}[b]{0.3\textwidth}
				\centering
				\begin{tikzpicture} 
					\filldraw [color=bleudefrance, thick, fill=bleudefrance!30] (0,0) -- (0,6) -- (4,8) -- (4,2) -- (0,0); 
					\coordinate[label={$A$}] (A2)  at ($(2,4)+1.2*(45: 1 and 2)$) []{};
					\coordinate[label={$A$}]  (B2) at ($(2,4)+1.15*(150: 1 and 2)$) {};
					\coordinate [label={below:$A$}]  (C2) at ($(2,4)+1.2*(225: 1 and 2)$) {};
					\coordinate [label={below:$A$}]  (D2) at ($(2,4)+1.15*(330: 1 and 2)$) {};
					\coordinate [label={left:$F^4~$}]  (V) at ($(2,4)$) {};
					\draw [very thick,dashed] (A2)--(V);
					\draw [very thick,dashed] (D2)--(V);
					\draw [very thick,dashed] (B2)--(V);
					\draw [very thick,dashed] (C2)--(V);
					\filldraw (A2) circle (2pt);
					\filldraw (B2) circle (2pt);
					\filldraw (C2) circle (2pt);
					\filldraw (D2) circle (2pt);
					\filldraw [red] (V) circle (3pt);
				\end{tikzpicture}
			\end{minipage}
			\qquad\qquad\qquad
			\begin{minipage}[b]{0.3\textwidth}
				\centering
				\begin{tikzpicture} 
					\filldraw [color=bleudefrance, thick, fill=bleudefrance!30] (0,0) -- (0,6) -- (4,8) -- (4,2) -- (0,0); 
					\coordinate[label={$A$}] (A2)  at ($(2,4)+1.2*(45: 1 and 2)$) []{};
					\coordinate[label={$A$}]  (B2) at ($(2,4)+1.15*(150: 1 and 2)$) {};
					\coordinate [label={below:$A$}]  (C2) at ($(2,4)+1.2*(225: 1 and 2)$) {};
					\coordinate [label={below:$A$}]  (D2) at ($(2,4)+1.15*(330: 1 and 2)$) {};
					\coordinate [label={left:$D^2F^4~$}]  (V) at ($(2,4)$) {};
					\draw [very thick,dashed] (A2)--(V);
					\draw [very thick,dashed] (D2)--(V);
					\draw [very thick,dashed] (B2)--(V);
					\draw [very thick,dashed] (C2)--(V);
					\filldraw (A2) circle (2pt);
					\filldraw (B2) circle (2pt);
					\filldraw (C2) circle (2pt);
					\filldraw (D2) circle (2pt);
					\filldraw [red] (V) circle (3pt);
				\end{tikzpicture}
			\end{minipage}
		}
		\caption{Some higher-derivative supervertices (represented by the red dots) that potentially contribute to the four-gluon amplitude. The distinct color structures have been suppressed in the diagrams and the corresponding momentum scalings are $s^2$ and $s^3$ respectively.}
		\label{fig:higherderAAAA}
	\end{figure}
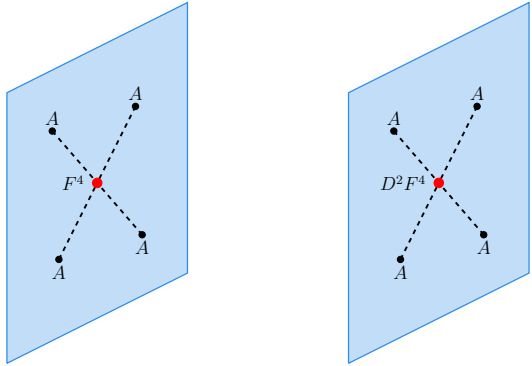

	Beyond the leading low energy limit, the momentum expansion of the full 4-gluon superamplitude is expected to take the form, where the dependence on the super-spinor-helicity variables are implicit, 
	\ie\label{asupfull}
	\mathfrak{A} &= \mathfrak{A}_{F^2} + \mathfrak{A}_{F^2|F^2}+  \mathfrak{A}_{F^4} 
	+  \left[ \mathfrak{A}_{R}   +  \mathfrak{A}_{F^2|F^2|F^2} + \mathfrak{A}_{D^2F^4} \right]+\cdots \,,
	\fe
	where the various terms are interpreted as follows.  
	First, $\mathfrak{A}_{F^2}$ is the leading low-energy amplitude (\ref{af2super}) with $s^0$ overall momentum scaling (first diagram in Figure~\ref{fig:AAAA}). $\mathfrak{A}_{F^2|F^2}$, given in (\ref{a1loop}), represents a ``1-loop'' term of $s^{3\over 2}$ overall momentum scaling, that is completely determined by its unitarity cut into a pair of $\mathfrak{A}_{F^2}$ amplitudes \cite{Bern:1996fj, Dixon:1996wi}  (second diagram in Figure~\ref{fig:AAAA}). The first higher derivative ``corrections'' that are not fixed by supersymmetry and perturbative unitarity considerations arise at order $s^2$ (first diagram in Figure~\ref{fig:higherderAAAA}), and take the form
	\ie\label{af4coup}
	\mathfrak{A}_{F^4} = \delta^7(P) \delta^8(Q) \Big[ \kappa_{[F^2]^2} {\mathtt A}^{a_1a_2a_3a_4} + \kappa_{F^4}{\mathtt B}^{a_1a_2a_3a_4} + ({\rm cyclic~permutations~on~}2,3,4) 
	\Big] \, ,
	\fe
	where we have introduced another color factor by
	\ie 
	{\mathtt A}^{abcd}\equiv \Tr(T^a T^b)\Tr(T^cT^d)\,.
	\label{colorA}
	\fe
	This amplitude may be viewed as the contribution from the supersymmetric completion of a term in an effective Lagrangian of the schematic form $\kappa_{[F^2]^2} (\Tr F^2)^2+\kappa_{F^4}\Tr F^4$,
	whose Lorentzian indices are contracted by the rank-eight $t_8$ tensor which is dictated by supersymmetry \cite{Green:1987mn,Polchinski:1998rr}.
	We will specify these effective couplings unambiguously through their contribution (\ref{af4coup}) to the local (i.e.~analytic in momenta) part of the amplitude (\ref{asupfull}).

	At order $s^3$ in  momentum scaling, there are several new structure that appear in (\ref{asupfull}). $\mathfrak{A}_{R}$ can be viewed as a ``tree-level'' gluon amplitude that involves the exchange of a bulk graviton (third diagram of Figure \ref{fig:AAAA}). The momentum power counting is unusual here: each cubic vertex contributes $+2$, the graviton propagator contributes $-2$, and the integration over the transverse momentum $p_\perp$ of the graviton contributes $+4$. From the perspective of Wilsonian effective theory, such a tree diagram is a priori subject to a UV divergence, schematically of the form \cite{Goldberger:2001tn, Michel:2014lva}
	\ie
	\int d^4 p_\perp {p_\parallel^4\over p_\parallel^2+p_\perp^2} \sim p_\parallel^6 \log{\Lambda\over p_\parallel} +  p_\parallel^4 \Lambda^2,
	\fe
	where the cutoff dependence can be absorbed into local terms. The superamplitude $\mathfrak{A}_R$ is most conveniently determined, up to local terms, by its unitarity cut into a pair of gluon-gluon-graviton vertices (\ref{2dRF2}), to be (cf.~(\ref{arloopint}))
	\ie\label{arloop}
	\mathfrak{A}_{R} &=- {1\over 2} \kappa^2 \delta^7(P)\delta^8(Q) \Big[ {\mathtt A}^{a_1a_2a_3a_4}  k{s \over 16\pi^2}\log(- \ell_p^2 s)+ ({\rm cyclic~permutations~on~}2,3,4) 
	\Big],
	\fe
	where the factor of $k$ is due to the sum over $\mathbb{Z}_k$ images in the bulk graviton propagator.
	
	The second term $\mathfrak{A}_{F^2|F^2|F^2}$ with the same momentum scaling in \eqref{asupfull}   is, up to local terms, the two-loop gluon superamplitude given in  (\ref{A2loop}) (see last diagram in Figure~\ref{fig:AAAA}). In contrast to the formal 7d SYM 2-loop amplitude which is logarithmically divergent \cite{Bern:1998ug}, we define $\mathfrak{A}_{F^2|F^2|F^2}$ by cutting off the loop integrals in (\ref{A2loop}) at Planck scale $\Lambda\sim 1/\ell_p$, giving a finite result with an IR non-analyticity similar to that of (\ref{arloop}). The curious fact that the gluon 2-loop amplitude arises at the same order in momentum scaling as the bulk graviton tree-level exchange, which is tied to the cubic relation between gravitational and Yang-Mills coupling $\kappa\sim g_{\rm YM}^3$, is reminiscent of (an  ``M-theory upgrade'' of) the relationship between 1-loop open and tree-level closed string amplitudes in open/closed duality.
	
	The remaining local terms of $s^3$ momentum scaling in (\ref{asupfull}) are constrained by supersymmetry Ward identities to be of the form  (see second diagram in Figure~\ref{fig:higherderAAAA})
	\ie\label{ad2f4term}
	\mathfrak{A}_{D^2F^4} = \delta^7(P) \delta^8(Q) \Big[ \kappa_{D^2[F^2]^2} {\mathtt A}^{a_1a_2a_3a_4} s + \kappa_{D^2 F^4}{\mathtt B}^{a_1a_2a_3a_4} u + ({\rm cyclic~permutations~on~}2,3,4) 
	\Big].
	\fe
	From the Wilsonian perspective, we expect an effective theory with cutoff at the Planck scale, whose coupling coefficients cannot depend on any parameters other than $k$ which specifies the M-theory background. Thus $\kappa_{D^2[F^2]^2}$ and $\kappa_{D^2F^4}$, related to coefficients of terms in the effective Lagrangian of the schematic form $D^2(\Tr F^2)^2$ and $D^2\Tr F^4$ (where $D$'s stand for gauge-covariant derivatives, suitably distributed on the $F_{\mu\nu}$'s), are proportional to $\ell_p^6$ with possibly $k$-dependent coefficients.

	\begin{figure}[!htb]
		\centering
		\scalebox{.6}{		\begin{minipage}[b]{0.2\textwidth}
				\centering
				\begin{tikzpicture} 
					\filldraw [color=bleudefrance, thick, fill=bleudefrance!30] (0,0) -- (0,6) -- (2.5,8) -- (2.5,2) -- (0,0); 
					\coordinate[label={$g$}] (A2)  at ($(2,4)+1.2*(45: 1 and 2)$) []{};
					\coordinate[label={$A$}]  (B2) at ($(2,4)+1.15*(150: 1 and 2)$) {};
					\coordinate [label={below:$A$}]  (C2) at ($(2,4)+1.2*(225: 1 and 2)$) {};
					\coordinate [label={below:$g$}]  (D2) at ($(2,4)+1.15*(330: 1 and 2)$) {};
					\coordinate (V1) at ($(2,4)+(0.,0.6)$) {};
					\coordinate (V2) at ($(2,4)-(0.0,0.6)$) {};
					\draw [decorate, very thick, decoration={snake, segment length=2.5mm, amplitude=1mm}] (A2)--(V1);
					\draw [very thick,dashed] (B2)--(V1);
					\draw [very thick,snake it] (D2)--(V2);
					\draw [very thick,dashed] (C2)--(V2);
					\draw [very thick,dashed] (V2)--(V1);
					\filldraw (A2) circle (2pt);
					\filldraw (B2) circle (2pt);
					\filldraw (C2) circle (2pt);
					\filldraw (D2) circle (2pt);
				\end{tikzpicture}
			\end{minipage}
			\qquad\qquad\qquad
			\begin{minipage}[b]{0.2\textwidth}
				\centering
				\begin{tikzpicture} 
					\filldraw [color=bleudefrance, thick, fill=bleudefrance!30] (0,0) -- (0,6) -- (2.,8) -- (2.,2) -- (0,0); 
					\coordinate[label={$g$}] (A2)  at ($(2,4)+1.2*(45: 1 and 2)$) []{};
					\coordinate[label={$A$}]  (B2) at ($(2,4)+1.15*(150: 1 and 2)$) {};
					\coordinate [label={below:$A$}]  (C2) at ($(2,4)+1.2*(225: 1 and 2)$) {};
					\coordinate [label={below:$g$}]  (D2) at ($(2,4)+1.15*(330: 1 and 2)$) {};
					\coordinate (V1) at ($(2,4)+(0.6,0.2)$) {};
					\coordinate (V2) at ($(2,4)-(0.6,0.2)$) {};
					\draw [very thick,snake it] (A2)--(V1);
					\draw [very thick,snake it] (D2)--(V1);
					\draw [very thick,dashed] (B2)--(V2);
					\draw [very thick,dashed] (C2)--(V2);
					\draw [very thick,snake it] (V2)--(V1);
					\filldraw (A2) circle (2pt);
					\filldraw (B2) circle (2pt);
					\filldraw (C2) circle (2pt);
					\filldraw (D2) circle (2pt);
				\end{tikzpicture}
			\end{minipage}
			\qquad\qquad\qquad
			\begin{minipage}[b]{0.2\textwidth}
				\centering
				\begin{tikzpicture} 
					\filldraw [color=bleudefrance, thick, fill=bleudefrance!30] (0,0) -- (0,6) -- (2.7,8) -- (2.7,2) -- (0,0); 
					\draw [very thick, dashed] (2,4) ellipse (0.5 and 1); 
					\coordinate [] (A1) at ($(2,4)+(50: 0.5 and 1)$) {};
					\coordinate [label={$g$}] (A2)  at ($(2,4)+1.4*(45: 1 and 2)$) []{};
					\coordinate [] (B1) at ($(2,4)+(150: 0.5 and 1)$) {};
					\coordinate [label={$A$}]  (B2) at ($(2,4)+1.15*(150: 1 and 2)$) {};
					\coordinate[] (C1) at ($(2,4)+(225: 0.5 and 1)$) {};
					\coordinate[label={below:$A$}]  (C2) at ($(2,4)+1.2*(225: 1 and 2)$) {};
					\coordinate [] (D1) at ($(2,4)+(330:0.5 and 1)$) {};
					\coordinate[label={below:$g$}]  (D2) at ($(2,4)+1.35*(330: 1 and 2)$) {};
					\draw [very thick,snake it] (A1)--(A2);
					\draw [very thick,dashed] (B1)--(B2);
					\draw [very thick,dashed] (C1)--(C2);
					\draw [very thick,snake it] (D1)--(D2);
					
					\filldraw (A2) circle (2pt);
					\filldraw (B2) circle (2pt);
					\filldraw (C2) circle (2pt);
					\filldraw (D2) circle (2pt);
				\end{tikzpicture}
			\end{minipage}
			\qquad\qquad\qquad
			\begin{minipage}[b]{0.2\textwidth}
				\centering
				\begin{tikzpicture} 
					\filldraw [color=bleudefrance, thick, fill=bleudefrance!30] (0,0) -- (0,6) -- (2,8) -- (2,2) -- (0,0); 
					\coordinate [] (A1) at ($(2,4)+1.2*(45: 0.5 and 1)$) {};
					\coordinate [label={$g$}] (A2)  at ($(2,4)+1.15*(45: 1 and 2)$) []{};
					\coordinate [] (B1) at ($(2,4)+1.2*(150: 0.5 and 1)$) {};
					\coordinate [label={$A$}]  (B2) at ($(2,4)+1.15*(150: 1 and 2)$) {};
					\coordinate[] (C1) at ($(2,4)+1.2*(225: 0.5 and 1)$) {};
					\coordinate[label={below:$A$}]  (C2) at ($(2,4)+1.2*(225: 1 and 2)$) {};
					\coordinate [] (D1) at ($(2,4)+1.2*(330:0.5 and 1)$) {};
					\coordinate[label={below:$g$}]  (D2) at ($(2,4)+1.15*(330: 1 and 2)$) {};
					\draw [very thick,snake it] (A1)--(A2);
					\draw [very thick,dashed] (B1)--(B2);
					\draw [very thick,dashed] (C1)--(C2);
					\draw [very thick,snake it] (D1)--(D2);
					\draw [very thick,snake it] (A1)--(B1);
					\draw [very thick,snake it] (D1)--(A1);
					\draw [very thick,snake it] (C1)--(D1);						\draw [very thick,dashed] (C1)--(B1);
					
					\filldraw (A2) circle (2pt);
					\filldraw (B2) circle (2pt);
					\filldraw (C2) circle (2pt);
					\filldraw (D2) circle (2pt);
				\end{tikzpicture}
			\end{minipage}
		}
		\caption{Diagrams that exhibit unitarity cuts of two-gluon-two-graviton amplitudes into minimal coupling vertices, of momentum scalings $s,s,s^{5\over 2}, s^{9\over 2}$ respectively. Note that the last diagram is beyond the derivative orders consider in this paper.}
		\label{fig:AAgg}
	\end{figure}
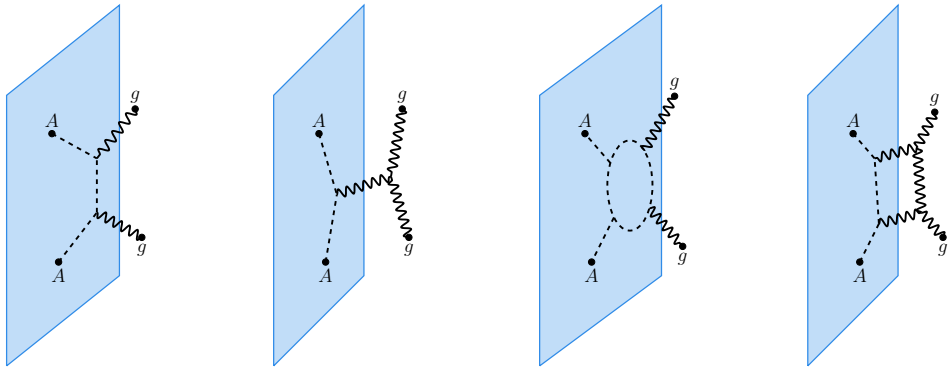

	The mixed 4-point superamplitude of two gluon multiplets and two graviton multiplets can be organized in a similar momentum expansion
	\ie\label{amixed}
	\widetilde{\mathfrak A} = \widetilde{\mathfrak A}_R + \widetilde{\mathfrak A}_{RF^2} + \widetilde{\mathfrak A}_{F^2|R} + \widetilde{\mathfrak A}_{R^2F^2} + \cdots \,,
	\fe
	where the various terms are interpreted as follows.  
	The leading low-energy term $\widetilde{\mathfrak A}_R$ may be viewed as tree-level exchange of a gluon or a graviton, and it is determined by its residues at the gluon or graviton pole, which factorize into a pair of cubic minimal coupling vertices (first two diagrams of Figure \ref{fig:AAgg}). For comparison with the holographic correlators considered in this paper, it suffices to restrict to the case where the gravitons do not carry transverse momenta, where the result can be expressed in a relatively simple manner using the 7d  super-spinor-helicity notation of  Appendix \ref{sec:superspinorhelicity},\footnote{This expression is derived by embedding the $\cN=1$ graviton and vector multiples into an $\cN=2$, i.e.~maximally supersymmetric, graviton multiplet, and projecting from the tree-level scattering amplitude of the latter.\label{fnote:embeddingtrick}}
	\ie\label{arsuper}
	\left.\widetilde {\mathfrak A}_R \right|_{p_3^\perp = p_4^\perp=0} &= \kappa^2 {\rm Tr}(T^{a_1} T^{a_2}) \delta^7(P) {\delta^8(Q) \over stu} \left. \partial_{\widehat\eta_1}^2\partial_{\widehat\eta_2}^2 \prod_{\A=1}^8 \left( \lambda^\A_{1I} \widehat\eta_1^I+\lambda^\A_{2I} \widehat\eta_2^I + \widetilde q_3^\A + \widetilde q_4^\A \right) \right|_{\widetilde\eta_3^2|_{\bf 5}\widetilde\eta_4^2|_{\bf 5}} \, ,
	\fe
	where $\{p_i, \lambda_i, \eta_i,a_i\}_{i=1,2}$ are the momenta, super-spinor-helicity variables, and gauge indices of the gluons, $\{p_i, \lambda_i, \eta_i, \widetilde \eta_i,\tilde q_i\}_{i=3,4}$ are the 7d momenta,  super-spinor-helicity variables  and auxiliary supermomenta \eqref{tq} of the gravitons,  and $\widehat \eta_1^I, \widehat \eta_2^I$ are auxiliary Grassmann variables, with $\partial_{\widehat\eta}^2\equiv \Omega^{IJ} {\partial\over \partial\widehat \eta^I} {\partial\over \partial\widehat \eta^J}$.

	\begin{figure}[!htb]
		\centering
		\scalebox{.6}{		\begin{minipage}[b]{0.3\textwidth}
				\centering
				\begin{tikzpicture} 
					\filldraw [color=bleudefrance, thick, fill=bleudefrance!30] (0,0) -- (0,6) -- (2.,8) -- (2.,2) -- (0,0); 
					\coordinate[label={$g$}] (A2)  at ($(2,4)+1.2*(45: 1 and 2)$) []{};
					\coordinate[label={$A$}]  (B2) at ($(2,4)+1.15*(150: 1 and 2)$) {};
					\coordinate [label={below:$A$}]  (C2) at ($(2,4)+1.2*(225: 1 and 2)$) {};
					\coordinate [label={below:$g$}]  (D2) at ($(2,4)+1.15*(330: 1 and 2)$) {};
					\coordinate (V1) at ($(2,4)+(0.6,0.2)$) {};
					\coordinate  [label={left:$D_\perp^2 D^{-2}RF^2$}]  (V2)  at ($(2,4)-(0.6,0.2)$) {};
					\draw [very thick,snake it] (A2)--(V1);
					\draw [very thick,snake it] (D2)--(V1);
					\draw [very thick,dashed] (B2)--(V2);
					\draw [very thick,dashed] (C2)--(V2);
					\draw [very thick,snake it] (V2)--(V1);
					\filldraw (A2) circle (2pt);
					\filldraw (B2) circle (2pt);
					\filldraw (C2) circle (2pt);
					\filldraw (D2) circle (2pt);
					\filldraw[red] (V2) circle (3pt);
					%		\node[cross=4pt, color=bleudefrance, very thick] at (V2) {};
				\end{tikzpicture}
			\end{minipage}
			\qquad\qquad\qquad
			\begin{minipage}[b]{0.3\textwidth}
				\centering
				\begin{tikzpicture} 
					\filldraw [color=bleudefrance, thick, fill=bleudefrance!30] (0,0) -- (0,6) -- (2.5,8) -- (2.5,2) -- (0,0); 
					\coordinate[label={$g$}] (A2)  at ($(2,4)+1.2*(45: 1 and 2)$) []{};
					\coordinate[label={$A$}]  (B2) at ($(2,4)+1.15*(150: 1 and 2)$) {};
					\coordinate [label={below:$A$}]  (C2) at ($(2,4)+1.2*(225: 1 and 2)$) {};
					\coordinate [label={below:$g$}]  (D2) at ($(2,4)+1.15*(330: 1 and 2)$) {};
					\coordinate  [label={below left:$D_\perp^2 D^{-2}RF^2$}]  (V1) at ($(2,4)+(0.,0.8)$) {};
					\coordinate (V2) at ($(2,4)-(0.0,0.6)$) {};
					\draw [very thick,snake it] (A2)--(V1);
					\draw [very thick,snake it] (D2)--(V2);
					\draw [very thick,dashed] (B2)--(V1);
					\draw [very thick,dashed] (C2)--(V2);
					\draw [very thick,dashed] (V2)--(V1);
					\filldraw (A2) circle (2pt);
					\filldraw (B2) circle (2pt);
					\filldraw (C2) circle (2pt);
					\filldraw (D2) circle (2pt);
					\filldraw[red] (V1) circle (3pt);
					%		\node[cross=4pt, color=bleudefrance, very thick] at (V2) {};
				\end{tikzpicture}
			\end{minipage}
			\qquad\qquad\qquad
			\begin{minipage}[b]{0.3\textwidth}
				\centering
				\begin{tikzpicture} 
					\filldraw [color=bleudefrance, thick, fill=bleudefrance!30] (0,0) -- (0,6) -- (2.5,8) -- (2.5,2) -- (0,0); 
					\coordinate[label={$g$}] (A2)  at ($(2,4)+1.3*(45: 1 and 2)$) []{};
					\coordinate[label={$A$}]  (B2) at ($(2,4)+1.15*(150: 1 and 2)$) {};
					\coordinate [label={below:$A$}]  (C2) at ($(2,4)+1.2*(225: 1 and 2)$) {};
					\coordinate [label={below:$g$}]  (D2) at ($(2,4)+1.25*(330: 1 and 2)$) {};
					\coordinate  [label={left:$R^2F^2$}]  (V) at ($(1.8,4)$) {};
					\draw [decorate, very thick, decoration={snake, segment length=2.5mm, amplitude=1mm}] (A2)--(V);
					\draw [very thick,dashed] (B2)--(V);
					\draw [very thick,snake it] (D2)--(V);
					\draw [very thick,dashed] (C2)--(V);
					\filldraw (A2) circle (2pt);
					\filldraw (B2) circle (2pt);
					\filldraw (C2) circle (2pt);
					\filldraw (D2) circle (2pt);
					\filldraw[red] (V) circle (3pt);
					%		\node[cross=4pt, color=bleudefrance, very thick] at (V2) {};
				\end{tikzpicture}
			\end{minipage}
		}
		\caption{Diagrams that exhibit possible unitarity cuts of the amplitude of two gluons and two gravitons that involve higher-derivative supervertices (represented by the red dots), with momentum scaling $s^2,s^2,s^3$ respectively.}
		\label{fig:higherderAAgg}
	\end{figure}
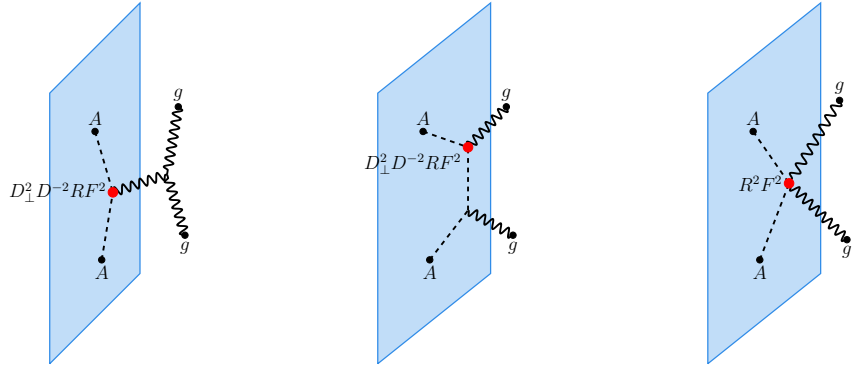

	The first possible correction to the minimal gluon-graviton coupling that is compatible with supersymmetry arises at 4-derivative order in the effective Lagrangian, of the schematic form $\kappa_{RF^2} R\Tr F^2 + \cdots$, corresponding to the cubic supervertex \cite{Lin:2015ixa}  
	\ie\label{arf2}
	\kappa_{RF^2}  {\mf V}_{D_\perp^2 D^{-2} RF^2} ,
	\fe
	where ${\mf V}_{D_\perp^2 D^{-2} RF^2}$ is given by the $k=1$ case of (\ref{RF2hder}).
	This supervertex vanishes when the transverse momentum $p_\perp$ of the graviton vanishes. On the other hand, one expects at the same derivative order a 2-gluon-2-graviton superamplitude $\widetilde {\mathfrak A}_{RF^2}$ that is rational in momenta, whose factorization through its gluon or graviton pole involves the cubic vertex (\ref{arf2}) (see first two diagrams in Figure~\ref{fig:higherderAAgg}). The vanishing of the latter at $p_\perp=0$ implies that $\widetilde {\mathfrak A}_{RF^2}$, when restricted to gravitons with vanishing transverse momenta ($p_3^\perp = p_4^\perp=0$ in the notation of (\ref{arsuper})), is free of poles. A closely related feature of the Mellin amplitude in ${\rm AdS}_4\times S^7/\mathbb{Z}_k$ will be observed in Section~\ref{sec:mellinmixed}.
	
	The next correction in the momentum expansion (\ref{amixed}) is the 1-loop term $\widetilde {\mathfrak A}_{F^2|R}$, which can be determined by its unitarity cut into tree amplitudes of gluons and gravitons (third diagram in Figure~\ref{fig:AAgg}). It can be viewed as the 1-loop superamplitude of the (2-derivative) SYM$+$SUGRA, which is free of logarithmic divergences and has $s^{5\over 2}$ overall momentum scaling.
	
	Finally, at 6-derivative order (momentum scaling $s^3$), we have the possibility of an effective coupling of the schematic form $\kappa_{R^2F^2} R^2 \Tr F^2 + \cdots$, corresponding to the supervertex  given by the $k=0$ case of \eqref{R2F2ver} (see last diagram in Figure~\ref{fig:higherderAAgg}),
	\ie\label{ar2f2s}
	\widetilde{\mathfrak A}_{R^2F^2} = \kappa_{R^2F^2} \delta^7(P) \delta^8(Q) {\rm Tr}(T^{a_1} T^{a_2}) \widetilde q_{34}^2\, .
	\fe
	
	To summarize, the effective theory of the M-theory orbifold is characterized by a sequence of higher derivative coupling coefficients $\kappa_{[F^2]^2}$, $\kappa_{F^4}$, $\kappa_{D^2[F^2]^2}$, $\kappa_{D^2F^4}$, $\kappa_{RF^2}$, $\kappa_{R^2F^2}$ up to momentum scaling $s^3$ and so forth for higher order corrections, that are thus far unconstrained by supersymmetry, perturbative unitarity, and locality considerations.

	\subsection{Constraints on the Coulomb branch}
	\label{sec:coulomb}
	
	It is useful to compare the aforementioned non-Abelian effective theory to the low-energy expansion of the $U(1)^{k-1}$ effective gauge theory on the Coulomb branch of the M-theory orbifold vacua, where the $\vec x_i$'s in (\ref{alemetric}) are separated. In particular, the BPS-protected terms in the Abelian effective action, of the schematic form 
	\ie\label{ffourfsix}
	f_{abcd}^{(4)}(\vec x_i) F^a F^b F^c F^d~~~~{\rm and}~~~~f_{ab,cd}^{(6)}(\vec x_m) D^2(F^a F^b ) F^c F^d,
	\fe
	where $F^a$ stands for the field strength in the $a$th Cartan $U(1)$ factor, are constrained by supersymmetry in such a way that $f^{(4)}$ and $f^{(6)}$ obey certain second-order differential equations on the Coulomb branch moduli space \cite{Paban:1998ea,Paban:1998qy,Sethi:1999qv,Maxfield:2012aw,Lin:2015ixa},\footnote{See \cite{Lin:2015zea, Lin:2015dsa} for closely parallel analyses of similar constraints in 6d effective theories in the context of superstring compactifications.} and thereby can be determined given enough data at large values of $|\vec x_{ij}|$ where the low energy expansion of the M-theory effective action in weakly curved spacetimes is applicable. 
	
	It is also possible to determine the coefficient functions appearing in (\ref{ffourfsix}) by considering further compactifications, where dual string theoretic descriptions are available, and combine the supersymmetry constraints with known results of the low energy expansion in a regime captured by string perturbation theory \cite{Wang:2015jna, Lin:2015ixa, Wang:2015aua}. For instance, via a circle compactification to type IIA string theory, a set of arguments similar to those given in Section~5 of \cite{Lin:2015dsa} determine $f^{(4)}_{abcd}$ in (\ref{ffourfsix}) to be identical to the coefficient of such a term in the 1-loop effective action of the 7d $SU(k)$ super-Yang-Mills theory on its Coulomb branch. In particular, $f^{(4)}_{abcd}$ vanishes in the limit $|\vec x_i|\to \infty$, and diverges at the origin $\vec x_i=0$.
	
	It is not entirely straightforward, however, to relate the ${\rm tr}F^4$ and  $({\rm tr} F^2)^2$ effective coupling at the origin of Coulomb branch to the Abelian $F^4$ effective coupling away from the origin of the Coulomb branch. In particular, the latter contribution by itself is singular at the origin of the Coulomb branch, and only after summing up an infinite set of (a priori BPS-unprotected) higher-derivative contributions to the Abelian effective theory on the Coulomb branch can we recover the amplitude at the origin of Coulomb branch. It is plausible, nonetheless, that a connection between the non-Abelian $\Tr F^4$, $(\Tr F^2)^2$ couplings at the origin and the Abelian $F^4$ effective coupling away from the origin can be made in the framework of a non-Abelian Wilsonian effective gauge theory. Starting with a Wilsonian effective action of the form $S_{\rm SYM} + \int d^4x \left( c_\Lambda \Tr F^4+\cdots \right)$, where the Wilsonian coefficients $c_\Lambda$, etc.~depend on the cutoff scale $\Lambda$,\footnote{Strictly speaking, the argument that follows requires a Wilsonian cutoff scheme that respects both gauge invariance and supersymmetry. An explicit construction of such a scheme, which is possible through the Batalin-Vilkovisky formalism \cite{Costello:2007ei, Costello:2011np}, is beyond the scope of this paper. } we can calculate both the $\Tr F^4$ term in the 4-gluon amplitude at the origin of the Coulomb branch and the $F^4$ term in the Abelian amplitude away from the origin. The structure of the supersymmetric amplitudes computed via the unitarity method \cite{Bern:1996fj, Dixon:1996wi} and power counting suggests that at the 8-derivative order, both the non-Abelian amplitude at the origin and the Abelian amplitude away from the origin should only receive a 1-loop contribution from the SYM theory with Wilsonian cutoff $\Lambda$ and a tree-level contribution that is linear in $c_\Lambda$. This would then imply that the $\Tr F^4$ term in the 4-gluon amplitude vanishes, given that we already know the Abelian $F^4$ term coincides with the 1-loop SYM amplitude in the $\Lambda\to \infty$ limit.

	In the rest of this paper, we will pursue a different strategy that allows for determining the non-Abelian effective couplings directly at the origin of the Coulomb branch, namely from the CFT data of a holographic dual. 
	
	\subsection{AdS deformation and the holographic dual}
	
	The holographic duality of interest arises from the decoupling limit of the system of $N$ M2-branes probing the M-theory orbifold singularity $\mC^2/\mZ_k\times \mR^{1,6}$, or equivalently $N$ D2-branes lying within the world-volume of a stack of $k$ D6-branes in type IIA string theory \cite{Ferrara:1998vf,Gomis:1998xj,Entin:1998ub,Pelc:1999ms,Cherkis:2002ir}. The near horizon limit of the gravity dual leads to ${\rm AdS}_4\times S^7/\mathbb{Z}_k$, where the $\mathbb{Z}_k$ action leaves fixed an $S^3\subset S^7$ and results in an orbifold singularity locus in the quotient space. Explicitly, we can parameterize the $S^7$ using four complex coordinates $z_i$ subject to the constraint $\sum_{i=1}^4 \abs{z_i}^2 = 1$, whereas the generator of the $\Z_k$ acts by $(z_1, z_2, z_3, z_4) \mapsto (z_1 e^{2 \pi i / k}, z_2 e^{-2 \pi i / k}, z_3, z_4)$  as in \eqref{Zkorb} and the fixed $S^3$ is located at $z_1=z_2=0$.  Note that this is different from the case of ABJM theory \cite{Aharony:2008ug} engineered from M2-branes probing a different $\mZ_k$ orbifold singularity $\mC^4/\mZ_k\times \mR^{1,2}$ where the $\mZ_k$ acts all four complex coordinates $z_{1,2,3,4}$ in the same way (i.e. $z_i\to z_i e^{2\pi i /k}$). After taking the decoupling limit, the holographic dual of the ABJM theory involves a fixed-point-free $\mZ_k$ quotient of $S^7$ as the internal manifold. 
	
	The low energy dynamics of the massless D2-D2 and D2-D6 open strings is governed by a three-dimensional ${\cal N}=4$ supersymmetric $U(N)$ gauge theory with an adjoint hypermultiplet and $k$ fundamental hypermultiplets, which admits $SU(2)_H\times SU(2)_C$ R-symmetry as well as $G_F= SU(k) \times SU(2)_F \times U(1)$ flavor symmetry, where the $SU(k)$ factor transforms the fundamental hypermultiplets into one another, the $SU(2)_F$ factor acts on the fields of the adjoint hypermultiplet, and the $U(1)$ factor is a topological symmetry under which the only charged operators are monopole operators. The ${\cal N}=4$ SCFT of interest, which arises at the infrared fixed point of this gauge theory, is the holographic dual of M-theory on the aforementioned ${\rm AdS}_4\times S^7/\mathbb{Z}_k$ orbifold.\footnote{In the special case $k=2$, the $U(1)$ topological symmetry is enhanced to yet another $SU(2)$ factor, and the full flavor symmetry is $SU(2) \times SU(2) \times SU(2)$. In another case $k=1$, supersymmetry is enhanced to ${\cal N} = 8$ \cite{Bashkirov:2010kz}, and the resulting SCFT is equivalent \cite{Bashkirov:2011pt} to the $U(N)_1 \times U(N)_{-1}$ ABJM theory of \cite{Aharony:2008ug}.\label{fn:ABJM}}   See for instance \cite{Benini:2009qs,Bashkirov:2010kz,Mezei:2013gqa,Grassi:2014vwa,Mezei:2017kmw,Gaiotto:2020vqj,Chester:2020jay} for various aspects of this theory and its holographic dual.
	
	While the $SU(k)$ flavor symmetry is not realized geometrically, the $SU(2) _F\times U(1)$ as well as the $SU(2)_H \times SU(2)_C$ R-symmetry are.  When acting on the embedding coordinates $z_i$, the following pairs form doublets under the various $SU(2)$ symmetries: 
	\es{Doublets}{
		SU(2)_H:  \begin{pmatrix} 
			z_3 \\
			-z_4^*
		\end{pmatrix} \,, \begin{pmatrix} 
			z_4 \\
			z_3^*
		\end{pmatrix} \,, \quad
		SU(2)_C:    \begin{pmatrix} 
			z_1 \\
			-z_2^*
		\end{pmatrix} \,, \begin{pmatrix} 
			z_2 \\
			z_1^*
		\end{pmatrix}
		\,, \quad
		SU(2)_F:  \begin{pmatrix} 
			z_3 \\
			z_4
		\end{pmatrix} \,, \begin{pmatrix} 
			-z_4^* \\
			z_3^*
		\end{pmatrix} \,.
	}
	In addition, the $U(1)$ flavor symmetry acts with charge $+1$ on $z_1$ and charge $-1$ on $z_2$.

	The curvature radius $L$ of the ${\rm AdS}_4$, measured in units of the 11d Planck length $\ell_p$,  is related to the parameters $N$ and $k$ of the gauge theory via \cite{Benini:2009qs}
	\es{dict}{
		\frac{L^9}{\ell_p^9}=\frac{\pi^3\sqrt{2}}{4} (N k)^{3\over 2}\,,
	}
	at leading order at large $N$.  The equations of motion of 11d supergravity imply that the radius of curvature of $S^7 / \Z_k$ is $2 L$, at least to leading order at large $L / \ell_p$. In the $L\to \infty$ limit, the ${\rm AdS}_4\times S^7/\mathbb{Z}_k$ reduces to the locally flat $\mathbb{C}^2/\mathbb{Z}_k\times \mathbb{R}^{1,6}$ orbifold.
	
	\subsection{Relation between 7d fields and flavor current multiplets in the SCFT }
	
	Here we briefly discuss the holographic dictionary between the 7d fields (from the graviton and the vector multiplets) on the fixed-point locus ${\rm AdS}_4\times S^3$ of the AdS orbifold ${\rm AdS}_4\times S^7/\mathbb{Z}_k$ and operators in the flavor current multiplets (denoted as $\cD[1]$ in the following) in the dual 3d $\cN=4$ SCFT\@. We focus on the bosonic operators for simplicity.  
	Each $\cD[1]$  flavor multiplet contains a superconformal primary scalar $J$ of dimension $\Delta=1$ in the representation of $(\bf 3,1)$  with respect to the $SU(2)_H\times SU(2)_C$ R-symmetry (labeled by dimensions of the $SU(2)$ irreps). The entire multiplet transforms in the adjoint representation of a flavor group $G_F$. This is also known as the Higgs branch flavor multiplet, which contains conserved $G_F$ currents as superconformal descendants at level 2 together with dimension $\Delta=2$ scalars $K$ transforming as $(\bf 1,3)$ under the $SU(2)_H\times SU(2)_C$.
	There is one other type of flavor multiplet with respect to the 3d $\cN=4$ superconformal algebra (see e.g. \cite{Cordova:2016emh}), known as the Coulomb branch flavor multiplet, where the R-symmetry groups $SU(2)_H$ and $SU(2)_C$ switch roles (i.e.~scalar operators $J$ and $K$ switch $SU(2)_H$ and $SU(2)_C$ R-symmetry quantum numbers). 
	
	\begin{table}[!htb]
		\centering
		\begin{tabular}{
				|c|c|c|c |c |c|c|}
			\hline  
			Fields & Weight $\Delta$ & Spin $\ell$ & $SU(2)_H$ & $SU(2)_F$ & $SU(2)_C$& CFT Op \\  \hline  \hline
			$\varphi^{\bm a}$ & 2 & 0  & {\bf 1} & {\bf 1} & {\bf 3} & $J'$
			\\
			& 1 & 0  & {\bf 1} & {\bf 1} & {\bf 3} & $K$
			\\\hline
			$A_\m$ & $1$  & 0 &  {\bf 3} & {\bf 1} & {\bf 1} & $J$
			\\
			& $2$  & 0 & {\bf 3} & {\bf 1} & {\bf 1} & $K'$
			\\\hline
			$g_{\m\n} $& 1 & 0  & {\bf 3} & {\bf 3} & {\bf 1} & $\widetilde J$
			\\\hline
			$V_{\m}^{\bm a} $
			& 2 & 0  & {\bf 1} & {\bf 3} & {\bf 3} & $\widetilde K$
			\\\hline
		\end{tabular}
		\caption{KK modes of the 7d fields reduced on $S^3$ that correspond to scalar operators in the $\cD[1]$ flavor multiplets. The representations with respect to the $SU(2)_H\times SU(2)_C$ R-symmetry and $SU(2)_F$ Higgs branch flavor symmetry are listed. The two rows for $\varphi^{\bm a}$ and $A_\m$ differ in the choice of the asymptotic boundary condition on AdS$_4$. We also note the correspondence to the notation for CFT operators used in the later sections, where $J',K'$ denote Coulomb branch operators that we will not focus on.}
		\label{tab:fieldsops}
	\end{table}
	
	In Table~\ref{tab:fieldsops}, we list the 7d fields whose KK reductions on $S^3$ produce the ${\rm AdS}_4$ dual scalar fields for the $J$ and $K$ type operators of 3d $\cN=4$ flavor multiplets. Note that for scalar fields of mass $m$ in the range $-{9\over 4}< m^2 L^2 <-{5\over 4}$ there are two boundary conditions compatible with unitarity, subjected to further constraints from supersymmetry \cite{Amsel:2008iz,Freedman:2013oja,Freedman:2016yue,Mezei:2017kmw}. Here, the relevant 7d fields from the 11d supergravity reduced on the orbifold singularity $\mC^2/\mZ_k$  consist of the 7d $\cN=1$ graviton multiplet, $SU(k)$ vector multiplets from the massless modes supported at the singularity, and a $U(1)$ vector multiplet for the residue $U(1)$ isometry of the orbifold (for $k>2$, see Footnote~\ref{fn:ABJM} for a related comment). 
	The KK reduction of general vector multiplets on $S^3$ has been worked out in \cite{Imamura:2010sa}. 
	The leading KK modes  indeed give the holographic dual for the $J$ and $K$ type operators (see Table~3 in \cite{Imamura:2010sa}). Consistency with supersymmetry requires that the $SU(k)$ vector multiplet and the $U(1)$ vector multiplet have opposite boundary conditions on AdS$_4$, as they correspond to Higgs and Coulomb branch symmetries respectively \cite{Mezei:2017kmw}.
	
	As mentioned above, the SCFT admits in addition an $SU(2)_F$ flavor symmetry of the Higgs branch type, which originates from the 7d graviton multiplet in the bulk.  In turn, this 7d graviton multiplet can be obtained by reducing the 11d graviton multiplet on the $S^3$ transverse to the orbifold singularity.   Thus, we can infer the relevant scalar KK modes that listed in Table~\ref{tab:fieldsops} using the known results of the reduction of the 11d fields on $S^7$  \cite{Sezgin:1983ik,Casher:1984ym,Gunaydin:1985tc,deWit:1986oxb}, and then imposing the $\Z_k$ quotient.  In the $S^7$ reduction, \cite{Sezgin:1983ik,Casher:1984ym,Gunaydin:1985tc,deWit:1986oxb} found that the lowest supermultiplet of KK modes contains scalar operators of dimensions $\Delta = 1$ and $\Delta = 2$ transforming in different $35$-dimensional representations of the $SO(8)_R$ R-symmetry which arise from the reduction on $S^7$ of the 11d graviton and three-form gauge potential.  In the convention in which the $\cN=8$ supercharges transform in the ${\bf 8}_v$ of $SO(8)_R$, we can take the $\Delta = 1$ scalars to transform in ${\bf 35}_c$ and the $\Delta = 2$ scalars to transform in ${\bf 35}_s$.  These representations split under an $SU(2)_H\times SU(2)_C\times SU(2)_1\times SU(2)_2$ subgroup of $SO(8)_R$ as
	\ie 
	{\bf 8}_v &= ({\bf 2},{\bf 2},{\bf 1},{\bf 1}) \oplus  ({\bf 1},{\bf 1},{\bf 2},{\bf 2}) \,, \\
	{\bf 35}_c  &= ({\bf 3},{\bf 1},{\bf 3},{\bf 1})\oplus ({\bf 1},{\bf 3},{\bf 1},{\bf 3})\oplus ({\bf 2},{\bf 2},{\bf 2},{\bf 2})\oplus ({\bf 1},{\bf 1},{\bf 1},{\bf 1})\,,\\
	{\bf 35}_s  &= ({\bf 1},{\bf 3},{\bf 3},{\bf 1})\oplus ({\bf 3},{\bf 1},{\bf 1},{\bf 3})\oplus ({\bf 2},{\bf 2},{\bf 2},{\bf 2})\oplus ({\bf 1},{\bf 1},{\bf 1},{\bf 1})\,.
	\label{N8split}
	\fe
	Note that the $SU(2)_1$ subgroup above is identified with the 3d $\cN=4$ $SU(2)_F$ flavor symmetry whereas $SU(2)_2$ is broken by the orbifold to $U(1)$ (for $k>2$). For determining the scalar operators in the half-maximal supergravity in 7d, it suffices to restrict to $SU(2)_2$-invariant fields in the last two lines of \eqref{N8split}, leaving only the first and last summands in these equations.  The first summands on the RHS of the second and third lines correspond, respectively, to the desired $J$ and $K$ operators in the $SU(2)_F$ flavor multiplet.  The last summand in the second line corresponds to the superconformal primary scalar in the 3d $\cN=4$ stress tensor multiplet.

	\section{Superconformal kinematics}
	\label{4point}

	The main observables of our interest in the 3D ${\cal N}=4$ SCFT are the 4-point correlation functions of flavor multiplets, which are dual to amplitudes of gravitons and gluons in the ${\rm AdS}_4\times S^7/\mathbb{Z}_k$ orbifold.

	\subsection{Correlators and multiplet structure}
	
	We will work with the $\Delta=1$ scalar superconformal primaries in the flavor multiplets of type $\cD[1]$ associated with the $SU(k)$ and $SU(2)_F$ symmetries,
	\ie\label{yydef}
	& J^a(\vec x,y) \equiv \phi^a_{\A\B}(\vec x) y^\A y^\B,
	\\
	& \widetilde J^A(x,\vec y) \equiv \widetilde\phi^A_{\A\B}(\vec x) y^\A y^\B,
	\fe
	where $a$ is an $SU(k)$ adjoint index, $A$ an $SU(2)_F$ adjoint index, $\A$ an $SU(2)_H$ doublet index, and $y^\A$ is an auxiliary ``spinor polarization'' variable with respect to $SU(2)_H$. In particular, the $SU(k)$ current $j^a_\mu$ is a superconformal descendant of $\phi^a_{\A\B}$, whereas the $SU(2)_F$ current $\widetilde j^A_\mu$ is a descendant of $\widetilde\phi^A_{\A\B}$.  We normalize the operators \eqref{yydef} such that their two-point functions are
	\es{TwoPtJ}{
		\langle J^a(\vec x_1,y_1) J^b(\vec x_2,y_2) \rangle &= \frac{\langle y_1,y_2\rangle^2}{|\vec x_{12}|^2}  \tr(T^a T^b) \,, \\
		\langle \widetilde J^A(\vec x_1,y_1) \widetilde J^B(\vec x_2,y_2) \rangle &= \frac{\langle y_1,y_2\rangle^2}{|\vec x_{12}|^2} \delta^{AB} \,,
	}
	where $\langle y_1, y_2\rangle \equiv \epsilon_{\A\B} y_1^\A y_2^\B$ and $\vec x_{12} \equiv \vec{x}_1 - \vec{x}_2$.
	
	The correlators of interest are constrained by conformal and  R-symmetries to be\footnote{We will not consider the 4-point function of only $\widetilde J$'s in this work.}
	\es{4point1}{
		\langle J^a(\vec x_1,y_1) J^b(\vec x_2,y_2) J^c(\vec x_3,y_3) J^d(\vec x_4,y_4) \rangle &=
		\frac{\langle y_1,y_2\rangle^2\langle y_3,y_4\rangle^2}{|\vec x_{12}|^2|\vec x_{34}|^2}\mathcal{G}^{abcd}(U,V;\alpha)\,,\\
		\langle J^a(\vec x_1,y_1) J^b(\vec x_2,y_2) \widetilde J^C(\vec x_3,y_3) \widetilde J^D(\vec x_4,y_4) \rangle &=
		\frac{\langle y_1,y_2\rangle^2\langle y_3,y_4\rangle^2}{|\vec x_{12}|^2|\vec x_{34}|^2}\mathcal{G}^{abCD}(U,V;\alpha)\,,\\
	}
	where the conformal cross-ratios $U, V$ and the R-symmetry cross-ratio $\A$ are defined as
	\es{crossRatios}{
		U\equiv \frac{\vec x_{12}^2 \vec x_{34}^2}{\vec x_{13}^2  \vec x_{24}^2} \equiv z \bar z \,,\qquad  V\equiv \frac{\vec x_{14}^2 \vec x_{23}^2}{\vec x_{13}^2  \vec x_{24}^2}\equiv (1-z)(1-\bar z) \,,\qquad \alpha=\frac{ \langle y_1,y_3\rangle  \langle y_2,y_4\rangle}{ \langle y_1,y_2\rangle  \langle y_3,y_4\rangle}\,.
	}
	To avoid cluttering the notation, the two types of correlators are both denoted $\cG$, distinguished only through their flavor indices ($abcd$ versus $abCD$). 
	
	With this way of representing the correlators, the superconformal Ward identities can be expressed very compactly as \cite{Dolan:2001tt}
	\es{3D}{
		\left(z\partial_z-\frac{1}{2}\alpha\partial_\alpha\right)\mathcal{G}(U,V;\alpha)\big\vert_{\alpha=z^{-1}}=0\,,
	}
	where $z$ is defined as in (\ref{crossRatios}), and $\cG$ stands for either $\cG^{abcd}$ or $\cG^{abCD}$. The superconformal block decomposition of the correlator takes the form
	\es{blockExp}{
		\cG(U,V;\alpha)=\sum_{(\cM, R)\in {\rm Sym}^2(\cD[1] , {\bf adj})} {\mathtt T}_R \lambda^2_{\cM,R}\mathfrak{G}_\cM(U,V;\alpha)\,,
	}
	where ${\cal M}$ and $R$ label irreducible representations of the superconformal group and the flavor group that appear in the OPE\@. ${\mathtt T}_R$ are invariant tensors of the flavor group that carry the same suppressed flavor indices as $\cG$ does. The corresponding OPE coefficients are denoted by $\lambda_{\cM, R}$, and $\mathfrak{G}_\cM$ are superconformal blocks.
	
	The tensor product of a pair of $\cD[1]$ flavor multiplets can be decomposed into irreducible superconformal and flavor group representations, in the notation of \cite{Chang:2017xmr}, as
	\es{OPE}{
		{\rm Sym}^2 (\mathcal{D}[1], {\bf adj}) &=1+(\mathcal{D}[1],{\bf adj})+\sum_{R\in\text{sym}} (\mathcal{D}[2], R)+(\mathcal{B}[0], 1) +\sum_{R\in\text{asym}}\sum_{\text{odd } j } (\mathcal{B}[1]_j,R)
		\\
		&~~~ +\sum_{R\in\text{sym}}\sum_{\text{even } j }\sum_{\Delta>j+1} (\mathcal{L}[0]_{\Delta,j}, R)\,.
	}
	Here, 1 stands for the trivial (identity) representation. $\mathcal{B}[0]$ (denoted as $\cB[0]_0$ in  \cite{Chang:2017xmr}) is the stress tensor multiplet, whose superprimary is a $\Delta=1$ scalar.\footnote{For free theories we would also have the $\cB[0]_\ell$ multiplets in the notation of \cite{Chang:2017xmr}), corresponding to conserved currents of higher spins $\ell+2$.} $(\cD[2], R)$ are half-BPS multiplets, whose superprimaries are $\Delta=2$ scalars in flavor irreps $R$ in the symmetric product of two adjoint representations. $(\cB[1]_j,R)$ are $\Delta=j+2$ semi-short multiplets of odd integer spin $j$, and flavor irrep $R$ in the anti-symmetric product of two adjoints. Finally, $(\mathcal{L}[0]_{\Delta,j}, R)$ are long multiplets of even integer spin $j$ and flavor irrep $R$ in the symmetric product of two adjoints, subject to the unitarity bound $\Delta>j+1$. 
	
	The explicit formulae for the superblocks $\mathfrak{G}_{\cM}$ of the BPS multiplets $\cM=\cD[1],\cD[2], \cB[0]$, relevant for the observables extracted in this paper, are given in Appendix \ref{fourApp}. 
	
	\subsection{Central charges}
	
	The central charges $c_J, c_{\widetilde J}$ and $c_T$ are defined through the 2-point functions of the canonically normalized\footnote{Note that the normalization of $j^a_\mu$ here differs from that of \cite{Chang:2019dzt} by a factor $2^{-{1\over 2}}$ due to our convention of the $SU(k)$ generators $T^a$.} flavor currrents $j^a_\mu$, $\widetilde j^A_\mu$, and the stress tensor $T_{\mu\nu}$\footnote{We use the conventions where $c_T=1$ for a free scalar or Majorana fermion. This differs from the $c_T$ normalization of \cite{Chang:2017xmr} by the factor $3\over 2$.} 
	\ie\label{cJcT}
	&\langle j_\mu^a(x) j_\nu^b(0)\rangle=\frac{c_J}{(4\pi)^2}\frac{{\rm Tr}(T^a T^b) I_{\mu\nu}(x)}{x^4}\,,\qquad \langle \widetilde j_\mu^A(x) \widetilde j_\nu^B(0) \rangle=\frac{c_{\widetilde J}}{(4\pi)^2}\frac{\delta^{AB}I_{\mu\nu}(x)}{x^4} \,,\\
	& \langle T_{\mu\nu}(x)T_{\sigma\rho}(0) \rangle=\frac{c_T}{(4\pi)^2}\frac{\left[\left(\frac12 I_{\mu\sigma}(x) I_{\nu\rho}(x)+ I_{\mu\rho}(x) I_{\nu\sigma}(x) \right)-\frac{\delta_{\mu\nu}\delta_{\sigma\rho}}{3}   \right]}{x^6}\,,\\
	\fe
	where $I_{\mu\nu}\equiv\delta_{\mu\nu}-2\frac{x_\mu x_\nu}{x^2}$. 
	
	\subsection{Kinematics of the $SU(k)$ current 4-point function $\cG^{abcd}$}
	
	The decomposition (\ref{blockExp}) for the $SU(k)$ current 4-point function $\cG^{abcd}$ involves the representations (\ref{OPE}) with $R$ labeling one of the following $SU(k)$ irreps\footnote{Here, ${\bf S}$ is the irrep with Dynkin label $[20\cdots02]$ for $k>2$ and $[4]$ for $k=2$, and ${\bf S}'$ is the irrep with Dynkin label $[010\cdots010]$ for $k>4$ and $[020]$ for $k=4$, and ${\bf A}$ is the irrep with Dynkin label $[20\cdots010]$ for $k>3$ and $[30]$ for $k=3$.}
	\es{adjoint}{
		{\bf adj}\otimes {\bf adj}={\bf 1}_s\oplus {\bf S}_s \oplus{\bf S}'_s\oplus{\bf A}_a\oplus{\bf \bar A}_a   \oplus {\bf adj}_a\oplus {\bf adj}_s \,,
	}
	for $k\geq 4$, where the subscript $s$ or $a$ refers to irreps appearing in the symmetric or anti-symmetric tensor product, respectively. In the special case $k=3$ we should remove ${\bf S}'_s$, which does not exist, whereas in the case $k=2$ we keep only ${\bf 1}_s$, ${\bf adj}_a$, and ${\bf S}$. 
	
	It will be convenient to adopt a notation in which all values of $k$ are treated on equal footing by rewriting \eqref{blockExp} as
	\es{newBasis}{
		\cG^{abcd}(U,V;\alpha)&=\sum_{\cM\in \cD[1]\times\cD[1]} \sum_{R={\bf1},{\bf2},{\bf3},{\bf4},{\bf adj}_a,{\bf asym}_a}{\mathtt T}^{abcd}_R \lambda^2_{\cM,R}\mathfrak{G}_\cM(U,V;\alpha)\,,\\
	}
	where, by a slight abuse of notation, we label by $R$ the following invariant tensors ${\mathtt T}_R$:
	\es{newBasist}{
		&{\mathtt T}_{\bf1}^{abcd}={\mathtt A}^{abcd}\,,\qquad {\mathtt T}_{\bf2}^{abcd}={\mathtt A}^{acdb}+{\mathtt A}^{adbc}\,,\quad  {\mathtt T}_{\bf3}^{abcd}={\mathtt B}^{abcd}+{\mathtt B}^{acdb}\,,\quad {\mathtt T}_{\bf4}^{abcd}={\mathtt B}^{adbc}\,, 
		\\
		&  {\mathtt T}_{{\bf adj}_a}^{abcd}={\mathtt B}^{abcd}-{\mathtt B}^{acdb}\,, \qquad {\mathtt T}_{{\bf asym}_a}^{abcd}={\mathtt A}^{abcd}-{\mathtt A}^{acdb}\, ,
	}
	where ${\mathtt A}^{abcd}$ and ${\mathtt B}^{abcd}$ are the color factors defined in (\ref{colorA}) and (\ref{colorB}).

	In this convention, the flavor current and stress tensor OPE coefficients are related, by conformal Ward identities, to the central charge $c_J$ and $c_T$ as \cite{Osborn:1993cr} 
	\es{OPEnorm2}{
		\lambda^2_{\mathcal{D}[1],{\bf adj}_a}=\frac{16}{c_J}\,,\qquad
		\lambda^2_{\mathcal{B}[0],{\bf1}}=\frac{64 }{c_T}\,  .\\
	}

	\subsection{Kinematics of the $SU(k)$-$SU(2)_F$ mixed 4-point function $\cG^{abCD}$}
	
	The decomposition (\ref{blockExp}) for the mixed current multiplet 4-point function $\cG^{abCD}$, in the $ab\to CD$ channel, involves only singlets of the flavor group (``$R=1$''). This immediately excludes the superconformal representations $\cB[1]_j$ and $\cD[1]$ from the OPE. The only flavor invariant tensor that appear in this OPE channel is
	\ie\label{tonecd}
	{\mathtt T}_{\bf1}^{abCD}=\Tr(T^a T^b) \delta^{CD} \, ,
	\fe 
	with the $\mathcal{B}[0]$ channel coefficient 
	\ie\label{doneopec}
	\lambda^2_{\cB[0]} = {64 \over c_T}\,.
	\fe
	In the $aC\to bD$ or $aD\to bC$ channel, on the other hand, $R$ is necessarily the adjoint representation with respect to $SU(k)\times SU(2)_F$, and the superconformal representations $\cB[1]_j$, rather than $\cB[0]$, appear in the OPE.

	\section{Mellin amplitudes in the large $N$ expansion}
	\label{largeN}
	
	The holographic interpretation of the CFT correlators of interest is most easily understood through its Mellin transform \cite{Mack:2009mi,Mack:2009gy, Penedones:2010ue}, whose $1/N$ expansion is naturally organized through the derivative expansion of a bulk effective field theory \cite{Heemskerk:2009pn}.

	To begin with, at strictly $N=\infty$ the correlators ${\cal G}^{abcd}$ and ${\cal G}^{abCD}$ reduce to those of the generalized free field theory (GFFT),
	\es{GFFT}{
		{\cal G}^{abcd}_\text{GFFT}(U, V; \alpha) 
		= {\mathtt A}^{abcd} + \alpha^{2} U{\mathtt A}^{dbca}+(1-\alpha)^2 \frac{U}{V} {\mathtt A}^{cbad} \,,\quad {\cal G}^{abCD}_\text{GFFT}(U, V; \alpha)= {\mathtt T}_{\bf 1}^{abCD}\,.
	}
	The (reduced) Mellin amplitudes $M^{abcd}(s,t;\A)$ and $M^{abCD}(s,t;\A)$ are defined by first subtracting from $\cG(U,V;\A)$ the GFFT correlator, and rewriting the ``connected correlator'' $\mathcal{G}_\text{conn}\equiv\mathcal{G}-\mathcal{G}_\text{GFFT}$ as the integral transform
	\es{mellinDef}{
		\mathcal{G}_\text{conn}(U,V;\alpha)=\int\frac{ds\, dt}{(4\pi i)^2} U^{\frac s2}V^{\frac t2-1}M(s,t;\alpha)\Gamma^2\left[1-\frac s2\right]\Gamma^2\left[1-\frac t2\right]\Gamma^2\left[1-\frac u2\right]\,,
	}
	where $u \equiv 4-s-t$, and the $s$, $t$ integration contours run parallel to the imaginary axis with ${\rm Re}(s),{\rm Re}(t),{\rm Re}(u)<2$, so that all poles of the Gamma functions lie on one side or the other of the contours. 
	$M(s,t;\A)$ is expected to be well-defined, order by order in the $1/N$ expansion, with suitable analyticity assumptions that are motivated from the bulk effective field theory. These assumptions, combined with the superconformal Ward identities and crossing invariance, impose nontrivial constraints on the Mellin amplitude at each order in $1/N$, dubbed the ``analytic bootstrap'' \cite{Rastelli:2017udc}.

	\subsection{Analytic bootstrap conditions and the large radius limit}

	The superconformal Ward identities \eqref{3D} can be translated into the Mellin representation via the replacement rule
	\es{wardM}{
		{\cal G}(U,V;\alpha)\to M(s,t)\,,~~~U \partial_U \to \widehat{U\partial_U}\,,~~~ V\partial_V \to \widehat{V\partial_V}\,,~~~ U^mV^n \to \widehat{U^mV^n}\,.
	}
	Here the hatted operators are defined to act on the Mellin amplitude $M(s,t)$ as \cite{Zhou:2017zaw}
	\es{3DMellin}{
		\widehat{U\partial_U} {M}(s,t)&=\frac s 2  {M}(s,t)\,,\\
		\widehat{V\partial_V} {M}(s,t)&=\left[\frac t 2-1\right] {M}(s,t)\,,\\
		\widehat{U^mV^n} {M}(s,t)&={M}(s-2m,t-2n)\left(1-\frac{s}{2}\right)_m^2\left(1-\frac{t}{2}\right)_n^2\left(1-\frac{u}{2}\right)_{-m-n}^2\,,
	}
	where $(x)_n\equiv \Gamma(x+n)/\Gamma(x)$ is the Pochhammer symbol.
	
	Crossing invariance of the CFT correlators amounts to 
	\ie\label{crossMel}
	&M^{abcd}(s,t;\alpha)=M^{bacd}(s,u;1-\alpha) 
	=(\alpha-1)^2M^{cbad}(t,s;\alpha/(\alpha-1))\,,\\
	&M^{abCD}(s,t;\alpha)=M^{baCD}(s,u;1-\alpha)\,.
	\fe
	
	At each order in the $1/N$ expansion, the leading large $s,t$ asymptotic behavior of the Mellin amplitude is controlled by the derivative expansion of the bulk effective field theory. Their precise relation is captured by the large radius limit \cite{Penedones:2010ue}, where the Mellin amplitude turns into a simple integral transform of a suitable ``flat space'' scattering amplitude
	\es{FlatLimit}{
		32\pi^{9\over 2}\lim_{L\to \infty} L^3 M^{abcd}(L^2  s, L^2  t;\alpha) &= (u+s\alpha)^2\int_0^\infty d \beta\, \beta^{3\over 2} e^{-\beta} {\cal A}^{abcd} \left( {2 \beta} s, {2 \beta}  t \right)  \, ,\\
		\frac{128\pi^{13\over 2}}{3k}  \lim_{L\to \infty} L^7 M^{abCD}(L^2  s, L^2  t;\alpha) &=\delta^{CD}(u+s\alpha)^2 \int_0^\infty d \beta\, \beta^{3\over 2} e^{-\beta} \widetilde{\cal A}^{ab} \left( {2 \beta} s, {2 \beta}  t \right)  \, .
	}
	On the LHS, $L$ is the AdS radius, related to $N$ by \eqref{dict}. Note importantly that $M(s,t;\A)$ itself has, of course, highly nontrivial $N$-dependence which is not shown explicitly in (\ref{FlatLimit}). The power of $L$ is dictated by the number of transverse dimensions of the relevant scattering amplitudes, whereas the overall factor is fixed by comparison with the reduced SYM+SUGRA tree-level amplitudes  \eqref{treeanew} and \eqref{gravampol} presented in Section~\ref{sec:averaging}.
	
	On the RHS, $\cA^{abcd}(s,t)$ comes from the 7d scattering amplitude of gluons supported at the M-theory orbifold singularity, with momenta restricted to a 4d Minkowskian subspace and polarization vectors restricted to the transverse 3d. A special choice of polarizations \cite{Chester:2018aca, Alday:2021odx}, resulting from the flat space limit of the wave functions of the KK modes in ${\rm AdS}_4\times S^3$ produces the factor $(u+s\alpha)^2$.\footnote{An analogous factor for the scattering of graviton KK modes in M-theory on ${\rm AdS}_4\times S^7$ was derived in Appendix C of \cite{Chester:2018aca}.} 
	
	Similarly, $\widetilde\cA^{ab}(s,t)$ comes from the scattering amplitude of a pair of 7d gluons with a pair of 11d gravitons, whose momenta are restricted to 4d and polarizations in the transverse 3d. Their precise relation, which involves a choice of polarizations that corresponds to the integration over KK wave functions and produces the factor $\delta^{CD}(u+s\alpha)^2$, will be explained in the next subsection.

	\subsection{Relation to 7d amplitudes}
	\label{sec:averaging}
	
	We now explain the precise relation between the 7d gluon and graviton amplitudes in the locally flat M-theory orbifold, considered in Section \ref{sec:amps}, and $\cA^{abcd}(s,t)$, $\widetilde\cA^{ab}(s,t)$  that arise in the flat space limit (\ref{FlatLimit}) of Mellin amplitudes in ${\rm AdS}_4\times S^3$.
	
	We begin with the four-gluon superamplitude (\ref{asupfull}), and restrict to the bosonic gluon states with momenta $p_i$ in a 4d subspace, and polarization vectors $\epsilon_i$ restricted to the 3 transverse directions. The integration over the relevant KK wave functions in ${\rm AdS}_4\times S^3$ amounts to further specializing the polarization vector $\epsilon_i$ in a way that is correlated with the $SU(2)_H$ spinor polarization $y^\A$ in (\ref{yydef}), via \cite{Alday:2021odx}
	\ie
	(\epsilon_i)^{\A\B} = y_i^\A y_i^\B,
	\fe
	where we have used the 3d bispinor notation for $\epsilon_i$. In terms of the super helicity variables introduced in Appendix \ref{sec:superspinorhelicity}, this amounts to specializing to the super spinor helicities 
	\ie\label{lambdarestr}
	& \lambda_{i ({\mathfrak a},\A)(+, \gamma)} = \lambda_{i\mathfrak{a}} \epsilon_{\A\gamma} \,,~~~\lambda_{i ({\mathfrak a},\A)(-, \gamma)} =\lambda_{i (\dot{\mathfrak a},\A)(+, \gamma)} = 0 \,,~~~
	\lambda_{i (\dot{\mathfrak a},\A)(-, \gamma)} = \widetilde\lambda_{i\dot{\mathfrak a}} \epsilon_{\A\gamma} \,,
	\\
	&
	\eta_{i(\varkappa,\gamma)} = \eta_{i\varkappa} y_{i\gamma} \,,
	\fe
	where we have traded the $SO(1,6)$ Lorentz spinor index $\A$ with $SO(1,3)\times SO(3)$ spinor indices $({\mathfrak a}, \A)$ and $(\dot{\mathfrak a},\A)$, and $SO(5)$ little group spinor index $I$ with $SO(2)\times SO(3)$ indices $(\varkappa=\pm, \gamma)$. $(\lambda_{i{\mathfrak a}}, \widetilde\lambda_{i\dot{\mathfrak a}}, \eta_{i\varkappa})$ are the 4d super-spinor-helicity variables associated with the $i$-th particle. The super momentum of the $i$-th particle is now written as
	\ie
	q_{i({\mathfrak a}, \A)} = \lambda_{i \mathfrak{a}} \eta_{i+} y_{i\A},~~~ q_{i(\dot{\mathfrak a}, \A)} = \widetilde\lambda_{i \dot{\mathfrak a}} \eta_{i-} y_{i\A} \,,
	\fe
	resulting in the polarization factor below which can be fixed easily by going to a specific kinematic frame (see also \cite{Alday:2021odx})
	\ie\label{gluonreduce}
	\delta^8(Q)  
	\propto \left(\prod_{i=1}^4 \eta_{i+}\eta_{i-}\right) \det_{i, {\mathfrak a}\A}( \lambda_{i {\mathfrak a}} y_{i\A}) \det_{i, {\dot{\mathfrak a}}\B}(\widetilde\lambda_{i\dot {\mathfrak a}} y_{i\B}) =  \left(\prod_{i=1}^4 \eta_{i+}\eta_{i-}\right)  \la y_1,y_2\ra^2 \la y_3,y_4\ra^2(u + s\A)^2.
	\fe
	Therefore, up to an overall normalization convention, the reduced 4d amplitude $\cA^{abcd}$ appearing in (\ref{FlatLimit}) is obtained from the superamplitude $\mathfrak{A}$ (\ref{asupfull}) by simply stripping off the $\delta^8(Q)$ factor. We adopt the normalization convention in which the SYM tree-level contribution is
	\es{treeanew}{
		\cA^{abcd}_{F^2}(s,t)&= 2g_{\rm YM}^2 \left(\frac{{\mathtt B}^{abcd}}{st}+\frac{{\mathtt B}^{acdb}}{su}+\frac{{\mathtt B}^{adbc}}{tu}\right)\,.
	}
	
	Similarly, for the two-gluon-two-graviton amplitude (\ref{amixed}), we will restrict to bosonic gluon and graviton states with momenta $p_i$ in a 4d subspace, and polarization tensors restricted to 3 transverse directions along the orbifold locus. The effect of integration over KK wave functions on ${\rm AdS}_4\times S^3$ leads to, in the flat space limit, the restrictions (\ref{lambdarestr}), together with
	\ie
	\widetilde\eta_{i(\varkappa,\gamma)} = \widetilde\eta_{i\varkappa} \widetilde y_{i\gamma} \,,
	\fe
	where $\widetilde y_3, \widetilde y_4$ are $SU(2)_F$ polarization spinors. This results in the polarization factor below (for the restricted super-spinor-helicity variables),
	\ie\label{gravampol}
	&\delta^8(Q)  \left. \partial_{\widehat\eta_1}^2\partial_{\widehat\eta_2}^2 \prod_{\A=1}^8 \left( \lambda^\A_{1I} \widehat\eta_1^I+\lambda^\A_{2I} \widehat\eta_2^I + \widetilde q_3^\A + \widetilde q_4^\A \right) \right|_{\widetilde\eta_3^2|_{\bf 5}\widetilde\eta_4^2|_{\bf 5}} 
	\\
	&\propto \left(\prod_{i=1}^4 \eta_{i+}\eta_{i-}\right) \det_{i, {\mathfrak a}\A}( \lambda_{i {\mathfrak a}} y_{i\A}) \det_{i, {\dot{\mathfrak a}}\B}(\widetilde\lambda_{i\dot {\mathfrak a}} y_{i\B}) \cdot  \left(\prod_{i=3,4} \widetilde\eta_{i+}\widetilde\eta_{i-}\right) t u \langle \widetilde y_3, \widetilde y_4\rangle^2
	\\
	& = \, \left(\prod_{i=1}^4 \eta_{i+}\eta_{i-}\right )  \left(\prod_{i=3,4} \widetilde\eta_{i+}\widetilde\eta_{i-}\right)  \la y_1,y_2\ra^2 \la y_3,y_4\ra^2 \la \widetilde y_3, \widetilde y_4\rangle^2  t u ( u+ s\A)^2\,,
	\fe
	where the first line implements the projection of the four-point supersymmetric invariant $\D^{16}(Q)$  for the 7d maximal supergravity  (see \cite{Wang:2015aua} for details) to the  2-gluon-2-graviton component (see Footnote~\ref{fnote:embeddingtrick}). The second line of \eqref{gravampol} follows easily after using the $SO(5)_R$ invariance of the four-point invariant $\D^{16}(Q)$  with momentum conservation \cite{Wang:2015aua} which allows us to replace $\partial^2_{\hat \eta_2}$ by  $\partial^2_{\hat \eta_1}$ in the first line. To obtain the final equality in \eqref{gravampol} we have used \eqref{gluonreduce}.
	
	Applying (\ref{gravampol}) to $\widetilde{\mathfrak A}_R$ (\ref{arsuper}), we obtain the leading low energy contribution to the reduced 4d amplitude $\widetilde\cA^{ab}$ appearing in (\ref{FlatLimit}), 
	\es{treeA2new}{
		\widetilde \cA^{ab}_{R}(s, t)&=  - \kappa^2 {\rm Tr}(T^a T^b) \,{1\over s}\,. %factor 64 stripped off 
	}
	Note that this reduced amplitude no longer has the gluon pole in the $t$ or $u$ channel.

	\subsection{Structure of the four-gluon Mellin amplitude $M^{abcd}$}
	\label{sec:mabcd}
	
	The analytic bootstrap conditions constrain the $1/N$ or large radius expansion of the gluon Mellin amplitude $M^{abcd}(s,t;\alpha)$ to be of the form
	\es{gluon}{
		M^{abcd}&=\frac{M^{abcd}_{F^2}}{c_J}+\frac{M^{abcd}_{F^2|F^2}}{c_J^2}+\frac{\kappa_{[F^2]^2}M^{abcd}_{[F^2]^2}+\kappa_{F^4}M^{abcd}_{F^4}}{N^{7\over 6}}\\
		&{}+\Bigg[\frac{M^{abcd}_{R}}{c_T}+\frac{M^{abcd}_{F^2|F^2|F^2}}{c_J^3}+\frac{\widetilde\kappa_1 M^{abcd}_{[F^2]^2}+ \widetilde\kappa_2 M^{abcd}_{F^4}+  \widetilde\kappa_3 M^{abcd}_{D^2 [F^2]^2}+ \widetilde\kappa_4 M^{abcd}_{D^2 F^4}}{N^{3\over 2}}\Bigg]\\
		&{}+\frac{\log c_J}{c_J^3}\Big[\kappa'_{\log,[F^2]^2}M^{abcd}_{[F^2]^2}+\kappa'_{\log,F^4}M^{abcd}_{ F^4}+\kappa_{\log,D^2[F^2]^2}M^{abcd}_{D^2 [F^2]^2}+\kappa_{\log,D^2F^4}M^{abcd}_{D^2  F^4}\Big] \\
		&{}+O(N^{-{11\over 6}})\,,
	}
	where $c_J$ and $c_T$, defined by (\ref{cJcT}) with the explicit large $N$ expressions given in (\ref{cJ2}) and (\ref{cTisrrael}), scale with $N$ as $\sim N^{1\over 2}$ and $\sim N^{3\over 2}$ respectively. $M_{F^2}^{abcd}$, $M_{F^2|F^2}^{abcd}$, and  $M_{F^2|F^2|F^2}^{abcd}$ are the tree-level, one-loop, and (the finite part of) two-loop Yang-Mills Mellin amplitudes, respectively, with the Yang-Mills coupling constant stripped off. $M^{abcd}_{[F^2]^2}$ and $M^{abcd}_{F^4}$ are contributions to the gluon Mellin amplitude from supersymmetric effective couplings of the schematic form $(\Tr F^2)^2$ and $\Tr F^4$, respectively. Likewise, $M^{abcd}_{D^2 [F^2]^2}$ and $M^{abcd}_{D^2 F^4}$ capture the contribution from the analogous effective couplings with two extra derivatives. The coefficients $\kappa_{[F^2]^2}$, $\kappa_{F^4}$, etc., to be determined later, are by definition independent of $s, t$ and do not scale with $N$.
	
	The leading term on the RHS of (\ref{gluon}), namely the tree-level Mellin amplitude ${1\over c_J}M^{abcd}_{F^2}$, comes from the Witten diagram with a gluon exchange, shown in the first diagram of Figure~\ref{fig:AAAA}. In terms of the superconformal block expansion of the CFT correlator, it involves the flavor multiplet superblock $\mathfrak{G}_{\cD[1]}$ appearing on the RHS of (\ref{newBasis}), whose coefficient is $\lambda^2_{\cD[1]}\propto 1/c_J$ \eqref{OPEnorm2}. The explicit expression of $M^{abcd}_{F^2}$, derived in \cite{Alday:2021odx}, is
	\ie\label{mero}
	M^{abcd}_{F^2}(s,t;\A)
	&= \left[ {16} \left({\mathtt B}^{abcd} -{\mathtt B}^{acdb}\right)\left(P_1(2\alpha-1)G^{M}_{1,0}(s,t)-G^M_{2,1}(s,t)\right)+4{\mathtt B}^{abcd}\frac{2-\alpha^2-2\alpha}{\pi^2} \right]
	\\
	&~~~ + \left.\Big[ \cdots \Big]\right|_{(a\leftrightarrow b, ~t\leftrightarrow u,~\A\to 1-\A)}
	+ (\A-1)^2 \left.\Big[ \cdots \Big]\right|_{(a\leftrightarrow c, ~s\leftrightarrow t,~\A\to {\A\over \A-1})} \,,
	\fe
	where $P_n(x)$ is the $n$-th Legendre polynomial, and $G^M_{\Delta,j}(s,t)$ is the singular part of the Mellin space conformal blocks of weight $\Delta$ and spin $j$ \cite{Zhou:2017zaw}, given explicitly in (\ref{Mblocks}). The second line of (\ref{mero}) contains two terms related by crossing, where the $\left[\cdots\right]$ represents the same expression as in the first line.
	
	In particular, (\ref{mero}) contains poles in $s$ that correspond to the exchange of single trace operators in the flavor multiplet with all possible odd positive integer twists. Its large $s,t$ asymptotic expression is
	\es{F2}{
		M^{abcd}_{F^2} \to ~ &\frac{8}{\pi^2}(u+s\alpha)^2\left[ \frac{{\mathtt B}^{abcd}}{st} +\frac{{\mathtt B}^{acdb}}{su} +\frac{{\mathtt B}^{adbc}}{tu} \right],~~~s,t\to\infty.\\
	}
	The overall $s^0$ momentum scaling is expected from the first diagram of Figure \ref{fig:AAAA}, where each Yang-Mills vertex contributes 1 to the momentum power counting, whereas the propagator subtracts 2. Indeed, one can apply the 7d flat space limit \eqref{FlatLimit} to ${1\over c_J}M^{abcd}_{F^2}$, and recover the expected flat space tree-level gluon scattering amplitude (with specialized polarizations) \eqref{treeanew}.
	
	The second term on the RHS of (\ref{gluon}), ${1\over c_J^2}M^{abcd}_{F^2|F^2}$, is a 1-loop Mellin amplitude that admits a unitarity cut into a pair of tree-level Yang-Mills ``$F^2$'' amplitudes, as depicted in the third diagram of Figure \ref{fig:AAAA}. In principle, this 1-loop Mellin amplitude can be determined from the ``tree-level'', i.e. leading large $N$, CFT data, using the AdS unitarity cut method of \cite{Aharony:2016dwx}. Its explicit computation requires unmixing a rather involved set of degenerate tree-level data, which we will not pursue here.\footnote{We briefly discuss its expected flavor structure in Appendix \ref{loopApp}.} Its flat space limit is governed by (\ref{a1loop}), with $s^{3\over 2}$ overall momentum scaling.
	
	The third term on the RHS of (\ref{gluon}), of order $N^{-{7\over 6}}$, corresponds to Witten diagrams that involve contact interaction vertices, or effective couplings, of the form $\kappa_{[F^2]^2} (\Tr F^2)^2$ and $\kappa_{F^4} \Tr F^4$, where the precise contractions of tensor indices are fixed by supersymmetry. The large $N$ scaling follows from the AdS/CFT dictionary \eqref{dict} and derivative counting. In practice, the Mellin amplitudes $M^{abcd}_{[F^2]^2}$ and $M^{abcd}_{F^4}$ can be determined from the expectation that they are degree 2 polynomials in $s$ and $t$, and that they obey the superconformal Ward identities and crossing invariance. Their explicit expressions are 
	\es{d21}{
		& M^{abcd}_{[F^2]^2}(s,t;\alpha) = \Theta(s,t,\alpha)\left({\mathtt A}^{abcd}+{\mathtt A}^{acdb}+{\mathtt A}^{adbc}\right) \,,
		\\
		& M^{abcd}_{ F^4}(s,t;\alpha)= \Theta(s,t,\alpha) \left({\mathtt B}^{abcd}+{\mathtt B}^{acdb}+{\mathtt B}^{adbc}\right)\,,
	}
	where
	\ie\label{theta}
	\Theta(s,t,\alpha)=\frac{1}{3} \left(16 \alpha ^2-8 \alpha +3 (\alpha -1)^2 s^2-2 (\alpha -1) s (7 \alpha
	+3 t-5)+3 t^2-2 (2 \alpha +5) t+8\right)\,.\\
	\fe
	Note that in the large $s,t$ limit, $\Theta(s,t,\alpha)\to(u+s\alpha)^2$.
	The coefficients $\kappa_{[F^2]^2}$ and $\kappa_{F^4}$ are the same as those appearing in the flat space amplitude (\ref{af4coup}), and are thus far unconstrained by superconformal symmetry and the general structure of the large radius expansion. We will determine them in the next section with the additional input of supersymmetric localization data.
	
	In the second line of (\ref{gluon}), we find three different structures that arise at order $N^{-{3\over 2}}$: the tree-level Mellin amplitude with a graviton exchange ${1\over c_T}M^{abcd}_{R}$, the gluon two-loop Mellin amplitude ${1\over c_J^3}M^{abcd}_{F^2|F^2|F^2}$, and additional contributions that can be viewed as due to contact interactions. The graviton exchange term $M^{abcd}_{R}$ contains the stress tensor multiplet superblock $\mathfrak{G}_{\cB[0]}$ in the CFT correlator (\ref{newBasis}), whose overall coefficient is $\lambda^2_{\cB[0]}\propto 1/c_T$ \eqref{OPEnorm2}, together with an infinite tower of long multiplet superblocks whose coefficients also scale like $1/c_T$.\footnote{The $\mathcal{N}=8$ half-BPS KK modes on ${\rm AdS}_4\times S^7$ appear in rank $p$ traceless symmetric irreps of $SO(8)_R$ with $\Delta=p$. Upon reduction to $\mathcal{N}=4$ multiplets of ${\rm AdS}_4\times S^7/\mathbb{Z}_k$, only even $p$ modes include a long multiplet that will contribute to $M^{abcd}_{R}$.}  The gluon 2-loop term $M^{abcd}_{F^2|F^2|F^2}$ can in principle be fixed by AdS unitarity cuts and lower order amplitudes, up to polynomial ambiguities which can be absorbed into the coefficients $\widetilde\kappa_i$ of the contact Mellin amplitudes. 
	
	Up to local terms, the flat space limit of $M^{abcd}_R$ is governed by (\ref{arloop}), whereas the flat space limit of $M^{abcd}_{F^2|F^2|F^2}$ is governed by (\ref{A2loop}), both having $s^3 \log s$ scaling at large momenta.
	However, unlike the flat space graviton exchange and gluon 2-loop amplitudes whose log divergences are cutoff at Planck scale $1/\ell_p$, $M^{abcd}_R$ and $M^{abcd}_{F^2|F^2|F^2}$ involve only the AdS radius scale $1/L$. The comparison with the flat space limit indicates that there must be additional contact interaction terms in the Mellin amplitude of order $N^{-{3\over 2}} \log (L/\ell_p)$ (or equivalently $\log c_J\over c_J^3$), that are proportional to a pair of degree 3 polynomials $M^{abcd}_{D^2[F^2]^2},M^{abcd}_{D^2 F^4}$ corresponding to the $D^2(\Tr F^2)^2$ and $D^2\Tr F^4$ supervertices in the flat space limit, up to degree 2 ambiguities proportional to $M^{abcd}_{[F^2]^2} ,M^{abcd}_{D^2 F^4}$ (\ref{d21}). We can determine the explicit expression of $M^{abcd}_{D^2[F^2]^2}$ and $M^{abcd}_{D^2 F^4}$ by solving the analytic bootstrap conditions, 
	\ie\label{d31}
	M^{abcd}_{D^2 [F^2]^2}(s,t;\alpha)&={\mathtt A}^{abcd}\left[
	(\alpha -1)^2 s^3+\frac{s}{5}  \left(4 \left(7 \alpha ^2+4 \alpha -4\right)+5 t^2-24 \alpha
	t\right)\right.\\
	&\left.\qquad+\frac{4}{5}  (t-2) (-2 \alpha +2 t-3)-\frac{2}{5}  (\alpha -1) s^2 (12 \alpha +5
	t-4)
	\right]+({\rm cross~terms}),
	\\
	M^{abcd}_{D^2  F^4}(s,t;\alpha)&=-\frac{ s+t-2}{5}{\mathtt B}^{abcd}\left[
	28 \alpha ^2+5 (\alpha -1)^2 s^2-2 (\alpha
	-1) s (12 \alpha +5 t-4)+5 t^2-8 (2 \alpha +1) t
	\right]
	\\
	&~~~~~~ + ({\rm cross~terms}) ,
	\fe
	where the cross terms are related by the same manner as in (\ref{mero}). The large $s, t$ asymptotic expressions of (\ref{d31}) are
	\ie{}
	M^{abcd}_{D^2[F^2]^2} &\to (u+s\alpha)^2\left( s{\mathtt A}^{abcd}+u{\mathtt A}^{acdb}+t{\mathtt A}^{adbc} \right),
	\\
	M^{abcd}_{D^2 F^4} &\to (u+s\alpha)^2 \left( u{\mathtt B}^{abcd}+t{\mathtt B}^{acdb}+s{\mathtt B}^{adbc} \right).
	\fe
	Without the explicit expressions of $M^{abcd}_R$ and $M^{abcd}_{F^2|F^2|F^2}$, we cannot meaningfully determine the $\widetilde \kappa$ coefficients in the second line of (\ref{gluon}). We can, however, unambiguously determine the order $\log c_J\over c_J^3$ term coefficients $\kappa'_{\log,[F^2]^2}$, $\kappa'_{\log,F^4}$, $\kappa_{\log,D^2[F^2]^2}$, $\kappa_{\log,D^2F^4}$ appearing in the third line of (\ref{gluon}), as will be explained in Section~\ref{MtheoryCon}.

	\subsection{Structure of the mixed gluon-graviton Mellin amplitude $M^{abCD}$}
	\label{sec:mellinmixed}
	
	We next consider the mixed gluon-graviton Mellin amplitude $M^{abCD}(s,t;\alpha)$, whose large radius expansion is constrained by the analytic bootstrap to be of the form
	\es{gluonM}{
		M^{abCD}&=\frac{M^{abCD}_{R}}{c_T}+\frac{\kappa_{RF^2}M_{RF^2}^{abCD}}{N^{11\over 6}}+\Bigg[\frac{M^{abCD}_{F^2|R}}{c_J c_T}+\frac{\widetilde\kappa M^{abCD}_{RF^2}}{N^{2}}\Bigg]\\
		&+\frac{\kappa'_{RF^2}M^{abCD}_{RF^2}+ \kappa_{R^2F^2}M^{abCD}_{R^2F^2}}{N^{13\over 6}}+O(N^{-{15\over 6}})\,.
	}
	Here $M^{abCD}_{R}$ and $M^{abCD}_{F^2|R}$ are tree-level and one-loop gluon-graviton Mellin amplitudes respectively, whereas $M^{abCD}_{RF^2}$ and $M^{abCD}_{R^2F^2}$ (given in (\ref{polyM})) are degree 2 and 3 polynomials in $s, t$. The coefficients $\kappa_{RF^2}$, $\kappa_{R^2F^2}$ etc., to be determined, are independent of $s, t$ and do not scale with $N$.
	
	The tree-level term ${1\over c_T} M^{abCD}_{R}$, with the explicit expression
	\es{mero3}{
		M^{abCD}_{R}(s,t;\alpha)&= {\mathtt T}_{\bf 1}^{abCD} \Big[ {128}  \left(G^{M}_{1,0}(s,t)-2 P_1(2\alpha-1)G^M_{2,1}(s,t)+G^M_{3,2}(s,t)\right)\\
		&+\frac{8}{\pi^2}\left(\left(-8 \alpha ^2+16 \alpha -7\right) s+4 \left(4 \alpha ^2-12 \alpha +(4
		\alpha -2) t+3\right)\right) \Big]\,,\\
	}
	contains graviton exchange poles (in the $s$-channel), corresponding to the second diagram of Figure \ref{fig:AAgg}. From the point of view of superconformal block decomposition of the mixed correlator  $\cG^{abCD}$\eqref{4point1} in the $ab\to CD$ channel, the graviton exchange involves the stress tensor ${\cal B}[0]$ superblock, whose  coefficient is $\lambda^2_{\cB[0]}\propto 1/c_T$ \eqref{OPEnorm2}. In the large radius limit, using \eqref{FlatLimit}  one can recover from $\frac{1}{c_T}M^{abCD}_{R}$ the flat space tree-level mixed gluon-graviton amplitude (with specialized polarizations) \eqref{treeA2new}.
	
	Note that $M^{abCD}_{R}$ does not contain gluon exchange poles (in the $t$- or $u$-channel), which one might have naively expected from the first diagram of Figure \ref{fig:AAgg}. However, as seen in Section~\ref{sec:averaging}, the kinematic restriction of the corresponding flat space amplitude is such that the gluon poles drop out. From the perspective of the superconformal block decomposition of $\cG^{abCD}$ in the cross channels $aC\to bD$ or $aD\to bC$, the tree-level Mellin amplitude involves the twist-2 $\cB[1]_j$ superblocks, which do not have poles in $t$ or $u$ as explained in \cite{Zhou:2017zaw}.

	The possibility of a flat space $R\Tr F^2$ 4-derivative effective coupling \cite{Lin:2015ixa}  suggests a potential contribution to $M^{abCD}$ at order $N^{-{11\over 6}}$. Naively, one might expect that such a Mellin amplitude contains gluon or graviton poles. However, as explained following (\ref{arf2}), the $R\Tr F^2$ 3-point vertex in fact vanishes when the graviton momentum is restricted to 7d, and thus we expect the $R\Tr F^2$ contribution to the Mellin amplitude, denoted by $M^{abCD}_{RF^2}$, to be free of poles and takes the form of a degree 2 polynomial in $s$ and $t$. The only candidate expression that solves the superconformal Ward identities is 
	\ie\label{polymtwo}
	M^{abCD}_{RF^2}(s,t;\alpha)= {\mathtt T}_{\bf 1}^{abCD} \Theta(s,t,\alpha) \, ,
	\fe
	where $\Theta$ is defined in (\ref{theta}).

	At order $N^{-2}$, we expect that the only contribution to the gluon-graviton Mellin amplitude arises from the 1-loop Witten diagram of SYM+SUGRA (third diagram of Figure \ref{fig:AAgg}), as the latter is free of logarithmic divergences, and there should be no independent contact interactions that contribute at this order. In practice, one may determine the non-polynomial part of the 1-loop contribution, ${1\over c_J c_T} M^{abCD}_{F^2|R}$, via AdS unitarity cuts, in terms of tree-level data. Note that it has $s^{5\over 2}$ overall momentum scaling in the large radius limit. Such a unitarity cut computation would leave a polynomial ambiguity, in this case of degree 2 in $s$ and $t$, which is necessarily proportional to $M^{abCD}_{RF^2}$.\footnote{An analogous ambiguity exists for the 1-loop amplitude in 3d $\mathcal{N}=8$ ABJM theory, which in that case was fixed using localization to a value such that the corresponding CFT data is analytic for all spins \cite{Alday:2021ymb,Alday:2022rly}.}

	At order $N^{-{13\over 6}}$, one encounters potential contributions from contact interactions of the form $R^2 \Tr F^2$, which take the form of a degree 3 polynomial in $s$ and $t$, and are further determined by superconformal Ward identities to be a linear combination of $M^{abCD}_{RF^2}$ (\ref{polymtwo}) and 
	\es{polyM}{
		& M^{abCD}_{R^2F^2}=
		{\mathtt T}_{\bf 1}^{abCD}\Big[  (\alpha -1)^2 s^3+\frac{s}{5}  \left(4 \left(7 \alpha ^2+4 \alpha -4\right)+5 t^2-24 \alpha
		t\right)\\
		&\qquad\qquad\qquad+\frac{4}{5}  (t-2) (-2 \alpha +2 t-3)-\frac{2}{5}  (\alpha -1) s^2 (12 \alpha +5
		t-4) \Big] \,.
	}
	The coefficient of the latter, $\kappa_{R^2F^2}$, is the same as the coefficient appearing in the flat space amplitude (\ref{ar2f2s}).

	\section{BPS-protected CFT data}
	\label{loc}
	
	\subsection{The 1d topological sector}
	
	Every 3d $\cN=4$ SCFT contains two ``1d topological sectors,'' each generated by a special family of half-BPS operators \cite{Chester:2014mea,Beem:2016cbd,Dedushenko:2016jxl,Dedushenko:2017avn,Dedushenko:2018icp}.  The 1d topological operators of interest to this work belong to one of the two sectors and are constructed as 
	\es{1dJDef}{
		\widehat {J^a}(\tau) &=  {\sqrt{c_J}\over 16\sqrt{2} \pi}\left.J^a(\vec x;y) \right|_{\vec x = (0,\tau,0), y = (1,\tau)},
		\\
		\widehat {\widetilde J^A}(\tau) &=  {\sqrt{c_{\widetilde J}}\over 16\sqrt{2} \pi}\left.\widetilde J^A(\vec x;y) \right|_{\vec x = (0,\tau,0), y = (1,\tau)}\,,
	}
	where the normalization was chosen such that these operators couple in a simple way to the mass parameters that we will introduce later.  As shown in \cite{Chester:2014mea}, the variable $\tau$ may be viewed as valued in $\mathbb{R}\cup \{\infty\}\simeq S^1$, such that the correlators of $\widehat {J^a}$ and $\widehat {\widetilde J^A}$ depend only on the cyclic ordering of their $\tau$-coordinates.
	
	Indeed, the 1d topological 4-point function of $\widehat {J^a}$ and $\widehat {\widetilde J^A}$ can be obtained from $\cG^{}(U,V;\alpha)$  (defined as in (\ref{4point1})) by simply setting
	\ie\label{toprest}
	\eta=1, ~~~~\alpha=1/\sqrt{U},
	\fe
	where the $\eta$ variable is defined in (\ref{reta}). This is such that only the half-BPS superblocks survive, and that the dependence on the remaining cross-ratio drops out.
	
	The 1d topological correlator associated with $\cG^{abcd}$ is given, for $k\geq 4$, by\footnote{The $c_J^2$ factor outside the parantheses comes from the noncanonical two-point function \eqref{cJcT}, and the extra factors of $\frac12$ and $\frac38$ come from the $P_r\left(2\A -1\right)$ in the definition of the superblocks \eqref{superblocks}, where only the conformal block associated with the superconformal primary survives.} 
	\es{1d4}{
		\left\langle \widehat {J^a}(\tau_1) \widehat {J^b}(\tau_2) \widehat {J^c}(\tau_3) \widehat {J^d}(\tau_4) \right\rangle=&\frac{c_J^2}{(512\pi^2)^2}\Bigg[{\mathtt T}_{\bf1}^{abcd}+\frac12 \sgn(\tau_{12}\tau_{13}\tau_{24}\tau_{34})\lambda^2_{\mathcal{D}[1],{\bf adj}_a} {\mathtt T}_{{\bf adj}_a}^{abcd}\\
		&+\frac38 \Big(   \lambda^2_{\mathcal{D}[2],{\bf 1}}{\mathtt T}_{\bf1}^{abcd}+  \lambda^2_{\mathcal{D}[2],{\bf2}} {\mathtt T}_{\bf2}^{abcd}+ \lambda^2_{\mathcal{D}[2],{\bf3}}{\mathtt T}_{\bf3}^{abcd}+\lambda^2_{\mathcal{D}[2],{\bf4}} {\mathtt T}_{\bf4}^{abcd} \Big)\Bigg]\,,
	}
	where $\tau_{ij}\equiv \tau_i - \tau_j$, and the various OPE coefficients and flavor tensors are defined as in \eqref{newBasis} and \eqref{newBasist}. The superconformal multiplets that appear in the OPE include the identity, the flavor multiplet $\cD[1]$, as well as four $\cD[2]$ multiplets that come in the four $SU(k)$ irreps that appear in the symmetric part of the tensor product \eqref{adjoint}. In the special case $k=3$, (\ref{1d4}) still holds with the term associated with the representation $(\cD[2],{\bf 4})$ dropped, whereas in the case $k=2$ both $(\cD[2],{\bf 3})$ and $(\cD[2],{\bf 4})$ terms are dropped.
	
	Swapping $(a,\tau_1)\leftrightarrow (c,\tau_3)$ on the LHS of  (\ref{1d4}) leads to the nontrivial crossing equations\footnote{Here we have substituted $\lambda^2_{\cD[1], {\bf adj}_a}$ with its expression in terms of $c_J$ \eqref{OPEnorm2}. }
	\es{1dcross}{
		&k\geq2:\quad 0=\frac{8}{3}+\lambda^2_{\mathcal{D}[2],{\bf1}}-\lambda^2_{\mathcal{D}[2],{\bf2}},
		\\
		&k\geq4:\quad0=\lambda^2_{\mathcal{D}[2],{\bf3}}-\lambda^2_{\mathcal{D}[2],{\bf4}}-\frac{64}{3c_J}\,,\\
		& k=3:\quad0=\lambda^2_{\mathcal{D}[2],{\bf3}}-\frac12\lambda^2_{\mathcal{D}[2],{\bf1}}-\frac{4}{3}+\frac{64}{9c_J}\,,
	}
	so there is just one crossing equation for $k=2$, but two otherwise.
	
	The 1d topological correlator associated with $\cG^{abCD}$ can be derived similarly, and gives 
	\es{1d42}{
		&\left\langle \widehat {J^a}(\tau_1) \widehat {J^b}(\tau_2) \widehat {\widetilde J^C}(\tau_3) \widehat {\widetilde J^D}(\tau_4) \right\rangle= \frac{c_Jc_{\widetilde J}}{(512\pi^2)^2}   {\mathtt T}_{\bf 1}^{abCD}\big(1+\frac38\lambda^2_{\mathcal{D}[2]}\big)\,,
	}
	where $\lambda_{\cD[2]}$ is the OPE coefficient of the flavor singlet $\cD[2]$ multiplet that appears in (\ref{blockExp}) and (\ref{OPE}), and the flavor tensor ${\mathtt T}_{\bf 1}^{abCD}$ is defined as in (\ref{tonecd}). Note that if we considered the $aC\to bD$ or $aD\to bC$ channels, where the only flavor channel is the adjoint, then we would not have an identity block contribution.
	
	After taking into account \eqref{OPEnorm2} and (\ref{1dcross}), there remain two independent OPE coefficients that appear in the 1d topological correlator (\ref{1d4}) in the case $k\geq 4$, or one independent OPE coefficient in  (\ref{1d4}) in the case $k=2,3$, and one OPE coefficient that appear in the mixed 1d topological correlator (\ref{1d42}). These will be determined using supersymmetric localization.
	
	\subsection{Relation to the sphere partition function $Z$}
	
	A 3d $\cN=4$ gauge theory with hypermuiltiplet matter can be defined on $S^3$ via an action of the schematic form $S = S_{\rm hyper}[{\cal H}, {\cal V}] + S_{\rm YM}[{\cal V}]$ \cite{Dedushenko:2016jxl}, where the hypermultiplet coupling $S_{\rm hyper}[{\cal H}, {\cal V}]$ preserves $osp(4|4)$ superconformal symmetry, while the Yang-Mills action $S_{\rm YM}[{\cal V}]$, which preserves half of the supersymmetries, is irrelevant in the IR\@. One can further deform the theory on $S^3$ by hypermultiplet mass terms in way that preserves 8 out of the 16 supercharges. The partition function $Z$ of this theory, on one hand, can be evaluated exactly via supersymmetric localization, and on the other hand, captures the 1d topological correlators through its dependence on the mass parameters.
	
	Specializing to the 3d ${\cal N}=4$ $U(N)$ gauge theory of interest, namely one with 1 adjoint and $k$ fundamental hypermultiplets, let $m_a$ be the mass parameters for the fundamental hypermultiplets, where $a$ is an adjoint index of the $SU(k)$ flavor symmetry, and let $M_A$ be the mass parameters for the adjoint hypermultiplet, where $A$ is an adjoint index of $SU(2)_F$. The $S^3$ partition function of this mass deformed theory is denoted $Z(m,M)$. It relates to the central charges $c_J$ and $c_{\widetilde J}$ by \cite{Closset:2012vg}
	\es{int2}{
		\frac{\partial_{m_a} \partial_{m_b} Z}{Z}\Big\vert_{m,M=0} = - {\pi^2\over 8} c_J {\rm Tr}(T^a T^b) \,,\qquad \frac{\partial_{M_A} \partial_{M_B} Z}{Z}\Big\vert_{m,M=0} = - {\pi^2\over 8} c_{\widetilde J} \delta^{AB} \,.
	}
	
	The partition function $Z(m, M)$ can also be related to correlation functions of the topological operators \eqref{1dJDef}, as follows.   A Weyl transformation from $\R^3$ to a round $S^3$ shows that the 1d topological sector, originally defined in \eqref{1dJDef} on a straight line in $\R^3$ that passes through the origin, can also be defined on $S^3$.  Since a straight line passing through the origin gets mapped to a great circle on $S^3$, the topological sector on $S^3$ will be defined on this great circle.   As shown in \cite{Dedushenko:2016jxl} in a more general setup that includes the theory of interest here, one can use supersymmetric localization for the theory on $S^3$ to show that the 1d sector is described by a gauged quantum mechanics theory with the same partition function $Z(m, M)$ as the 3d theory.  Furthermore, if one deforms the $S^3$ theory by real mass parameters $m_a$ and $M_A$, the 1d gauged quantum mechanics gets modified by the addition of the operator
	\es{OpInsertion}{
		4 \pi i \int d \varphi\,  \left( m_a \widehat {J^a}(\varphi) + M_A \widehat {J^A}(\varphi) \right)
	}
	to the 1d action.\footnote{For a general proof that does not rely on supersymmetric localization, see \cite{Bomans:2021ldw,Guerrini:2021zuk}.} The normalizations of the operators $\widehat {J^a}$ and $\widehat {J^A}$ in \eqref{1dJDef} were chosen precisely such that \eqref{OpInsertion} holds without additional prefactors.  One can check that this is the case by reproducing \eqref{int2} from the point of view of the 1d theory.  Indeed, we have 
	\es{twoptAgain}{
		\frac{\partial_{m_a} \partial_{m_b} Z}{Z}\Big\vert_{m,M=0} &= (4 \pi i)^2 \int d \varphi_1\, d\varphi_2 \, 
		\left\langle \widehat {J^a}(\varphi_1) \widehat {J^b}(\varphi_2)  \right\rangle \,, \\
		\frac{\partial_{M_A} \partial_{M_B} Z}{Z}\Big\vert_{m,M=0} &= (4 \pi i)^2 \int d \varphi_1\, d\varphi_2 \, 
		\left\langle  \widehat {\widetilde J^A}(\varphi_1) \widehat {\widetilde J^B}(\varphi_2) \right\rangle \,.
	}
	The definition \eqref{1dJDef} and the normalization condition \eqref{TwoPtJ} imply that the two-point functions of the 1d operators are $\left\langle \widehat {J^a}(\varphi_1) \widehat {J^b}(\varphi_2)  \right\rangle = \frac{c_J}{512 \pi^2} \tr (T^a T^b)$ and $\left\langle  \widehat {\widetilde J^A}(\varphi_1) \widehat {\widetilde J^B}(\varphi_2) \right\rangle = \frac{c_{\widetilde J}}{512 \pi^2} \delta^{AB}$.  Plugging these expressions into \eqref{twoptAgain} and performing the integrals, one can recover \eqref{int2}. 
	
	Using a similar expression to \eqref{twoptAgain} for the four-point function gives  
	\ie\label{int4}
	\frac{\partial_{m_a}\partial_{m_b}\partial_{m_c}\partial_{m_d}Z}{Z}\Big\vert_{m,M=0}
	&= (4\pi)^4 \int_0^{2\pi} d\varphi_1 d\varphi_2 d\varphi_3 d\varphi_4 \left\langle \widehat {J^a}(\varphi_1) \widehat {J^b}(\varphi_2) \widehat {J^c}(\varphi_3) \widehat {J^d}(\varphi_4) \right\rangle
	\\
	&= {\pi^4c_J^2\over 64}\left[ {\mathtt T}_{\bf 1}^{abcd}+{1\over 6}\lambda^2_{\mathcal{D}[1],{\bf adj}_a} {\mathtt T}^{abcd}_{{\bf adj}_a} +  {3\over 8} \sum_{{\bf i}={\bf 1},{\bf 2},{\bf 3},{\bf 4}} \lambda^2_{\mathcal{D}[2],{\bf i}} {\mathtt T}^{abcd}_{\bf i} \right]\,.
	\fe
	Recall that $\lambda_{\cD[1],{\bf adj}_a}$ has already been determined in (\ref{OPEnorm2}), and that $\lambda_{\cD[2],{\bf i}}$ obey the crossing constraints \eqref{1dcross}. For $k\geq 4$, (\ref{int4}) gives two new independent constraints on the OPE coefficients appearing on the RHS, associated with the two quartic Casimirs of $su(k)$. It suffices to consider mass deformation parameters $m_1$ and $m_3$ associated with the following two Cartan generators of $su(k)$, expressed in the fundamental representation as diagonal $k\times k$ matrices
	\es{gens}{
		(T^1)_I{}^J&=\frac12\text{diag}\{ 1,-1, 0, \cdots,0\}\,,\qquad    (T^3)_I{}^J=\frac{1}{2\sqrt{6}}\text{diag}\{1,1,1,-3,0\cdots,0\} \, ,
	}
	which lead to the independent constraints
	\es{CartChoice}{
		& {256\over \pi^4 c_J^2} \frac{\partial_{m_1}^4 Z}{Z}\Big\vert_{m,M=0}=1+\frac{3}{16}\left( 2\lambda^2_{\mathcal{D}[2],{\bf 1}} + 4\lambda^2_{\mathcal{D}[2],{\bf 2}} + 2\lambda^2_{\mathcal{D}[2],{\bf 3}} + \lambda^2_{\mathcal{D}[2],{\bf 4}} \right)\,,\\
		& {256\over \pi^4 c_J^2} \frac{\partial_{m_3}^4 Z}{Z}\Big\vert_{m,M=0}=1+\frac{3}{16}\left( 2\lambda^2_{\mathcal{D}[2],{\bf 1}} + 4\lambda^2_{\mathcal{D}[2],{\bf 2}} + {7\over 3} \lambda^2_{\mathcal{D}[2],{\bf 3}} + {7\over 6}\lambda^2_{\mathcal{D}[2],{\bf 4}} \right)\,.
	} 
	For $k=2,3$, (\ref{int4}) produces a single new independent constraint on the BPS-protected OPE coefficients, in addition to the two crossing constraints in the $k=3$ case, or the one crossing constraint in the $k=2$ case. 
	
	We can similarly extract constraints on the 1d mixed topological correlator \eqref{1d42} from the $S^3$ partition function by taking the derivative of the latter with respect to both $m_a$ and $M_A$,
	\ie\label{int42}
	\frac{\partial_{m_a}\partial_{m_b}\partial_{M_C} \partial_{M_D} Z}{Z}\Big\vert_{m,M=0}&= (4\pi)^4 \int_0^{2\pi} d\varphi_1 d\varphi_2 d\varphi_3 d\varphi_4 \left\langle \widehat {J^a}(\varphi_1) \widehat {J^b}(\varphi_2) \widehat {\widetilde J^C}(\varphi_3) \widehat {\widetilde J^D}(\varphi_4) \right\rangle
	\\
	& = {\pi^4c_J c_{\widetilde J}\over 64}{\mathtt T}_{\bf 1}^{abCD} \big(1+\frac38  \lambda^2_{\mathcal{D}[2]} \big)\,,
	\fe
	where as before we assumed the $ab\to CD$ channel, and if we considered the $aC\to bD$ or $aD\to bC$ channels, then we would not have an identity block contribution.
	
	\subsection{Evaluating $Z$ from supersymmetric localization}
	
	An essential property of the $S^3$ partition function $Z(m,M)$ is that its path integral representation can be reduced, 
	via the supersymmetric localization method \cite{Kapustin:2009kz,Jafferis:2010un} (for reviews, see \cite{Willett:2016adv,Pufu:2016zxm}), to a finite dimensional matrix integral. It suffices to restrict to mass parameters $m_\A$ associated with a set of $k-1$ Cartan generators $T^\A$, $\A=1,\cdots,k-1$, which we take to be diagonal $k\times k$ matrices, and write $m_\A(T^\A)_I{}^J \equiv m_{\A,I}\delta_I^J$. Similarly, we can restrict the mass parameters for the adjoint hypermultiplet to only one non-zero parameter, $M_3 = M$.  The localized matrix integral expression for $Z(m, M)$ for this choice of $m$ can be further reduced to an $N$-dimensional integral over eigenvalues,
	\es{Z}{
		Z(m,M)=\frac{1}{N!}\int \prod_{i=1}^N \frac{dx_i}{2^{k+1}}\prod_{I=1}^{k}\frac{1}{\cosh[\pi(x_i+\sum_{\alpha=1}^{k-1}m_{\alpha,I})]} \frac{\prod_{i<j} \sinh^2\left(\pi(x_{ij})\right)}{\prod_{i,j}  \cosh\left(\pi(x_{ij}-M)\right) }\,,
	}
	where $x_{ij}\equiv x_i-x_j$. The large $N$ expansion of this expression can be extracted from its interpretation as the partition function of a non-interacting Fermi gas of $N$ particles \cite{Marino:2011eh,Grassi:2014vwa,Nosaka:2015iiw}, leading to
	\es{Zmfinal2}{
		Z(m,M)=e^A C^{-\frac13}\text{Ai}\left[C^{-\frac13}(N-B)\right]+O(e^{-N})\,,
	}
	where $A, B, C$ are functions of $m_\A, M$, and $k$. The Fermi gas method \cite{Marino:2011eh,Nosaka:2015iiw}, summarized in Appendix \ref{locApp}, produces $A,B,C$ a priori as power series in $\hbar = {1\over 2\pi}$. Here we simply state the results for $B$ and $C$,
	\es{BC2}{
		C=\frac{2}{\pi^2 k(1+4M^2)}\,,\qquad B=\frac{\pi^2 C}{3}+\frac{k}{24}-\frac{1}{6k}-\frac{1}{1+4M^2}\Big(\frac{k}{6}+\sum_{\alpha=1}^{k-1}{m_\alpha^2}\Big) \,,
	}
	and give $A$ as an expansion in $m_\A$ to order $m^2$,\footnote{The exact formula for $A$ at $m=0$ was guessed in \cite{Nosaka:2015iiw}.}
	\es{Amixed}{
		A&=\frac{k^2}{4} \big[ \cA(1-2iM)+\cA(1+2iM) \big] 
		+\frac{\cA(k)}{2}
		+k\sum_{\alpha=1}^{k-1}{m_\alpha^2} \bigg[ \frac{1}{4M^2+1}-\frac{\pi
			^2}{72}
		\\
		&~~~ -\frac{\pi ^4 \left(12M^2-1\right)}{21600}-\frac{\pi^6(1-40M^2+80M^4)}{1270080}+\dots \bigg]+O(m^4)\,,
	}
	where $\cA(x)$ on the RHS is the ``constant map'', defined in \cite{Hanada:2012si} as
	\es{constantMap}{
		{\cal A}(x)=&\frac{2\zeta(3)}{\pi^2x}\left(1-\frac{x^3}{16}\right)+\frac{x^2}{\pi^2}\int_0^\infty dy\frac{y}{e^{xy}-1}\log\left(1-e^{-2y}\right)\\
		=&\frac{2\zeta(3)}{\pi^2 x}-\frac{x}{12}-\frac{x^3\pi^2}{4320}+\frac{x^5\pi^4}{907200}-\frac{x^7\pi^6}{50803200}+\frac{x^9\pi^8}{1437004800}+O(x^{13})\,.\\
	}
	We have also computed the order $m^4$ terms of $A$ at $M=0$, with the result 
	\es{m4}{
		\partial_{m_1}^4A\vert_{m,M=0}=4(6+k)\mathfrak b\,,\qquad \partial_{m_3}^4A\vert_{m,M=0}=4(6+{7\over 6} k)\mathfrak b\,,
	}
	where $\mathfrak b$ is the constant\footnote{This closed form expression is guessed from the series $\mathfrak b=\frac{\pi ^2}{12}+\frac{\pi ^4}{720}-\frac{\pi ^6}{60480}+\frac{\pi^8}{1814400}-\frac{\pi ^{10}}{31933440} +\cdots$.} 
	\es{quartic}{
		\mathfrak b = -\frac{\pi^2}{2}(x\cA(x))''\big\vert_{x=1}=-\frac{1}{2} \pi ^2 \left(\frac{\pi ^2}{32}-\frac{1}{2}\right)\,.
	}

	\subsection{The $1/N$ expansion of central charges}
	
	We can now evaluate the $1/N$ expansion of the central charges $c_J$ and $c_{\widetilde J}$ using (\ref{int2}) and  (\ref{Zmfinal2}). The lowest order terms are
	\es{cJ2}{
		c_{ J}&=\frac{8 \sqrt{2 Nk}}{ \pi }-\frac{16k}{\pi^2}\mathfrak a+O(N^{-{1\over 2}})\,,\\
		c_{\widetilde J}&=\frac{32 \sqrt{2} N^{3\over 2} \sqrt{k}}{3 \pi }+\frac{10 \sqrt{2} \sqrt{N} (4-k^2) }{3 \pi\sqrt{k} }+O(N^0)\,,\\
	}
	where the constant $\mathfrak a$ is evaluated using \eqref{Amixed} to be
	\es{a}{
		\mathfrak a = {1\over 2k} \partial_m^2 A\vert_{m,M=0} = 1-\frac{\pi ^2}{72}+\frac{\pi ^4}{21600}-\frac{\pi
			^6}{1270080}+\cdots\,.
	} 
	For the purpose of numerical evaluations, the expansion on the RHS of (\ref{a}) converges quickly. In the case $k=2$, we have tested the numerical integral of (\ref{Z}) in $N=2$ case against the large $N$ asymptotic formulae (\ref{cJ2}) together with (\ref{a}) (truncated to the order shown), and the results (shown in Table \ref{comp}) are very close already.
	
	\begin{table}[!h]
		\begin{center}
			\begin{tabular}{c||c|c|c}
				& asymptotic & exact & error ($\%$)\\
				\hline \hline
				$c_J$ &   $4.5824$   & $4.5894$ & $0.15$ \\
				\hline
				$c_{\widetilde J}$ &   $20.634$   & $20.589$ & $0.21$ \\
				\hline
				$\lambda^2_{\cD[2]_1}$ &  $0.4866$   & $0.48318$ &$1.12$  \\
				\hline
				$\lambda^2_{\cD[2]}$ &   $2.7626$  & $2.7819$ &  $0.70$\\
				\hline
			\end{tabular}
			\caption{Comparison of the large $N$ asymptotic formulae (\ref{cJ2}) against the exact values at $N=2$, in the $k=2$ case, for various CFT data computable from localization, with surprisingly good agreement. 
				\label{comp}}
		\end{center}
	\end{table}

	The large $N$ expansion of $c_T$ has been previously obtained from the squashed 3-sphere partition function \cite{Chester:2020jay}, with the result
	\es{cTisrrael}{
		c_T=\frac{64 \sqrt{2} N^{3\over 2} \sqrt{{k}}}{3 \pi }+\frac{4 \sqrt{2} \sqrt{N} \left(7
			{k}^2+8\right)}{3 \pi  \sqrt{{k}}}+O(N^0)\,.
	}
	Note that the ratio $c_T/c_{\widetilde J} = 2 + {\cal O}(N^{-1})$ has nontrivial dependence on $1/N$ at the subleading order.\footnote{In the case $k=1$, the supersymmetry enhances to $\mathcal{N}=8$ and the flavor multiplet combines with the stress tensor multiplet, leading to an exact relation $c_T=2c_{\widetilde J}$.}

	\subsection{The $1/N$ expansion of BPS-protected OPE coefficients}
	\label{sec:operesults}
	
	We can now determine, using the crossing relation \eqref{1dcross}, the relation between the sphere partition function $Z$ and the 1d topological correlator (\ref{CartChoice}), and the large $N$ expansion of $Z$ (\ref{Zmfinal2}), all of the protected OPE coefficients that appear on the RHS of (\ref{1d4}). The first few order terms in the $1/c_J \sim 1/\sqrt{N}$ expansion are\footnote{For $k=3$, the same formula holds with $\lambda^2_{\mathcal{D}[2],{\bf 4}}$ omitted, whereas for $k=2$ both $\lambda^2_{\mathcal{D}[2],{\bf 3}}$ and $\lambda^2_{\mathcal{D}[2],{\bf 4}}$ are to be omitted. }
	\es{sunZ2}{
		k\geq2:\quad \lambda^2_{\mathcal{D}[2],{\bf 1}}&= {8192  \over  \pi ^4} {\mathfrak b\over c_J^2}
		-\frac{8192}{3 \pi ^4} {k\over c_J^3} +\frac{131072 }{3 \pi ^6} {\mathfrak{a} k^2\over c_J^4}+O(c_J^{-5})\,,\\
		\lambda^2_{\mathcal{D}[2],{\bf 2}}&=\frac{8}{3}+\frac{8192}{
			\pi ^4} {\mathfrak{b} \over c_J^2} - \frac{8192
		}{3 \pi ^4} {k\over c_J^3}
		+\frac{131072 }{3 \pi ^6} {\mathfrak{a} k^2\over c_J^4}+O(c_J^{-5})\,,\\
		k=3:\quad\lambda^2_{\mathcal{D}[2],{\bf 3}}&=\frac{4}{3}+\frac{64}{9}{1\over c_J}+ {4096  \over  \pi ^4} {\mathfrak b\over c_J^2}
		-\frac{4096}{3 \pi ^4} {k\over c_J^3} +\frac{65536 }{3 \pi ^6} {\mathfrak{a} k^2\over c_J^4}+O(c_J^{-5})\,,\\
		k\geq4:\quad\lambda^2_{\mathcal{D}[2],{\bf 3}}&=\frac{64}{9}{1\over c_J}+\frac{8192 }{3\pi ^4} {\mathfrak{b} k\over c_J^2}+O(c_J^{-5})\,,\\
		\lambda^2_{\mathcal{D}[2],{\bf 4}}&=-\frac{128}{9 }{1\over c_J}+\frac{8192 }{3 \pi ^4} {\mathfrak{b} k\over c_J^2}+O(c_J^{-5})\,,\\
	}
	where $\mathfrak{a}, \mathfrak{b}$ are given in (\ref{a}), \eqref{quartic} respectively. Recall that for $k=3$, the second crossing equation in \eqref{1dcross} trivially relates $\lambda^2_{\mathcal{D}[2],{\bf 3}}$ to $\lambda^2_{\mathcal{D}[2],{\bf 1}}$ and $\lambda^2_{\mathcal{D}[2],{\bf 2}}$, while for $k\geq4$ they are independent.

	We can also determine from (\ref{int42}) the only protected OPE coefficient $\lambda_{\cD[2]}$ (in the $ab\to CD$ channel) that appears in the mixed correlator (\ref{1d42}). The first two terms in its $1/N$ expansion are
	\es{OPEmixed}{
		\lambda^2_{\cD[2]}&=\frac{4\sqrt{2}}{\pi\sqrt{k}N^{3\over 2}}+\frac{3}{4\pi^2 k N^2}(\mathfrak{a}+\mathfrak{d})+O(N^{-{5\over 2}})\,,\\
	}
	where the constant $\mathfrak d$ is obtained from \eqref{Amixed} to be
	\es{d}{
		\mathfrak d= {1\over 2} \partial_m^2 \partial_M^2 A\vert_{m,M=0}=-8-\frac{\pi^4}{900}+\frac{\pi^6}{15876}+\dots\, .
	}
	As discussed above, in the $aC\to bD$ or $aD\to bC$ channels the unique protected OPE coefficient would differ by the identity contribution $8\over 3$.

	\section{Back to M-theory via Mellin amplitudes}
	\label{sec:backtom}
	
	We now return to the question of determining the effective theory of massless degrees of freedom of M-theory in the orbifold background $\mathbb{R}^{1,6}\times \mathbb{C}^2/\mathbb{Z}_k$ by taking the large radius limit of Mellin amplitudes in ${\rm AdS}_4\times S^7/\mathbb{Z}_k$, as constrained by the known conformal data of the dual 3d ${\cal N}=4$ SCFT.

	\subsection{The leading low energy limit in the M-theory orbifold}
	
	To proceed, we will compare the $1/N$ expansion of the BPS-protected OPE coefficients (\ref{OPEnorm2}), (\ref{doneopec}), (\ref{sunZ2}),  and (\ref{OPEmixed}) against that of the gluon Mellin amplitude $M^{abcd}$ \eqref{gluon} and the mixed gluon-graviton Mellin amplitude $M^{abCD}$ (\ref{gluonM}). This is achieved by first expressing the Mellin transform of (the relevant term of) the Mellin amplitude as an expansion in conformal blocks $G_{\Delta, j}(U,V)$ and their derivatives $\partial_\Delta G_{\Delta,j}(U,V)$, taking the lightcone limit $U\to 0$ with fixed $V$, and then projecting onto a given lightcone block using orthogonality relations, thereby expressing the contribution to the OPE coefficient as a contour integral of the said Mellin amplitude that can be evaluated using residues. The algorithm, following \cite{Chester:2018lbz}, is outlined in Appendix~\ref{app:extraction}\@. Below we simply state the results.

	For the 4-gluon Mellin amplitude $M^{abcd}(s,t;\A)$ \eqref{gluon}, the leading SYM tree-level term ${1\over c_J} M_{F^2}^{abcd}$ (given explicitly in (\ref{mero})) contributes to $\lambda^2_{\cD[1],{\bf adj}_a}$, $\lambda^2_{\cD[2],{\bf 3}}$, and $\lambda^2_{\cD[2],{\bf 4}}$, producing precisely the order $1/c_J$ terms of these OPE coefficients. 
	While we have not determined the explicit $s,t$-dependence of the 1-loop term ${1\over c_J^2} M_{F^2|F^2}^{abcd}$, the $k$-dependence of its known general structure \cite{Alday:2021ajh} (similar to its flat space analog (\ref{a1loop})) matches that of \eqref{sunZ2} at order $1/c_J^2$.
	
	For the mixed gluon-graviton Mellin amplitude $M^{abCD}(s,t;\A)$ (\ref{gluonM}), the tree-level graviton exchange term ${1\over c_T} M_{R}^{abCD}$ contributes to $\lambda^2_{\cB[0]}$ (\ref{doneopec}) and $\lambda^2_{\cD[2]}$ (\ref{OPEmixed}), and the results precisely match at order $N^{-{3\over 2}}$.
	
	These results confirm, in a highly nontrivial manner, that the leading order low energy effective description of the M-theory orbifold, namely the 7d $SU(k)$ super-Yang-Mills theory coupled to 11d supergravity, is consistent with the holographic duality.
	
	\subsection{Constraining higher derivative corrections}
	
	Going beyond the leading low energy limit, we turn to the thus far undetermined coefficients on the RHS of \eqref{gluon} and (\ref{gluonM}).
	
	Comparing the 4-gluon Mellin amplitude \eqref{gluon} to the $1/c_J$ expansion of the OPE coefficients $\lambda^2_{\cD[2],i}$ \eqref{sunZ2}, we observe that the latter do not admit terms of order $N^{-{7\over 6}}$ that would a priori receive contribution from the Mellin amplitudes proportional to $M_{[F^2]^2}^{abcd}$ and $M_{F^4}^{abcd}$. This allows us to determine
	\ie
	\kappa_{[F^2]^2}= \kappa_{F^4}=0 \,.
	\fe
	Through the large radius limit (\ref{FlatLimit}), it follows that the momentum expansion of the gluon amplitude $\cA^{abcd}(s,t)$ (with specialized polarization and the factor $(u+s\A)^2$ stripped off) contains no terms of $s^0$ scaling, and consequently the gluon superamplitude $\mathfrak{A}^{abcd}$ in the M-theory orbifold background does not contain the 4-derivative supervertices (\ref{af4coup}) corresponding to $(\Tr F^2)^2$ or $\Tr F^4$ effective couplings.\footnote{This is in contrast to, for instance, the non-Abelian Born-Infeld action of D-branes which contains such effective couplings.}
	
	We further observe that the $1/c_J$ expansion of OPE coefficients \eqref{sunZ2} does not contain any terms with logarithmic dependence on $c_J$, and in particular there are no terms of order ${\log c_J\over c_J^3}$, in contrast to the Mellin amplitude (\ref{gluon}) that does contain order ${\log c_J\over c_J^3}$ terms. It follows that the coefficients $\kappa'_{\log,[F^2]^2}, \kappa'_{\log,F^4}, \kappa_{\log,D^2[F^2]^2}, \kappa_{\log,D^2F^4}$ must be such that the contribution from the third line of (\ref{gluon}) to the BPS-protect OPE coefficients \eqref{sunZ2} cancels, from which we deduce the linear relations
	\ie\label{kfixed}
	\kappa'_{\log,[F^2]^2} =-\frac{12}{5}\kappa_{\log,D^2[F^2]^2}, \qquad \kappa'_{\log,F^4} =-\frac{6}{5}\kappa_{\log,D^2F^4}\,.
	\fe
	We will determine the coefficients $\kappa_{\log,D^2[F^2]^2}, \kappa_{\log,D^2F^4}$ by comparison with the flat space amplitudes in the next subsection.
	
	Next, let us compare the mixed gluon-graviton Mellin amplitude (\ref{gluonM}) to the $1/N$ expansion of the OPE coefficient $\lambda_{\cD[2]}^2$ (\ref{OPEmixed}) appearing in the mixed correlator (\ref{1d42}). We observe that  (\ref{OPEmixed}) does not contain terms of order $N^{-{11\over 6}}$ nor $N^{-{13\over 6}}$. It follows that
	\ie\label{kfixed2}
	\kappa_{RF^2}=0\,,\qquad \kappa'_{RF^2}= -\frac{12}{5}\kappa_{R^2F^2}\,.
	\fe
	In particular, we conclude that the 4-derivative $R\Tr F^2$ coupling corresponding to the supervertex (\ref{arf2}) is absent in the M-theory orbifold background. On the other hand, the second relation in (\ref{kfixed2}) means that the known holographic CFT data do not pin down the $R^2\Tr F^2$ coupling coefficient $\kappa_{R^2F^2}$, corresponding to the supervertex (\ref{ar2f2s}).\footnote{It is likely that a non-renormalization property of the Coulomb branch effective theory similar to the one outlined in Section~\ref{sec:coulomb} can be used to argue that $\kappa_{R^2F^2}$ itself vanishes.}

	\subsection{The logarithmic threshold corrections}
	\label{MtheoryCon}

	Even though we have not explicitly computed the gluon 2-loop and graviton exchange contributions to the Mellin amplitude $M^{abcd}$, namely ${1\over c_J^3}M_{F^2|F^2|F^2}$ and ${1\over c_T}M_R$ in (\ref{gluon}), they are expected to behave as $N^{-{3\over 2}} s^3 \log(L^2 s)$ in the flat space limit. By contrast, the gluon 2-loop and graviton exchange terms in the flat space amplitude, namely $\mathfrak{A}_{F^2|F^2|F^2}$ and $\mathfrak{A}_R$ in (\ref{asupfull}), behave as $s^3 \log(-\ell_p^2 s)$, where the logarithmic dependence on $\ell_p$ follows from the expectation of a Wilsonian effective description with cutoff scale $\Lambda\sim 1/\ell_p$ (see the comment after (\ref{ad2f4term})). The difference between them is accounted for by the order ${\log c_J\over c_J^3}$ terms in the Mellin amplitude $M^{abcd}$ (third line of (\ref{gluon})), more precisely,
	\ie\label{logcj}
	{\log c_J\over c_J^3} \left[ \kappa_{\log,D^2[F^2]^2}M^{abcd}_{D^2 [F^2]^2}+\kappa_{\log,D^2F^4}M^{abcd}_{D^2  F^4} \right],
	\fe
	whose flat space limit is governed by the logarithmic threshold terms in $\mathfrak{A}_{F^2|F^2|F^2}+\mathfrak{A}_R$.
	
	The contribution of $\mathfrak{A}_{F^2|F^2|F^2}$ to the reduced amplitude $\cA^{abcd}(s,t)$ appearing in (\ref{FlatLimit}), as follows from (\ref{A2loop}) with the substitution (\ref{gluonreduce}), is
	\ie \label{2loop}
	\mathcal{A}^{abcd}_{F^2|F^2|F^2}(s,t)&= - 2 g_{\rm YM}^6  \left( s\mathtt e_{st} I_{1234}^P+s\mathtt f_{st} I_{1234}^{NP} +({\rm permutations~on~}2,3,4)\right)\,,
	\fe 
	% \ie\label{2loop}
	%  \mathcal{A}^{abcd}_{F^2|F^2|F^2}(s,t)&= -2 g_{\rm YM}^6 \big(s\mathtt e_{s_1}I_{1234}^P+s\mathtt e_{s_2}I_{1243}^P+t\mathtt e_{t_1}I_{1432}^P+t\mathtt e_{t_2}I_{1432}^P
	%  \\
	%  & ~~~ +u\mathtt e_{u_1}I_{1324}^P+u\mathtt e_{u_2}I_{1342}^P+2s\mathtt f_{s}I_{1234}^{NP}+2t\mathtt f_{t}I_{1423}^{NP}+2u\mathtt f_{u}I_{1324}^{NP}\big)\,,
	%  \fe
	where $\mathtt e$ and $\mathtt f$ are the flavor tensor structures defined in (\ref{strucs2loop}).  $I^P(s,t)$ and  $I^{NP}(s,t)$ are the planar $I^p(s,t)$ and non-planar $I^{NP}(s,t)$ 2-loop integrals defined in \cite{Bern:1998ug}, with UV cutoff $\Lambda\sim 1/\ell_p$. For our purpose of determining (\ref{logcj}), it suffices to evaluate the logarithmic ``divergence'' of the 2-loop integrals, giving
	\es{divs}{
		I_{1234}^P|_{\log}=-\log(\ell_p^2)\frac{\pi}{10(4\pi)^7}\,,\qquad  I_{1234}^{NP}|_{\log}=-\log(\ell_p^2)\frac{\pi}{15(4\pi)^7}\,.
	}
	The corresponding logarithmic dependence on $\ell_p$ of (\ref{2loop}) is given by (after substituting the explicit expressions of $\mathtt e,\mathtt f$ from \eqref{strucs2loop}) 
	\ie\label{2looplog}
	\mathcal{A}^{abcd}_{F^2|F^2|F^2}|_{\log}&= {g_{\rm YM}^6\over 2^{13}\pi^6} \log(\ell_p^2) \Big[ sk {\mathtt A}^{abcd}+uk {\mathtt A}^{acdb}+tk {\mathtt A}^{adbc}\\
	&\quad-(2+{k^2\over 10})u{\mathtt B}^{abcd}-(2+{k^2\over 10})t{\mathtt B}^{acdb}-(2+{k^2\over 10})s{\mathtt B}^{adbc}\Big]\, .
	\fe
	Similarly, from (\ref{arloop}) we deduce the logarithmic threshold contribution from the graviton exchange,
	\ie\label{looprred}
	\cA^{abcd}_R|_{\log} = -{\kappa^2\over 32\pi^2} { \log(\ell_p^2)} \left[ sk {\mathtt A}^{abcd}+uk {\mathtt A}^{acdb}+tk {\mathtt A}^{adbc} \right].
	\fe
	Using the relations $g_{\rm YM}^2=2(2\pi)^4 \ell_p^3$ and $\kappa^2 = {1\over 2} (2\pi)^8 \ell_p^9$, we find that the terms proportional to $\mathtt A^{abcd}$ in (\ref{2looplog}) and (\ref{looprred}) differ by a sign, so that the total logarithmic threshold vanishes. Comparing the sum of (\ref{2looplog}) and (\ref{looprred}) with the flat space limit of (\ref{logcj}), we determine 
	\es{logFin}{
		%& \kappa'_{\log,[F^2]^2} =-\frac{9216k}{\pi^4}\qquad \kappa'_{\log,F^4} =(20+k^2)\frac{2304}{5\pi^4}\,,
		%\\
		& \kappa_{\log,D^2[F^2]^2}=\kappa_{\log,D^2F^4}=0  \, .
	}

	\section{Conclusion}
	\label{conc}
	
	In this work we have introduced a framework to study the dynamic excitations of M-theory singularities beyond the leading low-energy limit. Our strategy is based on the holographic duality that maps M-theory in an AdS background, with singularity in its internal space, to a superconformal field theory whose OPE data are constrained by the conformal bootstrap with input from supersymmetric localization. In particular, the $1/N$ corrections to the CFT correlators translate to the higher derivative corrections to the scattering amplitudes in the AdS, including those of the graviton modes in the bulk, as well as of the quanta localized at the singularity. By taking the large radius limit, where the M-theory spacetime is locally flat away from the singularity, we can extract corrections to the effective theory of the M-theory singularity.
	
	We demonstrated this strategy in the example of M-theory on an ${\rm AdS}_4\times S^7/\mathbb{Z}_k$ orbifold, whose holographic dual is a certain 3d $\mathcal{N}=4$ CFT with $SU(k)\times SU(2)_F\times U(1)$ flavor symmetry. In the large radius limit, the M-theory background becomes an $\mathbb{C}^2/\mathbb{Z}_k \times \R^{1,6} $ orbifold. We analyzed the holographic correlators of $SU(k)\times SU(2)_F$ flavor currents to subleading orders in the $1/N$ expansion, combining analytic bootstrap methods and exact results of BPS-protected OPE coefficients from localization, and determined, in the large radius limit, the absence of 4-derivative effective couplings of the type $(\Tr F^2)^2$, $\Tr F^4$, $R\Tr F^2$ in the $SU(k)$ gauge theory of the M-theory orbifold.
	
	A curious feature of the scattering amplitude of $SU(k)$ gluons at the M-theory orbifold singularity is that there are logarithmic threshold corrections from the 2-loop gluon exchange and tree-level bulk graviton exchange at the same order in the momentum expansion. We have computed these threshold corrections applying unitarity cut methods of flat space amplitudes, and found that they exactly cancel each other. We then combined this with the localization results to show that the $N^{-{3\over 2}} \log N$ terms in the AdS gluon Mellin amplitude also vanish. It would be interesting to find a conceptual explanation for the vanishing of the logarithmic threshold.

	The next order higher derivative effective couplings of the M-theory orbifold, for instance the $D^2(\Tr F^2)^2$ and $D^2 \Tr F^4$ terms of the $SU(k)$ effective gauge theory, cannot be unambiguously separated from the gluon 2-loop and bulk graviton exchange amplitudes. To determine, say, the 6-derivative order gluon effective couplings from the dual CFT, requires knowing up to 2-loop corrections to the Mellin amplitudes. While the latter can in principle be fixed in terms of tree-level data via the AdS unitarity method \cite{Aharony:2016dwx}, the necessary computations have not been performed and will be left for future work. To begin with, the 1-loop correction to the gluon Mellin amplitude in ${\rm AdS}_4\times S^3$ which should be computable by the method of \cite{Alday:2021ajh, Alday:2022rly,Alday:2021ymb}, has no contact term ambiguity and should provide a highly nontrivial check against the localization constraints derived in this work.
	
	It would also be interesting to determine the tree-level bulk graviton exchange Mellin amplitude ($M_R^{abcd}$ in Section~\ref{sec:mabcd}), which receives contributions from an infinite tower of graviton KK modes. The lowest mode is in the BPS-protected stress tensor multiplet, while all higher modes are in unprotected long multiplets. To fix the relative coefficients between the single trace exchange diagrams for each mode, one would need to compute the 3-point functions of two gluon and one graviton KK mode from the ${\rm AdS}_4\times S^7/\mathbb{Z}_k$ geometry, which should be proportional to the inverse central charge $1/c_T$ for all KK modes. A consistency check is that the flat space limit must match the 7d amplitude determined in this paper. A similar graviton exchange diagram appeared in gluon scattering on branes in other examples of AdS/CFT for CFTs in four \cite{Alday:2021ajh,Behan:2023fqq, Behan:2022uqr}, five, and six dimensions \cite{Zhou:2018ofp}, which should be re-evaluated along similar lines.\footnote{The computation in \cite{Alday:2021ajh,Zhou:2018ofp} only considered the contribution from the leading KK mode, which by itself does not produce  the expected flat space limit of the graviton exchange amplitude.}

	Looking ahead, we would like to determine the unprotected higher derivative corrections, none of which are known in M-theory, with or without orbifold singularities. To this end, we can employ numerical bootstrap, rather than the current form of analytic bootstrap, combined with localization input. If we can pin down the CFT correlator/OPE data to sufficiently high numerical precision at finite values of $N$, it would be possible to numerically extract $1/N$ corrections in the large $N$ expansion. This has been partially achieved in  \cite{Agmon:2019imm,Chester:2014mea,Chester:2014fya} for the stress tensor multiplet correlator in the 3d $\mathcal{N}=8$ ABJM CFT \cite{Aharony:2008ug} where the CFT data, determined to high numerical precision at finite $N$, have been successfully matched with the $1/N$ expansion of Mellin amplitudes at both tree \cite{Chester:2018lbz} and loop level \cite{Alday:2021ymb}.\footnote{Without the input of localization data of ABJM theory, the analogous numerical bootstrap bounds are saturated by  pure supergravity in ${\rm AdS}_4$ \cite{Alday:2022ldo}.} For the 3d $\mathcal{N}=4$ SCFT considered in this work, the numerical bootstrap analysis for the flavor multiplet correlators has been initiated in \cite{Chang:2019dzt}, and should be improved using the BPS-protected OPE data that we have determined to all orders in $1/N$.

	Finally, the orbifold singularities $\mC^2/\mZ_k$ analyzed in this work are special cases of general half-BPS singularities $\mC^2/\Gamma$ in M-theory defined by finite subgroups $\Gamma\subset SU(2)$ which follow an $ADE$ classification (here we have focused on the $A_{k-1}$ type). The singular locus supports 7d non-Abelian gauge fields for the corresponding $ADE$ Lie algebra \cite{Acharya:1998pm}. It would be interesting to extend our analysis to general $D$- and $E$-type singularities, which can be achieved by studying 3d $\cN=4$ SCFTs  whose holographic duals contain such singularities. The 3d $\cN=4$ gauge theory studied here is also known as the rank $N$ 3d ADHM gauge theory \cite{Atiyah:1978ri} for $SU(k)$ which has a natural generalization for classical Lie algebras including the $D$-type \cite{deBoer:1996mp,deBoer:1996ck}, whose holographic dual contains the desired singularity  \cite{Mezei:2013gqa}. Alternatively, there is a class of 3d $\cN=4$ affine $ADE$ quiver gauge theories  \cite{deBoer:1996mp,deBoer:1996ck,Cremonesi:2014xha} that realize the $ADE$ singularities in their holographic duals and are related to the ADHM gauge theories of $A$- and $D$- types by mirror duality \cite{Seiberg:1996bd}. 
	In all examples mentioned above, the dual SCFTs can be engineered by M2 branes probing transversely the singularity of interest.
	Note that the singularity has to be non-isolated in the eight transverse directions to the M2 branes in order to survive the decoupling limit. 
	
	To probe singularities of more generic types in M-theory, which typically preserve fewer supersymmetries, we would need to study correlation functions in dual SCFTs with few supersymmetries.
	For example, complex three-fold Calabi-Yau singularities in M-theory can be analyzed in this way using 3d $\cN=2$ SCFTs. The simplest example is perhaps the conifold singularity $\cC\times \mR^{4,1}$ where the relevant 3d gauge theory can be either a modification of the 3d $\cN=8$ maximal super-Yang-Mills or the ABJM theory by adding certain 3d $\cN=2$ chiral multiplets with corresponding superpotentials \cite{Benini:2009qs,Jafferis:2009th}.
	Similarly, in order to extract the effective theory of real codimension-seven singularities preserving $G_2$ holonomy, we would need to study correlators in the dual 3d $\cN=1$ SCFTs. A class of $G_2$ singularities can be found in \cite{Acharya:2004qe} and the dual 3d $\cN=1$ CFTs can be potentially identified by repeating the analysis of \cite{Benini:2009qs,Jafferis:2009th} for $\cN=1$ gauge theories (see also \cite{Forcella:2009jj}). 
	In the cases of $\cN<4$, the constraints from supersymmetry are much weaker and numerical bootstrap will be important to analyze such SCFTs in order to extract the dynamics associated with the corresponding M-theory singularities. Nonetheless, we still expect localization constraints (if available) will play a crucial role in this analysis, especially when the singularities are constrained by isometry to a few parameters. This is possible in the  $\cN=2$ cases where certain integrated correlation functions can be determined from the supersymmetric $S^3$ partition function of the SCFT with squashing and mass deformations which can be employed to fix the parameters of the M-theory singularity.
	
	\section*{Acknowledgments} 
	
	We thank  Liam Fitzpatrick, Joao Penedones, and Juan Maldacena for useful conversations. This work was performed in part at Aspen Center for Physics, which is supported by National Science Foundation grant PHY-2210452.  SMC is supported by the Royal Society under the grant URF\textbackslash R1\textbackslash 221310.  SSP is supported in part by the US NSF under Grant No.~2111977\@. YW is
	supported in part by the NSF grant PHY-2210420 and by the Simons Junior Faculty Fellows program. XY is supported by DOE grant DE-SC0007870. 
	
	\appendix

	\section{7d $\cN=1$ superamplitudes}
	
	\subsection{7d $\cN=1$ super-spinor-helicity formalism}
	\label{sec:superspinorhelicity}
	
	Here we introduce the 7d super-spinor-helicity formalism to describe superamplitudes of massless particles in the 7d half-maximal supergravity (see \cite{Wang:2015aua} for the super-spinor-helicity formalism for 7d maximal supergravity and \cite{Lin:2015dsa,Lin:2015ixa} for such a formalism in half-maximal gauged supergravities in other dimensions). 
	
	We denote spinor-helicity variables by 7d symplectic Majorana fermions $\lambda_{\A I}$ where $\A=1,2,\dots,8$ is the $SO(1,6)$ spinor index and $I=1,2,3,4$ the spinor index for the $SO(5)$ little group. They satisfy the following quadratic relations with the 7d momentum $p_\m$:
	\ie 
	\lambda_{\A I}\gamma_\m^{\A\B}\lambda_{\B J}=p_\m \Omega_{IJ}\,,\quad \lambda_{\A I}\lambda_{\B J}\cC^{\A\B}=0\,,
	\quad 
	\lambda_{\A I} \lambda_{\B J}\Omega^{IJ}={1\over 2} p_\m \gamma^\m_{\A\B}
	\label{spinorhelicity}
	\fe 
	where $\cC^{\A\B}$ (similarly $\cC_{\A\B}$) is the symmetric charge conjugation matrix for $SO(1,6)$ spinors and $\Omega_{IJ}$ (similarly $\Omega^{IJ}$) is the  anti-symmetric invariant tensor of the little group $SO(5)$ and they are used to raise and lower the spinor indices as below
	\ie 
	\lambda_{\A I} =\Omega_{IJ} \lambda_\A^J=\cC_{\A\B}\lambda^\B_I\,,\quad \Omega^{IJ}\Omega_{JK}=\D^I_K\,,\quad \cC^{\A\B}\cC_{\B\gamma}=\D^\A_\gamma.
	\fe
	
	To take into account the supermultiplet structure, we introduce 4 Grassmann variables $\eta^I$ (Grassmann odd coordinates for on-shell superspace).
	The massless one-particle states are labeled by representations of $SO(5)\times SU(2)_R$ as $(\bf m,n)$ according to the dimensions of the irreps. 
	
	In this notation, 
	the $8$ bosonic and 8 fermionic states in the vector multiplet transform in the following representations of $SO(5)\times SU(2)_R$,
	\ie\arraycolsep=1pt\def\arraystretch{}
	\begin{array}{* {5} {c}}
		(\bf 5,1)&\oplus&
		(\bf 1, 3) &\oplus &  (\bf 4,2)
		\\
		\eta_I\eta_J|_{\bf 5} & & \eta^{2{\bm a}-2} & &     \eta_I \eta^{2{\bm i}-2}  \\
		A_{\m} & & \varphi^{\bm a} & & \chi_{\A}^{\bm i}  \end{array}
	\label{vectormultiplet}
	\fe 
	where ${\bm a}=1,2,3$ and ${\bm i}=1,2$ are adjoint and spinor indices for $SU(2)_R$ respectively. 
	We have listed  the monomials in the Grassmann variables representing these states in the second line above and the corresponding 7d fields in the third line. We have also defined the following little group invariant $\eta^2\equiv \eta_I \eta_J \Omega^{IJ}$ and covariant $\eta_I\eta_J|_{\bf 5}\equiv\eta_{[I}\eta_{J]}+{1\over 4}\Omega_{IJ}\eta^2$.
	
	Similarly the super-graviton multiplet consists of $40+40$ states in the following representations, 
	\ie\arraycolsep=.8pt\def\arraystretch{}
	\begin{array}{* {11} {c}}
		(\bf 14,1)&\oplus & (\bf 5,3)&\oplus &
		~(\bf 1,1)~&\oplus &
		(\bf 10,1)&\oplus& (\bf 16 ,2)&\oplus& (\bf 4,2) \\
		\eta_I\eta_J\widetilde\eta_K\widetilde\eta_L|_{\bf 14}   &&  \widetilde \eta_K \widetilde \eta_L|_{\bf 5} \eta^{2{\bm a}-2} && (\eta\widetilde\eta)^2 &&  \eta_I\eta_J\widetilde\eta_K\widetilde\eta_L|_{\bf 10} &&  \eta_I \widetilde \eta_K \widetilde \eta_L|_{\bf 16} \eta^{2{\bm i}-2}
		&&
		\eta_I \widetilde \eta^2 \eta^{2{\bm i}-2}
		\\
		g_{\m\n} & &V_\m^{\bm a} & & \phi & & B_{[\m\n]}& &\psi_{\A \m}^{\bm i} & & \xi_\A^{\bm i}
	\end{array}
	\label{gravitonmultiplet}
	\fe 
	where we have introduced an auxiliary Grassmann variable $\widetilde \eta_I$, which, together with $\eta_I$ can be used to write down the monomials representing the one-particle states\footnote{In the maximal $\cN=2$ supergravity, $\eta_I$ and $\widetilde \eta_I$ together describe the Grassmann-odd coordinates on the on-shell superspace \cite{Wang:2015aua}.} and the corresponding 7d fields are listed in the last line. Note that all representations in \eqref{gravitonmultiplet} are generated by tensoring \eqref{vectormultiplet} with $(\bf 5,1)$ which is represented as $\widetilde \eta_K\widetilde \eta_L|_{\bf 5}$.\footnote{The maximal supergraviton multiplet of 128+128 states decompose with respect to the 7d $\cN=1$ supersymmetry, into one super-graviton multiplet \eqref{gravitonmultiplet}, one vector multiplet \eqref{vectormultiplet}, and two gravitino multiplets. Each 7d $\cN=1$ gravitino multiplet contains 40+40 states and the bosonic field content includes 3 vector fields, 2 two-form fields and 5 scalar fields.}
	
	The $SU(2)_R$ symmetry acts on the one-particle states (as monomials in $\eta,\widetilde\eta$) by
	\ie 
	r_+=\Omega_{IJ} \eta^I \eta^J\,,\quad r_-= \Omega^{IJ}{\partial\over \partial \eta_I} {\partial\over \partial \eta_J}\,,\quad r_0=\eta_I {\partial \over \partial \eta_I}-2\,.
	\label{SU2R1ps}
	\fe
	Similarly the 16 supercharges (supermomenta) are represented as follows on one-particle states,
	\ie 
	q_\A=\lambda_{\A I} \eta^I\,,\quad \bar q_\A=\lambda_{\A I} {\partial \over \partial \eta_I}\,,
	\label{supercharge1ps}
	\fe
	which transform as a doublet with respect to $SU(2)_R$ given in \eqref{SU2R1ps}, and generate the 7d $\cN=1$ Poincare supersymmetry algebra,
	\ie 
	\{q_\A,\bar q_\B\}={1\over 2} p_\m \gamma^\m_{\A\B}\,,\quad \{q_\A,q_\B\}=\{\bar q_\A,\bar q_\B\}=0\,.
	\label{ossusy}
	\fe
	
	In a general $n$-point scattering amplitude, we associate to each particle a spinor-helicity variable $\lambda_{\A I}^i$ and Grassmann variables $\eta^i_I,\widetilde\eta^i_I$, together they define the super-spinor-helicity variables for the $i$th particle. Correspondingly,
	the $SU(2)_R$ and supersymmetry generators from \eqref{SU2R1ps} and \eqref{supercharge1ps} for each individual particle are denoted as $r^i_{\pm,0}$ and $q^i_\A,\bar q^i_\B$. We define the total R-symmetry generator and supercharge (supermomentum) by their sum,
	\ie 
	R_{\pm,0}=\sum_{i=1}^n r_{\pm,0}^i\,,\quad Q_\A =\sum_{i=1}^n q^{i}_\A \,,\quad \bar Q_\A =\sum_{i=1}^n \bar q^{i}_\A\,.
	\fe
	It is obvious that they obey the desired $SU(2)_R$ and supersymmetry algebras.
	
	The $n$-point superamplitude ${\mf A}_n(\lambda^i,\eta^i,\widetilde \eta^i)$
	is a Lorentz invariant and little group invariant function of the super-spinor-helicity variables for each external particle. It is
	a convenient way to package together the component amplitudes related by supersymmetry. Correspondingly, the superamplitude is constrained by the supersymmetry Ward identities
	\ie 
	Q_\A {\mf A}_n(\lambda^i,\eta^i,\widetilde \eta^i)=\bar Q_\A {\mf A}_n(\lambda^i,\eta^i,\widetilde \eta^i)=0\,,
	\label{SUSYWard}
	\fe
	which in particular imply momentum conservation as a consequence of the supersymmetry algebra. CPT conjugation (induced by its action on the one-particle states) in general acts nontrivially on the superamplitude, in particular changing its degree in the Grassmann variables $\eta^i_I$,
	\ie 
	{\rm deg}_\eta ({\mf A}_n^{\rm CPT})=4n-{\rm deg}_\eta ({\mf A}_n)\,,
	\label{CPT}
	\fe 
	which we will make use of later.

	We define supervertices ${\mf V}(\lambda^i,\eta^i,\widetilde \eta^i)$  as superamplitudes that are analytic in the super-spinor-helicity variables, as they capture local contact interactions related by supersymmetry \eqref{ossusy}. Via factorization and unitarity cuts, these supervertices are the basic building blocks for general superamplitudes. 
	
	Focusing on the supervertices, the general solutions to the Ward identities \eqref{SUSYWard} come in two families, one known as D-term supervertices,
	\ie 
	{\mf V} =\D^7(P) \D^8(Q) {\bar Q}^8 \cP(\lambda^i,\eta^i,\widetilde \eta^i)\,,\quad \D^8(Q)\equiv \prod_{\A=1}^8Q_{\A}\,,~\bar Q^8\equiv \prod_{\A=1}^8\bar Q_{\A}\,,
	\label{Dtermver}
	\fe
	where $\cP(\lambda^i,\eta^i,\widetilde \eta^i)$ is a general polynomial in the super-spinor-helicity variables that is Lorentz and little group invariant.
	Here $\D^7(P)$ is the usual momentum conservation delta function for the total momentum $P=\sum_{i=1}^n p_i$, while $\D^8(Q)$ and ${\bar Q}^8$ are the analogous delta functions for the super-momentum (supercharge) conservation. It solves the Ward identity \eqref{SUSYWard} immediately. We call \eqref{Dtermver} a D-term supervertex by analogy with the supersymmetry invariants constructed from full superspace integrals of superfields in the Lagrangian formalism.
	
	While the D-term supervertices proliferate at derivative order 8 and above, there is another type of supervertices that solve the Ward identities \eqref{SUSYWard} at low derivative orders. They are known as the F-term supervertices (again in analogy with F-term invariants in the supersymmetric Lagrangian), which take the following form,
	\ie 
	{\mf V}=\D^7(P) \D^8(Q)  \cF(\lambda^i,\eta^i,\widetilde \eta^i)\,,
	\label{Ftermver}
	\fe
	where $\cF(\lambda^i,\eta^i,\widetilde \eta^i)$ is a Lorentz and little group invariant polynomial in its variables such that \eqref{Ftermver} is annihilated by $\bar Q_\A$ as in \eqref{SUSYWard} and \eqref{Ftermver} cannot be written as a D-term supervertex as in \eqref{Dtermver}. In the following we will describe the F-term supervertices for both vector and graviton multiplets in the 7d $\cN=1$ supergravity.
	
	\subsection{Supervertices for vector multiplets}
	
	We start with the F-term supervertices for vector multiplets. As one would expect, this should capture the Yang-Mills interaction $\Tr F^2$ for non-Abelian gauge fields (and their super-partners) in 7d, together with the leading higher derivative corrections (such as $\Tr F^4$). 
	
	For $n$-point with $n\geq 4$, such supervertices are straightforward to write in the super-spinor-helicity formalism introduced previously. The leading F-term supervertices come at derivative order 4 and 6. For our purpose here, we list the corresponding 4-point vertices below (the generalization to $n>4$ is immediate),
	\ie \label{fvertices}
	{\mf V}_{F^4}=&\,\D^7(P)\D^8(Q)f^{(4)}_{a_1a_2 a_3a_4} \,,\quad 
	\\
	{\mf V}_{D^2F^4}=&\,\D^7(P)\D^8(Q)
	\left(
	f^{(6)}_{a_1a_2, a_3a_4} s 
	+
	f^{(6)}_{a_1a_3, a_2a_4} t
	+
	f^{(6)}_{a_1a_4 ,a_2a_3} u \right)\,,
	\fe
	where $a_i$ is the color index that labels the individual vector multiplets, $f^{(4)}_{a_1a_2a_3a_4},f^{(6)}_{a_1a_2,a_3a_4}$ are color structures and $s,t,u$ are the usual Mandelstam invariants. For notational simplicity, the color indices are suppressed for ${\mf V}_{D^{2k}F^4}$ on the LHS in \eqref{fvertices}.
	
	The three-point supervertices are more intricate due to the nontrivial kinematical constraints. As in \cite{Lin:2015dsa}, we will work in the frame where the three in-going momenta $p_1,p_2,p_3$ lie in a null-plane spanned by $e^0+e^1$ and $e^2+ie^3$ (where $e^\m$ denotes the usual orthonormal basis for $\mR^{1,6}$). Explicitly, we parametrize the external momenta by
	\ie 
	p_1=(p_1^+,p_1^+,0,0,\vec 0)\,,\quad p_2=(0,0,p_2^+,ip_2^+,\vec 0)\,,\quad p_3=-(p_1^+,p_1^+,p_2^+,ip_2^+,\vec 0)\,,
	\fe
	To solve the kinematic constraints on the spinor-helicity variables, it's convenient to use a different basis for the 7d spinor as $\A=(\pm\pm\pm)$ such that the $\pm$ labels the eigenvalue with respect to the rotation (boost) in the $01$, $23$ and $45$ plane respectively. We thus find from \eqref{spinorhelicity} that in this frame,
	\ie 
	\lambda^i_{--\pm}= \lambda^1_{-+\pm}=\lambda^2_{+-\pm}=0\,.
	\label{spinorhelicityconstraint}
	\fe
	Consequently $Q_{--\pm}$ and $\bar Q_{--\pm}$ vanishes in this frame and we are led to the following SUSY invariant (solving \eqref{SUSYWard})
	\ie 
	\prod_{s=\pm}\left( Q_{++ s}Q_{+-s}Q_{-+s} \right)\,.
	\fe
	The three-point supervertex for the vector multiplets is then given by
	\ie 
	\prod_{s=\pm}\left( Q_{++ s}Q_{+-s}Q_{-+s} \right)\cF_{a_1 a_2 a_3}(\lambda^i,\eta^i)
	\fe
	where $\cF_{a_1 a_2 a_3}$ must be annihilated by $\bar Q_{\A}$ up to terms proportional to $Q_{\A}$ and the dependence on the super-spinor-helicity variables are further constrained by Lorentz and little group invariance. Consistency with the CPT conjugation \eqref{CPT} demands that
	$\cF_{a_1 a_2 a_3}$ is degree 0 in $\eta$. Consequently, we arrive at the unique cubic supervertex (up to an overall normalization) below, which appears at the derivative order 1,
	\ie 
	{\mf V}_{F^2}=\D^7(P)
	{1\over p_1^+ p_2^+}\prod_{s=\pm}\left( Q_{++ s}Q_{+-s}Q_{-+s} \right)f_{a_1 a_2 a_3}\,,
	\fe 
	where $f_{a_1 a_2 a_3}$ denotes the cubic color structure. As before, the color indices are suppressed for ${\mf V}_{F^2}$ on the LHS.
	
	\subsection{Supervertices involving graviton multiplets}
	
	We now include graviton multiplets among the external legs of the supervertices. Once again, it is straightforward to derive the $n$-point F-term supervertices for $n\geq 4$. As was the case for the one-particle states, these supervertices are obtained from those for the vector multiplets by multiplying suitable monomials in the auxiliary Grassmann variables $\widetilde \eta_i^I$. For convenience, we introduce the following little group invariants as in \cite{Lin:2015dsa},
	\ie 
	\widetilde q_i^\A \equiv \lambda^\A_{i I}\widetilde \eta_i^I\,, \quad \widetilde q_{ij}\equiv \widetilde q_i^\A \widetilde q_j^\B \cC_{\A\B}\,.
	\label{tq}
	\fe 
	Note that $\widetilde q_{ij}$ is antisymmetric in $i,j$. The degrees of the supervertex in $\widetilde \eta_i^I$ are fixed, for each $i$, to be either 0 (for vector multiplet) or 2 (for graviton multiplet).
	
	There is  a unique (up to an overall normalization) 4-point  $F$-term mixed graviton-vector supervertex,
	\ie \label{R2F2ver}
	{\mf V}_{D^{2k}R^2F^2}=&\,\D^7(P)\D^8(Q) \delta_{a_3a_4} \widetilde q_{12}^2 s^k\,,
	\fe 
	at derivative orders $2k+6=6,8,10$  respectively, which correspond to the $D^{2k}R^2 \Tr F^2$ interactions in the supergravity Lagrangian. The D-term supervertices with the same external legs start proliferating at the 12-th derivative order (i.e. $D^6R^2 \Tr F^2$).
	
	For pure graviton scattering, there is a unique supervertex at derivative order $8$,
	\ie 
	{\mf V}_{R^4}=&\,\D^7(P)\D^8(Q)  (\widetilde q_{12}^2\widetilde q_{34}^2+\widetilde q_{13}^2\widetilde q_{24}^2+\widetilde q_{14}^2\widetilde q_{23}^2)\,,
	\fe 
	which corresponds to the $R^4$ interactions. Note that the D-term supervertices at derivative order $12$ (and higher) includes the $D^4R^4$ interactions which are ${1\over 4}$-BPS in maximal supergravity.
	
	At three-point, we work in the same kinematic frame as in the last subsection. A similar analysis shows that there is a unique (up to an overall normalization) graviton three-point supervertex at 2-derivative order,
	\ie 
	{\mf V}_{R}=\D^7(P){1\over (p_1^+)^3 p_2^+}\prod_{s=\pm}\left( Q_{++ s}Q_{+-s}Q_{-+s} \right)  \widetilde q_{1\A_1} \widetilde q_{1\A_2}\widetilde q_{2\A_3} \widetilde q_{2\A_4}\widetilde q_{3\A_5} \widetilde q_{3\A_6} {\hat\ep}^{\A_1\A_2\A_3\A_4\A_5\A_6}\,,
	\fe
	where ${\hat\ep}$ is defined by the rank-8 antisymmetric invariant tensor $\ep$ for the spinor representation of $SO(1,7)$ where two indices are restricted to $\A=--+$ and $\A=---$. The powers of $p_1^+$ and $p_2^+$ above are fixed by invariance under the boost in the 01 plane and rotation in the 23 plane.
	
	There is also a unique (up to an overall normalization) mixed graviton-vector-vector supervertex at 2-derivative order
	\ie
	{\mf V}_{D^{-2}RF^2}=\D^7(P){1\over (p_1^+)^2 }\prod_{s=\pm}\left( Q_{++ s}Q_{+-s}Q_{-+s} \right) \widetilde q_{1+-+} \widetilde q_{1 +--} \D_{a_2 a_3}\,,
	\label{2dRF2}
	\fe
	and  there are no graviton-graviton-vector supervertices.
	
	\subsection{Supervertices with 11d gravitons}
	So far we have focused on the supervertices for particles with momenta restricted to the 7d subspace, which is the fixed singular locus of the M-theory orbifold. More generally, the scattering amplitudes can involve particles carrying momenta transverse to this 7d subspace, in particular, from the 11d graviton multiplet. This gives rise to a larger family of supervertices that describe contact interactions between the 11d modes in the bulk and the 7d modes localized at the singularity, preserving the 7d $\cN=1$ supersymmetry. 
	
	Below we discuss such supervertices at leading derivative orders (of the F-term type) which contribute to perturbative superamplitudes even when all external momenta are restricted to the 7d locus (more specifically the four-point amplitudes of 7d graviton and vector multiplets). For our purpose, we are interested in amplitudes of derivative order 6 or less. It then suffices to focus on three-point supervertices as the four-point ones have already been classified in the previous subsections and two-point ones can only appear as an internal vertex and thus contribute at higher derivative order.\footnote{For example, the $R^2$ supervertex at the singularity, a close analog of the brane coupling in \cite{Lin:2015ixa}, would vanish when the external momenta are restricted to the 7d subspace and can first contribute as an internal vertex to the four-gluon scattering amplitude at the 12-derivative order, and to the mixed gluon-graviton four-point amplitude at the 8-derivative order.}
	
	By general argument, neither the graviton three-point and nor the gluon three-point supervertices receive higher derivative corrections. The mixed graviton-gluon supervertex \eqref{2dRF2} however has a simple higher derivative generalization following the analysis in \cite{Lin:2015ixa}, 
	\ie 
	{\mf V}_{D_\perp^{2k} D^{-2}RF^2}={\mf V}_{D^{-2}RF^2}\times (p_{1\perp}^2)^{k}\,,
	\label{RF2hder}
	\fe 
	where the 11d momentum of the graviton multiplet is split into longitudinal and transverse components to the singularity as in $p_1=(p_{1\parallel},p_{1\perp})$. It obviously solves the Ward identities \eqref{SUSYWard}. More precisely \eqref{RF2hder} is the mixed supervertex where the polarizations for the 11d  graviton multiplet are restricted to those in \eqref{gravitonmultiplet}. For general 11d polarizations, the supervertex is obtained from \eqref{RF2hder} by multiplying monomials in $\tilde \eta$ and imposing invariance under the 11d little group $SO(9)$.\footnote{We will not attempt to the write the full expression here as there is no convenient spinor-helicity formalism in 11d (see \cite{Wang:2015jna}).}
	The supervertex \eqref{RF2hder}
	contributes to the four-point functions of gluons and gravitons via tree diagrams. 
	
	\subsection{Loop amplitudes}
	\label{loopApp}
	
	The 1-loop 4-point superamplitude of 7d SYM is \cite{Bern:1996je} 
	\ie\label{a1loop}
	\mathfrak{A}_{F^2|F^2} &= i g_{\rm YM}^4 \delta^7(P) \delta^8(Q) \Big[ \mathtt{d}_{st} I(s,t)+\mathtt d_{su} I(s,u)+\mathtt d_{tu} I(t,u)
	\Big]\,
	\fe
	where $I(s,t)$ is the scalar box integral defined in \cite{Bern:1997nh}, and the flavor structure $\mathtt{d}$ is given in the case of $SU(k)$ gauge group by \cite{Bern:2010tq},\footnote{\label{convertt}Here we have converted the gauge indices $a_1, a_2, a_3, a_4$ to $a,b,c,d$.}
	\es{dst}{
		\mathtt{d}_{st}&=2({\mathtt A}^{abcd}+{\mathtt A}^{acdb}+{\mathtt A}^{adbc}+k {\mathtt B}^{abcd})\\
		\mathtt d_{su}&=2({\mathtt A}^{abcd}+{\mathtt A}^{acdb}+{\mathtt A}^{adbc}+k {\mathtt B}^{acdb})\\
		\mathtt d_{tu}&=2({\mathtt A}^{abcd}+{\mathtt A}^{acdb}+{\mathtt A}^{adbc}+k {\mathtt B}^{adbc})\,.\\
	}
	The 2-loop superamplitude is  \cite{Bern:1997nh,Bern:1998ug,Bern:2010tq}
	\ie\label{A2loop}
	\mathfrak{A}_{F^2|F^2|F^2} &=- 2 g_{\rm YM}^6 \delta^7(P) \delta^8(Q) \left( s\mathtt e_{st} I_{1234}^P+s\mathtt f_{st} I_{1234}^{NP} +({\rm permutations~on~}2,3,4)\right)\,,
	\fe
	where $I_{1234}^P$ and $I_{1234}^{NP}$ are 2-loop scalar planar and non-planar box-type integrals defined in \cite{Bern:1997nh}, $\mathtt e_{st}$ and $\mathtt f_{st}$ are the following invariant tensors (converting gauge indices as in footnote \ref{convertt})
	\es{strucs2loop}{
		\mathtt e_{st}&=3k {\mathtt A}^{abcd}+(2+k^2){\mathtt B}^{abcd}-4{\mathtt B}^{adbc}+2{\mathtt B}^{acdb}\,,\\
		\mathtt f_{st}&=k(2 {\mathtt A}^{abcd}-{\mathtt A}^{acdb}-{\mathtt A}^{adbc})+2{\mathtt B}^{abcd}+2{\mathtt B}^{acdb}-4{\mathtt B}^{adbc}\,.\\
	}
	%  \es{strucs2loop}{
	% \mathtt e_{s_1}&=3k {\mathtt A}^{abcd}+(2+k^2){\mathtt B}^{abcd}+2{\mathtt B}^{adbc}-4{\mathtt B}^{acdb}\,,\\
	%  \mathtt e_{s_2}&=3k {\mathtt A}^{abcd}+2{\mathtt B}^{abcd}+(2+k^2){\mathtt B}^{adbc}-4{\mathtt B}^{acdb}\,,\\
	%   \mathtt e_{t_1}&=3k {\mathtt A}^{adbc}+(2+k^2){\mathtt B}^{abcd}+2{\mathtt B}^{acdb}-4{\mathtt B}^{adbc}\,,\\
	%  \mathtt e_{t_2}&=3k {\mathtt A}^{adbc}+2{\mathtt B}^{abcd}+(2+k^2){\mathtt B}^{acdb}-4{\mathtt B}^{adbc}\,,\\
	%   \mathtt e_{u_1}&=3k {\mathtt A}^{acdb}+(2+k^2){\mathtt B}^{acdb}+2{\mathtt B}^{adbc}-4{\mathtt B}^{abcd}\,,\\
	%  \mathtt e_{u_2}&=3k {\mathtt A}^{acdb}+2{\mathtt B}^{acdb}+(2+k^2){\mathtt B}^{adbc}-4{\mathtt B}^{abcd}\,,\\
	%  \mathtt f_{s}&=2k {\mathtt A}^{abcd}-k{\mathtt A}^{acdb}-k{\mathtt A}^{adbc}+2{\mathtt B}^{abcd}+2{\mathtt B}^{acdb}-4{\mathtt B}^{adbc}\,,\\
	%  \mathtt f_{t}&=2k {\mathtt A}^{abcd}-k{\mathtt A}^{acdb}-k{\mathtt A}^{adbc}+2{\mathtt B}^{abcd}+2{\mathtt B}^{acdb}-4{\mathtt B}^{adbc}\,,\\
	%  \mathtt f_{u}&=2k {\mathtt A}^{abcd}-k{\mathtt A}^{acdb}-k{\mathtt A}^{adbc}+2{\mathtt B}^{abcd}+2{\mathtt B}^{acdb}-4{\mathtt B}^{adbc}\,.\\
	% }
	The 4-gluon amplitude with tree-level bulk graviton exchange can be determined by unitarity cut in a similar manner,\footnote{The relative sign between (\ref{arloopint}) and (\ref{A2loop}) is most easily determined by comparing the tree-level gluon exchange versus graviton exchange amplitudes, and using equation (3.1) of \cite{Bern:1998ug} relating the 2-loop gluon exchange to the tree amplitude. Note that the our convention for the ${\rm SU}(k)$ generator is related by $T^a={1\over \sqrt{2}} T^a_{\rm there}$ to those of \cite{Bern:1997nh,Bern:1998ug} and the normalizations of the polarizations are such that the supermomentum conservation delta function is related the color-ordered tree amplitude there by $\D^8(Q)=4 st A_4^{\rm tree}(1,2,3,4)$ . }
	\ie\label{arloopint}
	\mathfrak{A}_R = {1\over 2}\kappa^2 \delta^7(P)\delta^8(Q) \Big[ {\mathtt A}^{a_1a_2a_3a_4} k \int_{\mathbb{R}^4} {d^4p_\perp\over (2\pi)^4} {1\over -s + p_\perp^2} + ({\rm cyclic~ permutations~on~}2,3,4) 
	\Big],
	\fe
	which, up to local counter terms, evaluates to (\ref{arloop}). The appearance of the factor $k$ is due to the sum over $\mathbb{Z}_k$ images in the bulk graviton propagator in position space. The overall sign and normalization is fixed against the tree-level gluon exchange superamplitude (\ref{af2super}).

	\section{Formulae for superblocks and Mellin amplitudes}
	\label{fourApp}
	
	We will make use of the explicit superblocks for the flavor multiplet $\cD[1]$, the stress tensor multiplet $\cB[0]$, and the $\Delta=2$ half-BPS multiplet $\cD[1]$. These superblocks take the form \cite{Chang:2017xmr,Bobev:2017jhk,Chang:2019dzt}
	\es{superblocks}{
		\mathfrak{G}_{\mathcal{D}[1]} (U,V;\A) =&P_1\left({2}{\alpha}-1\right)G_{1,0}(U,V)-G_{2,1}(U,V)\,,\\
		\mathfrak{G}_{\mathcal{D}[2]}(U,V;\A)=&P_2\left({2}{\alpha}-1\right)G_{2,0}(U,V)-\frac43P_1\left({2}{\alpha}-1\right)G_{3,1}(U,V)+\frac{16}{45}G_{4,0}(U,V)\,,\\
		\mathfrak{G}_{\mathcal{B}[0]}(U,V;\A)=&G_{1,0}(U,V)-2P_1\left({2}{\alpha}-1\right)G_{2,1}(U,V)+G_{3,2}(U,V)\,,\\
	}
	where $P_r(x)$ is the $r$-th Legendre polynomial, and $G_{\Delta,j}(U,V)$ are ordinary 3d conformal blocks of primaries of weight $\Delta$ and spin $j$ (here we adopt the notation of \cite{Hogervorst:2013sma}). These conformal blocks can be efficiently expressed in terms of  the $r,\eta$ variables \cite{Hogervorst:2013sma,Kos:2013tga}, defined as 
	\es{reta}{
		U=\frac{16r^2}{(1+r^2+2r\eta)^2}\,,\qquad V=\frac{(1+r^2-2r\eta)^2}{(1+r^2+2r\eta)^2}\,,
	}
	and we normalize our blocks as $G_{\Delta,\ell}(r,\eta)\sim (4r)^{\Delta} P_{\ell}(\eta)$ as $r\to0$.

	\label{melApp}

	The singular part of the Mellin space conformal blocks, $G^M_{\Delta,j}(s,t)$, appearing in (\ref{mero}), defined with the same normalization convention as that of the position space blocks $G_{\Delta,j}(U,V)$,
	are
	\es{Mblocks}{
		G^M_{1,0}(s,t)=&\frac{\Gamma \left(\frac{1}{2}-\frac{s}{2}\right)}{4 \pi ^{5\over 2} \Gamma
			\left(1-\frac{s}{2}\right)}\,,\\
		G^M_{2,1}(s,t)=&-\frac{(s+2 t-4) \left(\frac{2 \sqrt{\pi }}{s}+\frac{\Gamma
				\left(\frac{1}{2}-\frac{s}{2}\right) \Gamma \left(-\frac{s}{2}\right)}{\Gamma
				\left(1-\frac{s}{2}\right)^2}\right)}{8 \pi ^{5\over 2}}\,,\\
		G^M_{3,2}(s,t)=&\frac{\left(s^2+8 s t-14 s+8 t^2-32 t+32\right) \left(-\frac{2 \sqrt{\pi
			}}{s}+\frac{\sqrt{\pi }}{s+2}+\frac{\Gamma \left(-\frac{s}{2}-1\right) \Gamma
				\left(\frac{1}{2}-\frac{s}{2}\right)}{\Gamma
				\left(1-\frac{s}{2}\right)^2}\right)}{16 \pi ^{5\over 2}}\,.
	}

	\section{Converting Mellin amplitudes to OPE coefficients}
	\label{app:extraction}
	
	The algorithm \cite{Chester:2018lbz} for extracting the contribution of (a component of) the Mellin amplitude $M_*(s,t;\A)$ to the OPE coefficient of a given superblock is as follows. We begin by noting that the contribution of $M_*(s,t;\A)$ of interest to the CFT correlator, i.e.~the Mellin transform of $M_*(s,t;\A)$ via (\ref{mellinDef}), admits an expansion in terms of superblocks and their derivatives of the form
	\es{Aexpansion}{
		\mathcal{G}_*(U,V;\alpha)= \sum_{\cM_{\Delta,j}} \left[ a^{(1)}_\cM \mathfrak{G}_\cM (U,V;\alpha)+ a_\cM^{(0)} \Delta_\cM^{(1)}\partial_{\Delta}\mathfrak{G}_\cM(U,V;\alpha)\right] ,
	}
	where the sum is over superconformal representations $\cM$ labeled by weight $\Delta$ and spin $j$ of superprimaries of the GFFT, $a_\cM^{(0)}$ are the OPE coefficients of the GFFT correlator, $a_\cM^{(1)}$ and $\Delta_\cM^{(1)}$ represent the first order corrections to the OPE coefficients and conformal weight due to the component of Mellin amplitude $M_*$.\footnote{Here we work only to first order in perturbation theory in the bulk effective couplings.  To this order, only tree-level Witten diagrams (but not restricted to those of minimal couplings) contribute to the Mellin amplitude.} The superblocks can be further expanded in bosonic conformal blocks as
	\ie
	\mathfrak{G}_\cM(U,V;\alpha) = \sum_{(\Delta,j,r)\in\cM} A_{\cM|\Delta,j,r} P_r(2\A-1) G_{\Delta,j}(U,V).
	\fe
	The lightcone expansion of the conformal block at small $U$ and finite $V$ takes the form
	\es{lightBlocksExp}{
		G_{\Delta,j}(U,V)=\sum_{k=0}^\infty U^{\frac{\Delta-j}{2}+k}g_{\Delta,j}^{[k]}(V) ,
	}
	where $g_{\Delta,j}^{[k]}(V)$ are lightcone blocks that behave as $(1-V)^{j-2k}$ in the $V\to 1$ limit. In particular, the leading lightcone block in our normalization convention is
	\es{lightconeBlock}{
		g_{\Delta,j}^{[0]}(V)&=\frac{\Gamma(j+{1\over 2})}{4^\Delta\sqrt{\pi}j!}(1-V)^j \,{}_2F_1\left(\frac{\Delta+j}{2},\frac{\Delta+j}{2},\Delta+j,1-V\right)\,.\\
	}
	We can evaluate the coefficient of $U^{\frac{\Delta-j}{2}+k}$ in the lightcone expansion of $\mathfrak{G}_\cM(U,V;\alpha)$ by taking the residue of $M_*(s,t;\A)$ at the pole $s=\Delta-j+2k$, and the remaining $t$-integral of the Mellin transform can be performed by summing over residues. Finally, we can project onto the leading lightcone block, as a function of $V$, using the orthogonality relation \cite{Heemskerk:2009pn}
	\es{ortho}{
		\delta_{x,x'}&=-\oint_{V=1} \frac{dV}{2\pi i}(1-V)^{x-x'-1} F_x(1-V) F_{1-x'}(1-V)\,,\\
		F_x(y)&\equiv {}_2 F_1(x,x,2x,y)\,,
	}
	where the integration contour encircles only the pole at $V=1$, thereby extracting the contribution to the OPE coefficient $a_\cM^{(1)}$.
	
	\section{Matrix model calculations}
	\label{locApp}
	
	In this appendix we will show how to compute the mass deformed sphere free energy using the Fermi gas method of \cite{Marino:2011eh}. We start by using the Cauchy determinant formula to rewrite the partition function in \eqref{Z} as 
	\es{Z2}{
		Z(k,N)=&\frac{1}{N!}\sum_{\sigma\in S_N}(-1)^\sigma\int d^Nx\prod_{i=1}^N\rho(x_i,x_{\sigma(i)})\,,\\
		\rho(x_1,x_2)=&\frac{1}{2\cosh[\pi (x_1-x_2-M)]}\prod_{I=1}^{k}\frac{1}{2\cosh[\pi(x_1+\sum_{\alpha=1}^{k-1}m_{\alpha,I})]}\,.
	}
	We can then use the canonical quantum operators
	\es{qp}{
		[\hat q,\hat p]=i\hbar\,,\qquad \hbar=\frac{1}{2\pi}\,,
	}
	to write the density as the expectation value 
	\es{rho2}{
		\rho(x_1,x_2)=\langle x_1|  \hat\rho |x_2\rangle\,,\qquad  \hat\rho=e^{-\frac {U(\hat q)}{2}}  e^{-T(\hat p)}  e^{-\frac {U(\hat q)}{2}}\,,
	}
	where we have conjugated so as to write $\rho$ in a symmetric form, and we define
	\es{UT}{
		U(q)\equiv\sum_{I=1}^{k}\log[2\cosh[\pi(q+\sum_{\alpha=1}^{k-1}m_{\alpha,I})]]\,,\qquad T(p)=\log[2\cosh[\pi p]]-2\pi i Mp\,.
	}
	The partition function now looks like a gas of non-interacting Fermions with the unconventional Hamiltonian in \eqref{UT}, which we can compute perturbatively in $\hbar$. To do this we introduce the grand potential $J(\mu)$ as
	\es{ZtoJ}{
		Z = \int\frac{d\mu}{2\pi i} e^{J(\mu)-\mu N}\,.
	}
	We will find that the grand potential takes the form
	\es{J}{
		J(\mu)=\frac{C}{3}\mu^3+B\mu+A+O(e^{-\mu})\,,
	}
	so that performing the $\mu$ integral gives the Airy function behavior in \eqref{Zmfinal2}, where $O(e^{-\mu})$ correspond to $O(e^{-N})$ corrections to $Z$ that we will ignore.
	
	A convenient way of computing the grand potential from the Fermi gas is to use the Mellin-Barnes representation \cite{Hatsuda:2015oaa}
	\es{phaseFull}{
		J(\mu)=-\int\frac{dt}{2\pi i}\Gamma(t)\Gamma(-t)\cZ(t)e^{t\mu}\,,
	}
	where the contour includes the right half of the complex plane for $\mu<0$. The single particle partition function here is
	\es{Zt}{
		\cZ(t)=\int\frac{dpdq}{2\pi \hbar} \left(\hat\rho^{-t}\right)_W\,,
	}
	where the Wigner transform of an operator $\cO$ is defined as
	\es{wigner}{
		(\cO)_W=\int dy\langle q-\frac y2|\cO|q+\frac y2\rangle e^{\frac{ipy}{\hbar}}\,.
	}
	We are interested in the perturbative in $\mu$ terms shown in \eqref{J}, which can be extracted from \eqref{phaseFull} by taking the pole $t=0$. There are two kinds of $\hbar$ corrections to $\left(\hat\rho^{-t}\right)_W$. The first comes from the Baker-Campbell-Hausdorff formula in the definition of $\hat\rho$ in \eqref{rho2}, including these corrections defines the quantum Hamiltonian $H_W$:
	\es{quantH}{
		H_W\equiv-\log_\star \rho_W=-\log_\star \left(e^{-\frac {U( q)}{2}}_W  \star e^{-T( p)}_W  \star e^{-\frac {U( q)}{2}}_W \right)\,,
	}
	where the BCH expansion here uses the Moyal star product
	\es{star}{
		\star =e^{\frac{i \hbar}{2}\left(\overset{\leftarrow}{\partial_q}\overset{\rightarrow}{\partial_p}-\overset{\leftarrow}{\partial_p}\overset{\rightarrow}{\partial_q}\right)}\,.
	}
	The second $\hbar$ correction comes from the fact that $H_W\star H_W\neq H_W^2$. If we decompose $e^{-t\hat H}=e^{-t H_W}e^{-t(\hat H-H_W)}$, then we can include these contributions by defining
	\es{kirk}{
		\mathcal{G}_r=\left((\hat H-H_W)^r\right)_W\,,
	}
	where this product is computed using
	\es{star2}{
		(\hat A\hat B)_W=A_W\star B_W\,.
	}
	These are called the Wigner-Kirkwood coefficients, and the lowest couple terms are
	\es{kirklow}{
		\mathcal{G}_2=&-\frac{\hbar^2}{4}\left[\frac{\partial^2 H_W}{\partial q^2}\frac{\partial^2 H_W}{\partial p^2}-\left(\frac{\partial^2 H_W}{\partial q\partial p}\right)^2\right]+O(\hbar^4)\,,\\
		\mathcal{G}_3=&-\frac{\hbar^2}{4}\left[  \left(\frac{\partial H_W}{\partial q}\right)^2  \frac{\partial^2 H_W}{\partial p^2}  + \left(\frac{\partial H_W}{\partial p}\right)^2  \frac{\partial^2 H_W}{\partial q^2}  -2\frac{\partial H_W}{\partial q}\frac{\partial H_W}{\partial p}\frac{\partial^2 H_W}{\partial q\partial p} \right]+O(\hbar^4)\,.\\
	}
	We can now write $\cZ(t)$ as
	\es{rho3}{
		\cZ(t)=\int\frac{dqdp}{2\pi \hbar}e^{-tH_W^{(0)}}e^{-t\sum_{s=2}\hbar^s H_W^{(s)}}\left(1+\sum_{r=2}^\infty\frac{(-t)^r}{r!} \sum_{s=2\lfloor\frac{r+2}{3}\rfloor}^\infty \hbar^s \mathcal{G}^{(s)}_r\right)\,,
	}
	where $\mathcal{G}^{(s)}_r$ and ${H}^{(s)}_W$ denotes the $O(\hbar^s)$ coefficient of each quantity.
	We then expand and collect the various terms that contribute at each order in $\hbar$. For instance, at quadratic order we have
	\es{quad}{
		\cZ(t)=\int\frac{dqdp}{2\pi \hbar}e^{-tH_W^{(0)}}\left(1-\hbar^2\left(tH_W^{(2)}-\frac{t^2}{2}\mathcal{G}_2^{(2)}+\frac{t^3}{6}\mathcal{G}_3^{(2)}\right)\right)+O(\hbar^4)\,.
	}
	After plugging in the explicit expression of the Hamiltonian from \eqref{rho2}, we find that it is difficult to compute the $p,q$ integrals for finite $m$, so we expand $U(q,m)$ to quartic order in mass for the masses $m_1$ and $m_3$, which is sufficient for the mass derivatives we consider in the main text. We find
	\es{Uexpand}{
		U(q,m_1)=&(k-2)\log[2\cosh[\pi q]]+\log[2\cosh[\pi (q+{m_1\over 2})]]+\log[2\cosh[\pi (q-{m_1\over 2})]]=\\
		=& k\log[2\cosh[\pi q]]+\frac{\pi^2m_1^2}{4}\sech[\pi q]^2+\frac{\pi^4 m_1^4}{16}\left( \frac13\sech[\pi q]^2-\frac12\sech[\pi q]^4 \right)+O(m_1^6)\,,\\
		U(q,m_3)=&(k-4)\log[2\cosh[\pi q]]+3\log[2\cosh[\pi (q+{m_3\over 2\sqrt{6}})]]+\log[2\cosh[\pi (q-{3m_3\over 2\sqrt{6}})]]\\
		=& k\log[2\cosh[\pi q]]+\frac{\pi^2m_3^2}{4}\sech[\pi q]^2+\frac23\sqrt{\frac{2}{3}}m_3^3\pi^3\csch[2\pi q]^3\sinh[\pi q]^4\\
		& +\frac{7\pi^4 m_3^4}{96}\left( \frac13\sech[\pi q]^2-\frac12\sech[\pi q]^4 \right)+O(m_3^6)\,.\\
	}
	The $m_3^3$ term will integrate to zero, so we can ignore it. Note that the only difference between the two masses is then a relative factor of ${7\over 6}$ for the quartic term, which is reflected in the result in \eqref{m4}. We now find that both the $p$ and $q$ integrals in \eqref{rho3} are of the form
	\es{pint}{
		&\int_{-\infty}^{\infty}dx \frac{e^{2 \pi i mt  x }}{(2\cosh(\pi x))^t}=\frac{\Gamma[{t\over 2}-imt]\Gamma[{t\over 2}+imt]}{2\pi\Gamma[t]}\,,\\
		&\int_{-\infty}^{\infty}dx \frac{e^{2 \pi i mt  x } \sinh(\pi x)}{(2\cosh(\pi x))^t}=imt\frac{\Gamma[{t-1\over 2}-imt]\Gamma[{t-1\over 2}+imt]}{2\pi\Gamma[t]}\,.\\
	}
	Using these identities we can compute \eqref{rho3} to a given order in $\hbar$, then take the $t=0$ pole in \eqref{phaseFull} to compute $A,B,C$ in \eqref{J}. We find that $B,C$ only receive quadratic corrections in $\hbar$, which is typically the case, and we in fact checked that \eqref{BC2} is exact in both $M$ and $m_\alpha$. The quantity $A$ receives corrections to all orders in $\hbar$, and computing the first few corrections gives \eqref{Amixed} and \eqref{m4}.

	\bibliographystyle{JHEP}
	\bibliography{3dn4Draft}

\end{document}